\documentclass[a4paper,10pt,english]{report}
\usepackage{babel,array,latexsym}
\usepackage{amsmath,amssymb,amsthm}
\include{epsf}
\newlength{\tw} 
\newlength{\thnew} 
\title{\vspace*{-4cm}{\Large  Universit\`{a} degli Studi di Napoli} \\ {\Large "Federico II"} \\~
\begin{figure}[ht]
\epsfxsize=1.3 in \centerline{\epsfxsize=1.3
in\epsffile{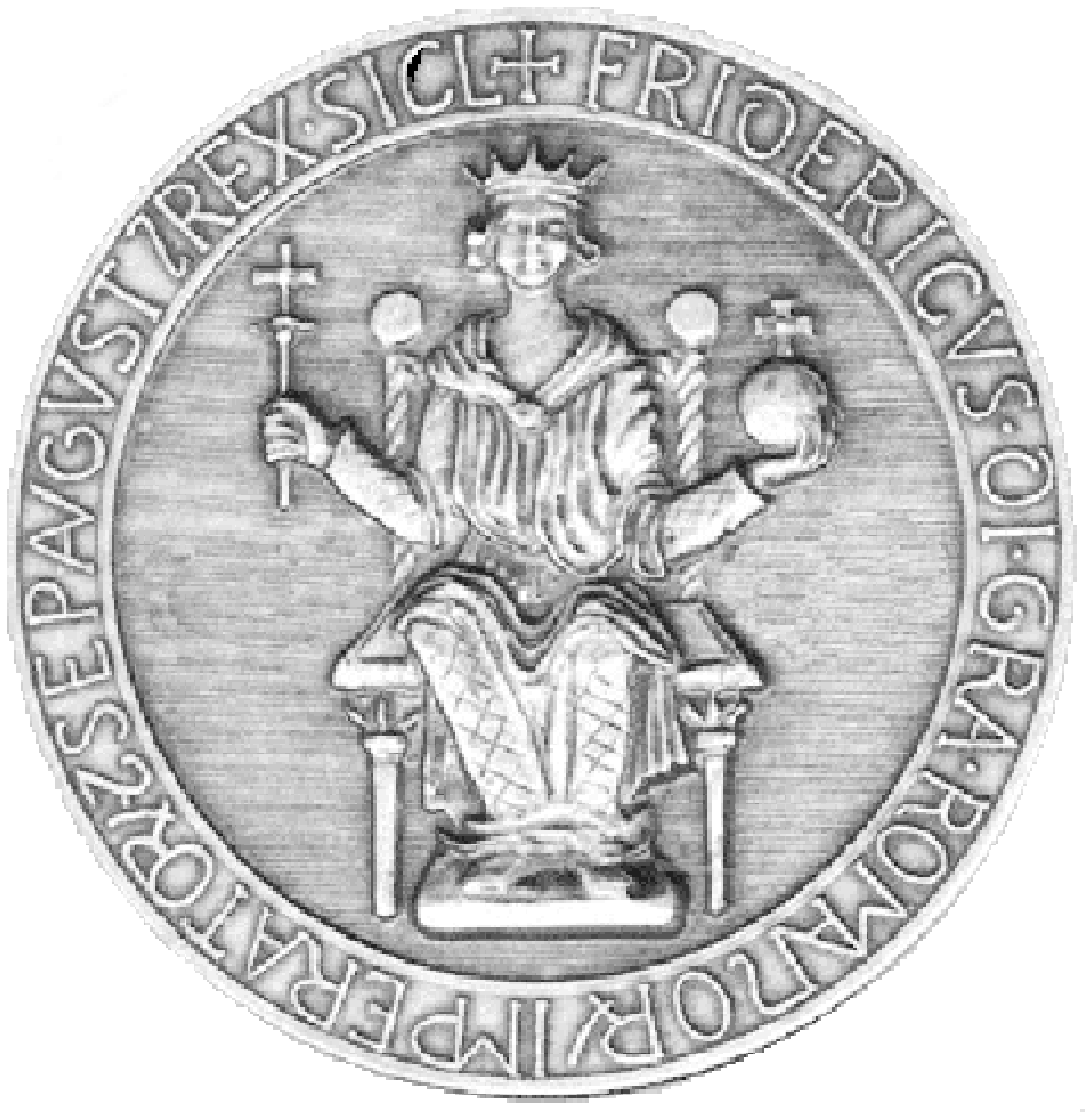}}
\end{figure}
\\~\\~
{\large Dottorato di Ricerca in Fisica Fondamentale ed Applicata}
\\ {\large XVII ciclo}
\\ ~\\ ~\\ ~\\
{\Large \bf Applications of}\\{\Large \bf the Weyl-Wigner
formalism to }\\{\Large \bf noncommutative~geometry}
\\~ \\~  
{\small Dissertation submitted for the degree of Philosophiae
Doctor}\\~
{\small November 2004}
\date{}
\author{}\\~ \\~ \\~  
\begin{flushleft}
  {\small candidate}
  \hfill {\small advisors}\\ {\large \bf Alessandro Zampini}   \hfill  {\large \bf  Prof. G. Marmo} \\ \hfill {\large \bf Prof. F. Lizzi}
\\~
\\~
\end{flushleft}}
\begin{document}
\newcommand{\bea}{\begin{eqnarray}}
\newcommand{\eea}{\end{eqnarray}}
\newcommand{\be}{\begin{equation}}
\newcommand{\ee}{\end{equation}}
\newcommand{\beqa}{\begin{eqnarray}}
\newcommand{\eeqa}{\end{eqnarray}}
\newcommand{\ba}{\begin{array}}
\newcommand{\ea}{\end{array}}
\newcommand{\beq}{\begin{equation}}
\newcommand{\eeq}{\end{equation}}
\newcommand{\til}{\tilde}
\newcommand{\pa}{\partial}
\newcommand{\law}{\leftarrow}
\newcommand{\raw}{\rightarrow}
\newcommand{\LAW}{\Leftarrow}
\newcommand{\RAW}{\Rightarrow}
\newcommand{\LRAW}{\Leftrightarrow}
\newcommand{\seq}{\simeq}
\newcommand{\disp}{\displaystyle}
\newcommand{\stac}{\stackrel}
\newcommand{\lab}{\label}
\newcommand{\gseq}{\stackrel{>}{\sim}}
\newcommand{\lseq}{\stackrel{<}{\sim}}
\newcommand{\la}{\langle}
\newcommand{\ra}{\rangle}
\newcommand{\del}{\partial}
\newcommand{\complex}{{\bb C}} 
\newcommand{\complexs}{{\bbs C}} 
\newcommand{\zed}{{\bb Z}} 
\newcommand{\nat}{{\bb N}} 
\newcommand{\real}{{\bb R}} 
\newcommand{\reals}{{\bbs R}} 
\newcommand{\zeds}{{\bbs Z}} 
\newcommand{\rat}{{\bb Q}} 
\newcommand{\mat}{{\bb M}} 
\newcommand{\mats}{{\bbs M}} 
\newcommand{\vecf}{\cal X}
\newcommand{\covecf}{\cal X^{*}}
\newcommand{\idop}{{\bf 1}} 
\newcommand{\der}{\cal D}
\font\mathsm=cmmi9 \font\mybb=msbm10 at 12pt
\def\bb#1{\hbox{\mybb#1}}
\font\mybbs=msbm10 at 9pt
\def\bbs#1{\hbox{\mybbs#1}}
\newcommand{\Hsta}{{\cal H}_{s}}
\newcommand{\Ao}{\hat{A}}
\newcommand{\Bo}{\hat{B}}
\newcommand{\qo}{\hat{q}}
\newcommand{\Do}{\hat{D}}
\newcommand{\Dod}{\hat{D}^{\dag}}
\newcommand{\No}{\hat{N}}\newcommand{\Ro}{\hat{R}}
\newcommand{\Ho}{\hat{H}}
\newcommand{\Uo}{\hat{U}}
\newcommand{\po}{\hat{p}}
\newcommand{\Vo}{\hat{V}}
\newcommand{\Po}{\hat{P}}
\newcommand{\Qo}{\hat{Q}}
\newcommand{\Go}{\hat{G}}
\newcommand{\So}{\hat{S}}
\newcommand{\Jo}{\hat{J}}
\newcommand{\Sod}{\hat{S}^{\dag}}
\newcommand{\Wod}{\hat{W}^{\dag}}
\newcommand{\Yo}{\hat{Y}}
\newcommand{\Uod}{\hat{U}^{\dag}}
\newcommand{\Fod}{\hat{F}^{\dag}}
\newcommand{\Wo}{\hat{W}}
\newcommand{\Fo}{\hat{F}}
\newcommand{\tW}{\tilde{W}}
\newcommand{\Lo}{\hat{L}}
\newcommand{\To}{\hat{T}}
\newcommand{\fo}{\hat{f}}
\newcommand{\go}{\hat{g}}\newcommand{\ouh}{{\cal U}\left({\cal H}\right)}
\newcommand{\baz}{\bar{z}}
\newcommand{\baw}{\bar{w}}
\newcommand{\md}{{\cdot}_{D}}
\newcommand{\omw}{\hat{\Omega}}
\newcommand{\omww}{\hat{\Omega}^{\left(w\right)}}
\newcommand{\oph}{{\cal O}p\left({\cal H}\right)}
\newcommand{\xo}{\hat{x}}
\newcommand{\yo}{\hat{y}}
\newcommand{\ao}{\hat{a}}
\newcommand{\aod}{\hat{a}^{\dag}}
\newcommand{\prm}{*_{\scriptscriptstyle M}}
\newcommand{\prmw}{*_{\scriptscriptstyle M\left(w\right)}}
\newcommand{\prmt}{*_{\scriptscriptstyle M}^{T}}
\newcommand{\fihm}{\frac{i\hbar}{2}}
\newcommand{\mqo}{{\hat{\cal Q}}}
\newcommand{\mpo}{{\hat{\cal P}}}
\newcommand{\bz}{\overline{z}}
\newcommand{\bu}{\overline{u}}
\newcommand{\bw}{\overline{w}}
\newcommand{\hil}{\mathcal{H}}
\newcommand{\fhil}{\mathcal{F}}
%
\maketitle
\begin{flushright}
{\bf Why noncommutative geometry?}
\end{flushright}

It is well known that the classical formulation for the dynamics of a
physical system is based on the notions of functions on a differentiable
manifold, of diffeomorphism group actions and of vector field.

Pure states are represented by points of a manifold, observables
by real functions on it. If this manifold is given both a
topological and a differentiable structure, then time evolution is
represented by a one parameter group of diffeomorphisms on it, whose
infinitesimal generator is a vector field.

This formalization originated from the analysis of the concepts of
position and velocity of a body, following the Newtonian's
assumptions. Classical mechanics developed its own language. This
language is differential geometry.

That this geometry were a branch of physics, quoting Einstein's
opinion, is perfectly clear looking at the way classical general
relativity and classical gauge theories describe gravitational and
electromagnetic interactions.

Quantum formalization for the dynamics of a physical system is
profoundly different. Pure states are represented by rays of a
separable Hilbert space; observables are represented by
self-adjoint operators on this space; time evolution is
represented by unitary transformations on the set of states.

In his book on the principle of quantum mechanics
\cite{diracprincipia}, Dirac wrote that one of the dominant
features of this scheme is that observables appear in it as
quantities which do not obey the commutative law of
multiplication. Moreover, noncommutativity among observables is
exactly the way uncertainty relations, whose appearance is one
among the most important differences between classical and quantum
physics, are introduced in the formalism.

The very first example of uncertainty relations is that related to
position and momentum observables for a quantum dynamics of point
particles. Its mathematical formulation is based on the definition
of canonical commutation relations: $$
\left[\qo^{a},\po_{b}\right]\,=\,i\hbar\delta^{a}_{b} $$ These
relations introduce a correlation among the dispersions of the
statistical distributions of measured values of positions and
momenta. In the paper \cite{dirac26} Dirac was led to consider the
possibility to describe quantum physics on the phase space
carrying a classical dynamics of point particles. The phase space
would have been recovered as the continuous spectra of a set of
fundamental quantum observables. He introduced the notion of
quantum algebra of functions, and of quantum analogue of classical
derivations, calling them quantum differentiations. Above all, he
was aware that the uncertainty relations drive to the
impossibility of an infinitely precise localization of points on
this phase space.

This example can be considered as the first noncommutative space.
The impossibility of such a perfect localization shows that the
``geometry'' of this space should be considered to have lost the
concept of point. ``Pointless geometry'' was exactly the name von
Neumann \cite{vNlandi} gave to the mathematical studies originated
by the analysis of the quantum formalism.

He started introducing the concept of rings of operators (nowadays
called von Neumann algebras) as a subalgebra of the algebra
$\mathcal{B}\left(\hil\right)$ of bounded operators on a Hilbert
space\footnote{A von Neumann algebra is a subalgebra of
$\mathcal{B}\left(\hil\right)$ which is closed under the
involution $\Ao\rightarrow\Ao^{\dagger}$, and sequentially
complete in the weak operator topology. This topology can be
defined by its notion of convergence. A sequence $\{\Ao_{n}\}$ of
bounded operators weakly converges to $\Ao$ when
$\la\psi\mid\Ao_{n}\mid\psi\ra\,\rightarrow\,\la\psi\mid
\Ao\mid\psi\ra$ for state vectors $\mid\psi\ra$ in the Hilbert
space $\hil$. This type of convergence is partly motivated by
quantum mechanics, in which $\la\psi\mid\Ao\mid\psi\ra$ represents
the expectation value of the observable represented by $\Ao$ when
the system is in the state represented by $\mid\psi\ra$, provided
that $\Ao$ is selfadjoint and $\la\psi\mid\psi\ra=1$.}. In this
context, for the first time he related a topological condition to
an algebraic one. His primary motivation came from the hope to
characterize quantum mechanical systems by algebraic conditions on
the observables, rather then topological analysis on the set of
states. This program would have been developed, and to some extent
realised, in algebraic quantum field theory, and in the study of
quantum statistical mechanics of infinite systems.

These research perspectives went on with the work of Gelfand who
combined studies of operator algebras with the theory of Banach
spaces. He introduced the notion of a Banach algebra in which
multiplication is separately continuous in the norm topology, and
formalised an intrinsic spectral theory. Then Gelfand and Neumark
defined what is now called a $C^{*}$-algebra, and proved the basic
theorem that each noncommutative $C^{*}$-algebra is isomorphic to
a norm closed $*$-subalgebra of bounded operators on a Hilbert
space. This theorem is now called GNS: its constructive proof was
developed by Segal, who established a connection between states
and representations. In \cite{segal47} there is a beautiful
account of an algebraic description of quantum mechanics.

These historical notes\footnote{A more refined and rigorous
account of these topics is in \cite{landsmann}.} are just to show
that quantum physics required the development of a specific
language. This language is noncommutative geometry, and
noncommutative topology.

This formalism is meant to give primary importance to the notion
of space (algebra) of observables, and to consider the state
vector as a derived object. The notion of pure states replaces that
of points, while derivations of the algebra replace vector fields.
The aim of noncommutative geometry is to reformulate the geometry
of a manifold in terms of features of the abelian algebra of
functions defined on it (there exists a version of GNS theorem
suited for commutative $C^{*}$-algebras), and then to generalize
the corresponding results of differential geometry to the case of
noncommutative algebras.

In the last years this program was evolved expecially by Connes,
who extended the notion of exterior calculus and of de Rham
cohomology to the noncommutative case. This, together with the
theory of Hilbert modules, generalising the notion of fiber
bundles, enables to study gauge theory in the noncommutative
setting. Moreover, if classical general relativity is beautifully
formalised using classical, say commutative, differential
geometry, noncommutative geometry appears, in the current research
activity, as the natural language to formalise quantum gravity,
and, more generally, a unified quantum description of fundamental
interactions.

\begin{flushright}
{\bf Why Weyl-Wigner formalism?}
\end{flushright}
After introducing the notions of quantum conditions (nowadays called commutation relations), Dirac went
on writing:
\begin{quote}
...The problem of finding quantum conditions is not of such a
general character...It is instead a special problem which presents
itself with each particular dynamical system one is called upon to
study. There is a fairly general method of obtaining quantum
conditions, applicable to a very large class of dynamical systems.
This is the method of \emph{classical analogy}... Those dynamical
systems to which this method is not applicable must be treated
individually and special considerations used in each case.

The value of classical analogy in the development of quantum
mechanics depends on the fact that classical mechanics provides a
valid description of dynamical systems under certain conditions,
when the particles and bodies composing the system are
sufficiently massive for the disturbance accompanying an
observation to be negligible. Classical mechanics must therefore
be a limiting case of quantum mechanics...
\end{quote}
His suggestion to consider a classical analogy is nowadays
referred to as a principle of correspondence. It is naively said
that a quantum system should be formalised in such a way that, in
the classical limit, formally obtained by letting
$\hbar\rightarrow 0$, the corresponding classical dynamics  could
be recovered. The Weyl-Wigner formalism is a setting convenient to
analyse the relations between the classical and the quantum
formalism, and to put in a more rigorous form the problem of
quantizing a classical dynamics, and of recovering the classical
limit of a quantum dynamics.

\begin{flushright}
{\bf and this dissertation...}
\end{flushright}

The theme of this dissertation is to use the Weyl-Wigner formalism
to study relations between quantum and classical physics, or, that
is the same, to study relations between quantum and classical
geometry.

In chapter \ref{chapone} the Weyl-Wigner formalism is introduced,
starting from a group theoretical interpretation of the cartesian
phase space as a carrier space for a classical dynamics. The
notion of Weyl system provides a more satisfactory account for the
definition of the quantum conditions, i.e. the commutation
relations for a certain class of quantum systems, and becomes a
way to analyse the formulation of the principle of classical
analogy.

Then the Weyl-Wigner map is introduced. It is a bijection between
a set of operators on a Hilbert space, and a set of functions on a
vector space. This map enables to write the noncommutative product
among quantum observables as a noncommutative product among
functions on this phase space. This is the Moyal product, and it
explicitly depends on $\hbar$, reducing to the standard pointwise
product in the limit $\hbar\rightarrow 0$. In this chapter it is
described how this formalism makes it possible to study both the
problem of quantizing certain classical systems, and the problem
of formalizing the quantum evolution in the space of functions
defined on the same phase space where classical observables are
represented. In this setting, a classical limit procedure is
written in a more rigorous form.

In chapter \ref{chapsecond} this formalism is extended to the case
of a quantum system corresponding to a classical dynamics whose
phase space is the cotangent bundle of a compact, simple, Lie group,
seen as a configuration space. Quantum conditions among
fundamental quantum observables are here mutuated by the Lie
algebra structure of the group. The novelty of this approach is
that a Weyl-Wigner isomorphism is now realised between operators
on a Hilbert space and functions which are defined not on the
"classical" phase space, namely that cotangent bundle of the Lie
group, but on the product of the group manifold, the configuration
space, with a discrete space. This can be seen as a sort of quantum phase
space. It explicitly depends on the global topological properties
of the group, and on its nonabelianess. These aspects play a role
to give the results of the specific harmonic analysis performed to
define the isomorphism.

Nevertheless, a different isomorphism between operators on a
Hilbert space and functions on the classical phase space can be
written. This eventually enables to study noncommutative algebras
of functions defined on the cotangent bundle of a compact simple
Lie group.

In chapter \ref{chapthree} the machinery developed in the first
part is fully used in the specific example of defining a new fuzzy
space, the fuzzy disc.

 A fuzzy space is a sequence of nonabelian algebras, more properly
 finite rank matrix algebras, approximating as "quantum metric
 spaces", the commutative algebra of functions on continuous 
manifolds on which field theory
models are defined. This approximation is seen to act as a
regulator for ultraviolet divergences in a class of field theory
obtained via a canonical quantization of classical fields.

The chapter opens with a description of what is the fuzzy sphere
(seen as a prototype of a fuzzy space) to describe what is the
meaning of this approximation, and what is the meaning of the
convergence of this sequence of nonabelian algebras towards an
abelian one. Then the fuzzy disc is introduced, as the first
example of a fuzzy approximation of a continuous space having a
boundary. It is developed starting from the analysis of the
noncommutative plane obtained via a standard Weyl-Wigner
isomorphism. The stress is put on the way a sequence of finite
rank matrix algebras is obtained, and the way the introduction of
derivatives and a "fuzzy" Laplacian operator leads to the concept
of "fuzzy Bessel functions". This notion extends that of fuzzy
spherical harmonics introduced in the case of the sphere. On the
fuzzy Bessels it is based the procedure by which the algebra of
functions on a disc can be given a fuzzy version. Moreover, this
approximation is seen to heal the ultraviolet divergences already
present in  noninteracting field theories on a disc.

At the end, some appendices recollect concepts used in the main
text. The first is devoted to briefly introduce some of the
algebraic concepts mentioned in the text. The second explains the
meaning of Fourier symplectic transform, used to perform the
harmonic analysis of the translation group. The third defines what
is a system of generalised coherent states, which is a unifying
scheme for the realization of the various kinds of Weyl-Wigner
isomorphisms developed in the text. The last appendix recollects 
the calculations performed to obtain the form 
of the nonabelian product among functions
defined on the quantum cotangent space defined in chapter
\ref{chapsecond}.

\tableofcontents
\chapter{An introduction to the Weyl-Wigner formalism}\label{chapone}

\section*{}

There is a profound difference between the classical and the
quantum formulation for the dynamics of a physical system. This
difference, together with the hypothesis - a principle of
correspondence - that the classical formalism were to be recovered
as a limiting case of the quantum one, originate the problem of
studying this comparison in a convenient setting.

The Weyl-Wigner formalism is a setting convenient for the analysis
of some aspects of this problem. It shows a procedure in
quantizing certain classical dynamics, and enables to give, for a
class of quantum dynamics, a more consistent meaning to the
formulation of the so called classical limit, often naively
considered as a formal manipulation obtained by letting
'$\hbar\,\rightarrow\,0$'.

This formalism is based on an application, the Weyl map, that
transforms functions defined on a real, even dimensional, vector
space, into operators on a separable Hilbert space: $$
\hat{\Omega}\,:\,\mathcal{F}\left(S\right)\,\mapsto\,Op\left(\hil\right)
$$ The Weyl map is invertible, and its inverse is called Wigner
map. They are introduced via an explicit use of the 
fundamental constant $\hbar$, and of a symplectic 2-form on the vector
space, so that this vector space can be identified with a phase
space carrying a classical dynamics. This bijection can then be seen
as a kind of map between the space of classical observables and
the space of quantum observables.

These two spaces are very different. The composition rule among
classical observables is abelian, while the product rule among
quantum observables is non abelian. Weyl-Wigner map enables to
translate the noncommutative product in the space of operators
into a noncomutative product in the set of functions on the phase
space carrying a classical dynamics. This means that the quantum
algebra can be represented in terms of functions on the classical
phase space, where a "$*$"-product is introduced: $$
f*g\,=\,\hat{\Omega}^{-1}\left(\hat{\Omega}\left(f\right)\cdot\hat{\Omega}\left(g\right)\right)
$$ This is also called a Moyal product, and it is a deformation of the
standard abelian pointwise product. Here deformation means that it
explicitly depends on $\hbar$, and reduces to the pointwise
product in the limit for $\hbar\,\rightarrow\,0$, where Planck's
constant is now seen as a parameter. The antisymmetrization of
this product gives a Moyal bracket: $$
\{f,g\}_{M}\,=\,\frac{i}{\hbar}\left(f*g-g*f\right) $$ The same
way the Moyal product represents the composition law among
operators, this application is the definition, in the space of
functions, of the notion of commutator of two operators. It is
bilinear, skewsymmetric, and satisfies both the Jacobi identity and
the Leibnitz rule. It is a deformation, in $\hbar$, of the Poisson
Bracket of two functions. This means that this formalism recovers
a more rigorous version of the analogy Dirac expressed, between
the commutators (among quantum observables) and Poisson Brackets
(among classical observables).

The first part of this chapter describes the notion of Weyl
system, which is the building block in the construction of the
Weyl and Wigner maps. The emphasys is given to the interpretation
of a
 vector phase space as the manifold carrying a realization of
the translation group, and of the Weyl system as a unitary
projective representation associated with this realization. The
phase factors are related to the symplectic structure defined on
the group. In the analysis of the covariance properties of this
representation for the action of the symplectic group, it is
possible to find a procedure of quantization for some classical
dynamics.

Then the Weyl and Wigner maps are presented, and the Moyal product is
introduced, to describe a setting where the quantum evolution can
be written in terms of equations on functions on the phase space.
In this setting the role of $\hbar$ is made explicit, and
recovering a classical limit is natural.

Afterwards two steps towards a generalization of the Weyl-Wigner
isomorphism are elucidated. The first is the full analysis of
the definition of this formalism in the case that classical
dynamics can be written in terms of a generic, though still
translationally invariant, symplectic form on the classical
cartesian phase space. The second describes how the Weyl
system's notion can be enlarged to the case where it is defined as
a more general projective representation of the same translation
group. The result is the introduction of the weighted Weyl maps,
that clarify how also the ordering problems, usually mentioned in
the formal quantization procedure, can be understood via this
formalism.

The chapter ends with the proof that the standard Weyl formalism
can be obtained without using the tools of harmonic analysis for
the translation group, but studying the definition of a system of
coherent states (the notion of generalized coherent states is
introduced in the Appendix) for the Heisenberg-Weyl-Wigner group.
It also shows one of the reason why, in this
harmonic analysis, the concept of symplectic Fourier transform has been used.
Even this concept is introduced in Appendix.

\section{Weyl systems}

\subsection{From classical mechanics to quantum conditions}

Dynamical evolution of a physical system can be represented as a
set of transformations, parametrised by time, in the space of
states of the system into itself \cite{MSSV}: $$
\mathcal{D}_{t}\,:\,\mathcal{S}\,\mapsto\,\mathcal{S} $$ These
transformations represent the half-line as a one parameter
semigroup $\left(\real^{+}_{0},+\right)$, because they satisfy a
composition rule, for every positive value of time parameter $t$:
$$ \mathcal{D}_{0}\,=\,\idop $$ $$
\mathcal{D}_{t^{\prime}+t^{\prime\prime}}\,=\,\mathcal{D}_{t^{\prime}}\,\circ\,\mathcal{D}_{t^{\prime\prime}}\,=\,
\mathcal{D}_{t^{\prime\prime}}\,\circ\,\mathcal{D}_{t^{\prime}} $$
If these $\mathcal{D}_{t}$ are bijective maps, then it is defined:
$$ \mathcal{D}_{t}^{-1}\,\equiv\,\mathcal{D}_{-t} $$ and time
evolution is seen as a one parameter group $\left(\real,+\right)$
of transformations of $\mathcal{S}$ into itself.

This section is devoted to the description of some aspects of the Poisson,
and of the symplectic formulation of classical dynamical systems.
The well established topics reviewed here\footnote{For the
concepts of manifold analysis and calculus, assumed in these pages,
an excellent textbook is \cite{AM}.} are meant to be steps of a
path driving to the definition of the so called \emph{quantum
conditions}.

In classical mechanics \cite{AM} the set of pure states is
formalised as a manifold $\mathcal{L}$ and observables are
represented by real functions on this manifold. With respect to
the topological and to the differentiable structure of
$\mathcal{L}$, the evolution $\mathcal{D}_{t}$ of a classical
dynamical system can be seen as a one parameter group of
diffeomorphisms of $\mathcal{L}$, whose infinitesimal generator is
a (complete) vector field. The integral curves of this vector
field represent the evolution of pure states, while the evolution
of observables can be written as $$f_{t}\,=\,f\circ\phi_{t}$$ or,
infinitesimally, as a solution of:
 \beq \frac{df}{dt}\,=\,X\cdot f\,=\,L_{X}f
\eeq
 ($f$ is the
function representing the observable, $X$ is the vector field
generating the dynamics, and $L_{X}f$ is the Lie derivative of $f$
along $X$, i.e. the directional derivative of $f$ along the
integral curves of $X$.)

Given a manifold $\mathcal{L}$, a Poisson bracket is a map that
associates a function on $\mathcal{L}$ to a pair of such
functions: $$
\mathcal{F}\left(\mathcal{L}\right)\,\times\,\mathcal{F}\left(\mathcal{L}\right)\,\mapsto\,
\mathcal{F}\left(\mathcal{L}\right) $$ which is bilinear,
skewsymmetric, and satisfies:
\begin{itemize}
\item
Jacobi identity, for every triple of functions:
$$
\{f,\{g,h\}\}\,+\{g,\{h,f\}\}\,+\{h,\{f,g\}\}\,=0
$$
\item
Leibnitz rule, for every triple of functions:
$$
\{fg,h\}\,=\,f\,\{g,h\}\,+\,\{f,h\}\,g
$$
\end{itemize}
with respect to the standard pointwise product among elements in
$\mathcal{F}\left(\mathcal{L}\right)$. The Poisson bracket is
regular if the requirement
$\{f,g\}=0\,\,\forall\,f\,\in\,\mathcal{F}\left(\mathcal{L}\right)$
implies that $g$ is constant.

A map $\phi\,:\,\mathcal{L}\,\mapsto\,\mathcal{L}$ is called
canonical if it preserves the Poisson tensor: $$
\{f,g\}\,\circ\,\phi\,\equiv\,\{f\,\circ\,\phi,g\,\circ\,\phi\} $$
$\forall\,f,g\,\in\,\mathcal{F}\left(\mathcal{L}\right)$. A
classical dynamics has a canonical formulation if time evolution
is represented by a one parameter group of canonical (with respect
to a given Poisson structure) transformations.

This formulation can be given an infinitesimal version. Since
Poisson bracket satisfies the Leibnitz rule, for a fixed
$H\,\in\,\mathcal{F}\left(\mathcal{L}\right)$ the map: $$
\{\cdot,H\}\,:\,\mathcal{F}\left(\mathcal{L}\right)\,\mapsto\,\mathcal{F}\left(\mathcal{L}\right)
$$ is a derivation in the abelian algebra
$\mathcal{F}\left(\mathcal{L}\right)$. In the theory of differentiable
manifolds, it is possible to prove that each smooth derivation on
$\mathcal{F}\left(\mathcal{L}\right)$ can be given the form of a
vector field. To the Poisson derivation defined by $H$  a vector field $X_{H}$ can be associated, such that: \beq
\{f,H\}\,=\,L_{X_{H}}f \eeq $X_{H}$ is called Hamiltonian vector
field of Hamiltonian function $H$ with respect to the given
Poisson bracket. The components of this Hamiltonian vector field
are, in a local coordinate chart $\{\xi^{a}\}$: $$
\dot{\xi}^{a}\,=\,\frac{d\xi^{a}}{dt}\,=\,\{\xi^{a},H\} $$ so that
the Poisson bracket of two functions can be written as: $$
\{f,g\}\,=\,X_{g}\cdot f\,=\,\{f,\xi^{b}\}\frac{\del g}{\del
\xi^{b}}\,=\,\frac{\del f}{\del\xi^{a}}\,
\{\xi^{a},\xi^{b}\}\frac{\del g}{\del\xi^{b}} $$ Poisson bracket
is defined by the components: \beq
\Lambda_{ab}\,\equiv\,\{\xi^{a},\xi^{b}\} \eeq of a skewsymmetric
twice contravariant tensor: \beq
\Lambda\,\equiv\,\Lambda^{ab}\,\frac{\del}{\del\xi^{a}}\,\otimes\,\frac{\del}{\del\xi^{b}}
\eeq such that\footnote{Jacobi identity is written as:
$$\Lambda^{sk}\left(\frac{\del}{\del\xi^{s}}\Lambda^{ab}\right)\,+\,\Lambda^{sa}\left(\frac{\del}{\del\xi^{s}}\Lambda^{bk}\right)\,+
\,\Lambda^{sb}\left(\frac{\del}{\del\xi^{s}}\Lambda^{ka}\right)\:=\:0$$}:
\beq \{f,g\}\,=\,\Lambda\left(df,dg\right) \eeq  A one parameter
group $\phi_{s}$ of canonical transformations on $\mathcal{L}$ is
generated by an infinitesimal canonical vector field $X$: \beq
L_{X}\Lambda\,=\,0\,\,\Leftrightarrow\,\,L_{X}\{f,g\}\,=\,\{L_{X}f,g\}\,+\,\{f,L_{X}g\}
\eeq An Hamiltonian vector field is a canonical vector field, so
time evolution generated by an Hamiltonian vector field is
canonical, i.e. it preserves the Poisson tensor $\Lambda$. The
infinitesimal form of the evolution equations for a dynamical
system in the Poisson formalism is: \beq
\frac{df}{dt}\,=\,\{f,H\}\eeq It is interesting to note that the
set of Hamiltonian vector field does not exhaust the set of
canonical vector field. There are canonical vector fields which
are not Hamiltonian \cite{marmosternibort}. This may happen either because 
the Poisson bracket is degenerate, or because the carrier space manifold
is not simply connected.

 If $\Lambda$ is a regular Poisson tensor whose matrix elements are
$\Lambda^{ab}$, that is globally invertible, then it is possible
to introduce a skewsymmetric twice covariant tensor on
$\mathcal{L}$. In this coordinate chart: $$
\omega\,=\,\omega_{ab}\,d\xi^{a}\,\otimes\,d\xi^{b} $$ where \beq
\Lambda^{ab}\,\omega_{bc}\,\equiv\,-\delta^{a}_{c}
\label{poissonsympl}\eeq This tensor is nondegenerate, and it can
be proved that \cite{marmosternibort} the requirement that Jacobi
identity is verified by $\Lambda$ is equivalent to the closedness
condition for this form: $d\omega=0$. The pair
$\left(\mathcal{L},\omega\right)$, with $\mathcal{L}$ a
differentiable manifold and $\omega$ a closed, nondegenerate,
skewsymmetric covariant 2-form is called a symplectic manifold,
and $\omega$ is a symplectic structure.

A vector field $X$ is called locally Hamiltonian if it is the
infinitesimal generator of symplectic transformations, those
preserving the symplectic tensor: \beq L_{X}\omega=0 \eeq The
symplectic vector fields are canonical with respect to the
associated Poisson tensor (\ref{poissonsympl}): \beq
L_{X}\Lambda\,=\,0\,\,\Leftrightarrow\,\,L_{X}\omega\,=\,0 \eeq
From Cartan's identity: \beqa L_{X}\omega&=&0\,\,\,\rightarrow
 \nonumber \\
i_{X}\,d\omega\,+\,d\,i_{X}\omega&=&0\,\,\,\rightarrow
\nonumber \\
d\,i_{X}\omega&=&0
\eeqa
Globally Hamiltonian vector fields are those fields for which the 1-form $i_{X}\omega$ is not only closed, but exact:
\beq
i_{X_{H}}\omega\,=\,dH
\label{gloham}\eeq
A classical dynamics has an Hamiltonian formulation if it is 
infinitesimally represented by an Hamiltonian vector field.

For example, if $\mathcal{Q}$ is the configuration space for the
evolution of a conservative Newtonian system, whose equations of
motions are written as ($q^{a}$ are position coordinates, $v^{a}$
are velocities coordinates, this evolution being formalised on
$T\mathcal{Q}$, the tangent bundle of $\mathcal{Q}$): \beqa
\dot{v}^{a}&=&F^{a}\left(q^{i},v^{j}\right) \nonumber \\
\dot{q}^{a}&=&v^{a} \label{newtoneq}\eeqa then, on the cotangent
bundle $\mathcal{L}=T^{*}\mathcal{Q}$ (where $q^{i}$ are still
coordinates on $\mathcal{Q}$, the basis of the bundle, labelling
position observables, while $p_{a}$ are local coordinates on the
fibers, labelling momenta), this evolution is symplectic via  the
definition of\footnote{This form of the symplectic structure is
very important, and it is called 'canonical'. The theorem of
Darboux \cite{AM} proves that, given a symplectic manifold
$\left(\mathcal{L},\omega\right)$, there exists a local coordinate
transformation on $\mathcal{L}$ such that the simplectic structure
locally acquires the canonical form.}: \beqa
\tilde{\Lambda}^{a}_{b}\,\equiv\,\{q^{a},p_{b}\}\,=\,\delta^{a}_{b}
\nonumber \\ \tilde{\Lambda}\,=\,\frac{\del}{\del
q^{a}}\wedge\frac{\del}{\del p_{a}} \nonumber \\
\tilde{\omega}\,=\,dq^{a}\wedge dp_{a} \eeqa because equations of
motions (\ref{newtoneq}) are written as\footnote{In these
expressions, indices are raised and lowered, to keep track of
labelling covariant or contravariant elements of the tensor
algebra on $\mathcal{L}$, by the metrics that is implicitly assumed in the
definition of the kinetic energy term in the Hamiltonian
function.}: \beqa
\dot{q}^{a}\,=\,p^{a}&=&\{q^{a},H\}\,=\,\frac{\del H}{\del p_{a}}
\nonumber \\ \dot{p}_{a}\,=\,-\,\frac{\del U}{\del
q^{a}}&=&\{p_{a},H\}\,=\,-\,\frac{\del H}{\del q^{a}}
\label{hamilton} \eeqa with
$H=\frac{1}{2}p^{a}p_{a}+U\left(q^{a}\right)$ where $U$ is the
potential energy whose gradient is the force field $F$:
$F_{a}=-\frac{\del U}{\del q^{a}}$.

It is possible to consider a symplectic manifold $\left(\mathcal{L},\omega\right)$,
such that it is an homogeneous space for a transitive and free action
of the group of translations. If the system of coordinates
$\left(q^{a},p_{b}\right)$ is global, and canonically adapted to
this action, then the coordinate functions are seen to generate
the Hamiltonian vector fields that represents this group of
translations: \beqa
q^{a}\,\rightarrow\,X_{\left(q^{a}\right)}&=&-\,\frac{\del}{\del
p_{a}} \nonumber \\
p_{a}\,\rightarrow\,X_{\left(p_{a}\right)}&=&\,\frac{\del}{\del
q^{a}} \label{displaceclass} \eeqa

\subsection{Standard Weyl systems}\label{staweylsys}

Quantum formulation for the dynamics of a physical system is
profoundly different. Pure state are represented by rays of a
separable Hilbert space; observables are represented by
self-adjoint operators on this space. Time evolution is
represented by unitary transformations on this set of states.

The principle of correspondence suggests to postulate that, in
quantum theory, the groups generated by the cartesian position and
momentum coordinates of a system of particles are the same as the
classical ones (\ref{displaceclass}), i.e. they displace,
respectively, in the momenta and the positions \cite{thirring}. The aim
of this section is to describe how the notion of \emph{Weyl
system} is able to fulfill this expectation, and to show how it can be
used to set up a procedure of quantization for a set of specific
classical dynamics.

The mathematical formulation of this basic postulate is the
introduction of the \emph{canonical commutation relations} among
quantum observables of positions and momenta: \beq
\left[\hat{q}^{a},\hat{p}_{b}\right]\,=\,i\hbar\idop\delta^{a}_{b}
\label{CCR} \eeq These canonical commutation relations were
introduced by Dirac to stress the formal analogy between the
properties of the Poisson bracket among classical observables, and
those of the commutator among quantum obsevables. Weyl \cite{weyl}
was the first to study the problems of a concrete realization of
these operators.

Given a pair of self-adjoint operators satisfying the canonical
commutation relations, it is possible to prove that they cannot be
both bounded. Let $\Ao$ and $\Bo$ two bounded operators, whose
commutator is a multiple of the identity: $$
\left[\Ao,\Bo\right]=c\idop $$ Boundedness of them both would
imply that: $$ \left[\Ao,\Bo^{n}\right]=cn\Bo^{n-1} $$ together
with triangle inequality for the operator norm, this would led to:
$$
cn\parallel\Bo\parallel^{n-1}\,=\,cn\parallel\Bo^{n-1}\parallel\,\leq\,2\parallel\Ao\parallel\parallel\Bo\parallel^{n}
$$ so one would finally have, $\forall\,n$: $$
cn\,\leq\,2\parallel\Ao\parallel\,\parallel\Bo\parallel $$ This is
a contradiction to the hypothesis of boundedness of both $\Ao$ and
$\Bo$. This result is known as Wintner's theorem. One of its
consequences is that the equality between l.h.s and r.h.s. of
eq.(\ref{CCR}) is strictly valid not on every element of the
Hilbert space on which those obsevables are represented. This is
usually referred to writing the canonical commutation relations
as: \beq
\left[\hat{q}^{a},\hat{p}_{b}\right]\,\subset\,i\hbar\idop\delta^{a}_{b}
\label{CCRsub} \eeq

Weyl suggested to look at them in a more general context. A
\emph{symplectic vector space} is a pair $\left(L,\omega\right)$
consisting of a real topological vector space $L$, equipped with a
continuous antisymmetric, nondegenerate bilinear form $\omega$ on
$L$. 'Nondegenerate' means that if
$\omega\left(z,u\right)=0\,\,\forall\,z\,\in\,L$, then $u=0$: this
is an example of a symplectic manifold. Let
$\left(L,\omega\right)$ be a symplectic vector space\footnote{In
this analysis, $L$ will always be finite dimensional. The
requirement that $\omega$ were nondegenerate forces then the space
to be even-dimensional. An interesting analysis of Weyl systems on
infinite dimensional vector spaces, suitable for a formalization
of field theories is in \cite{baezsz}}: a \emph{Weyl system} is
for $\left(L,\omega\right)$ is a map into the set of unitary
operators on a separable Hilbert space $\hil$: \beq
\hat{D}\,:\,L\,\mapsto\,\mathcal{U}\left(\mathcal{H}\right) \eeq
such that:
\begin{itemize}
\item $\Do$ is continuous in the strong operator topology,
\item for each pair of vectors $z$ and $u$ in $L$:
\beq
\Do\left(z+u\right)\,=\,e^{i\omega\left(z,u\right)/2\hbar}\Do\left(z\right)\Do\left(u\right)
\label{defws} \eeq
\end{itemize}
This condition is equivalent to: \beq
\Do\left(z\right)\Do\left(u\right)\,=\,e^{-i\omega\left(z,u\right)/\hbar}\,\Do\left(u\right)\Do\left(z\right)
\label{displcomm} \eeq If the vector space $L$ is identified with
the manifold representing the abelian Lie group of translations,
then a Weyl system can be seen as a unitary, projective
representation of the translation group. The phase factor of this
representation is related to the symplectic form on $L$. Linearity
of $\omega$ can be considered as the invariance of the this
symplectic tensor with respect to the action of the translation
group.  This also suggests the reason why operators $\Do$ are
called
 \emph{Displacement operators}.

On a one dimensional subspace of $L$, with $\alpha$ and $\beta$
real scalars, the phase factors cancel out:
\beq
\Do\left(\alpha z\right)\Do\left(\beta
z\right)\,=\,\Do\left(\left(\alpha+\beta\right)z\right)
\eeq
this
means that $\Do\left(\alpha z\right)$ is a strongly continuous, one parameter group of
unitary operators. The theorem of Stone says
that it can be considered as the exponentiation of a self-adjoint operator:
\beq
\Do\left(\alpha
z\right)\,=\,e^{i\alpha G\left(z\right)/\hbar}
\eeq
Moreover, it
can be seen that, up to additive terms like $2\pi n\idop$
in the identification of generators of a Weyl
system, one has:
\beq
\Go\left(\alpha
z\right)\,=\,\alpha\,\Go\left(z\right)
\label{lingen}
\eeq
These
generators have important properties. The defining relation (\ref{displcomm}) suggests that :
\beq
\Do\left(\alpha z\right)\Do\left(\beta
u\right)\,=\,e^{-i\omega\left(\alpha z,\beta
u\right)/\hbar}\,\Do\left(\beta u\right) \Do\left(\alpha z\right)
\eeq
If they are written in terms of generators:
\beq
e^{i\alpha
\Go\left(z\right)/\hbar}\,e^{i\beta
\Go\left(u\right)/\hbar}\,=\,e^{i\alpha\beta\omega\left(z,u\right)/
\hbar}\,e^{i\beta \Go\left(u\right)/\hbar}\,e^{i\alpha
\Go\left(z\right)/\hbar}
\label{expweylCCR}
\eeq
This relation shows, once more, that
in the Weyl approach every one dimensional subspace of $L$
corresponds to a one parameter group of unitary operators. They
will not suffer any of the problem of unbounded operators. It can
be seen as a global version, in terms of bounded operators, of the
canonical commutation rules.

If parameters $\alpha$ and $\beta$ are considered as
infinitesimal, one has
\begin{eqnarray*}
[\idop+i\alpha
\Go\left(z\right)/\hbar+o\left(\alpha^{2}\right)]
[\idop+i\beta \Go\left(u\right)/\hbar+o\left(\beta^{2}\right)]=
\\ {}  \left(1-i\omega\left(z,u\right)\alpha\beta/\hbar
+o\left(\left(\alpha\beta\right)^{2}\right)\right) [\idop+i\beta
\Go\left(u\right)/\hbar+o\left(\beta^{2}\right)]{\cdot}
[\idop+i\alpha \Go\left(z\right)/\hbar+o\left(\alpha^{2}\right)]
\end{eqnarray*}
and equating the coefficients of the first nonzero order in these
infinitesimals, one obtains: \beq
\left[\Go\left(z\right),\Go\left(u\right)\right]\,=\,i\hbar\omega\left(z,u\right)\idop
\label{weylCCR} \eeq This analysis is again performed as a formal
manipulation: it could give no more than a hint to prove a
rigorous result \cite{baezsz}. If $\Wo\left(z\right)$ defines a
Weyl system for $\left(L,\omega\right)$, whose generators are
selfadjoint $\Go\left(z\right)$ then, for arbitrary vectors $z$
and $u$ in $L$, a dense domain for the products of the two
generators, $Dom\left(\Go\left(z\right)\Go\left(u\right)\right)=
Dom\left(\Go\left(u\right)\Go\left(z\right)\right)$, can be found
and, for every element $\phi$ of the Hilbert space belonging to
this domain, one has: \beq
\left[\Go\left(z\right),\Go\left(u\right)\right]\phi\,=\,i\hbar\omega\left(z,u\right)\phi
\label{CCRexact} \eeq This is the exact form of eq.(\ref{CCRsub}).

So canonical commutation rules can be seen as the infinitesimal
version of (\ref{expweylCCR}). They involve the selfadjoint
generators of the group of displacements, recovered as
observables. Noncommutativity among quantum observables is
introduced via the symplectic form, and "measured" by $\hbar$.

The presence of the symplectic structure in the formalization of
the \emph{quantum conditions} is a first appearance of the
principle of \emph{analogy}. The quantum formalism is developed on
the geometric structures on which classical formalism is based.

\subsection{The Schr\"{o}dinger representation}\label{schrorep}

So far a Weyl system has been defined, and its properties have
been deduced at a formal level. Now it will be  explicitly
realized.

Let $M$ be a finite-dimensional real vector space, $M^{*}$ its
dual, the space of linear functions on $M$. Their direct sum
define an even dimensional real vector space $L\,\equiv\,M\oplus
M^{*}$, whose vectors $z$ are of the form $z=x\oplus \lambda$. A
symplectic form is given by
$\tilde{\omega}\left(z,z^{\prime}\right)\equiv
\lambda^{\prime}\left(x\right)-\lambda\left(x^{\prime}\right)$. If
$M$ is considered as a manifold, then $L$ is the cotangent bundle
of $M$: $L\simeq T^{*}M$, and vectors $z$ can be written in terms
of a global coordinate chart as $z=\left(q^{a},p_{b}\right)$. This
notation, with indices $a,b$ ranging from $1$ to the dimension of
$M$, makes it explicit that, in the identification of $L$ with
$T^{*}M$, $q^{a}$ coordinates label elements of the basis $M$,
while $p_{b}$ coordinates label elements of the fiber $M^{*}$. In
this coordinate chart, the symplectic tensor assumes the
canonical, Darboux form $\tilde{\omega}=dq^{a}\wedge dp_{a}$, so
that
 eq.(\ref{weylCCR}) gives the canonical
commutation relations (\ref{CCR}) for the generators of a Weyl
system.

Let $\hil=\mathcal{L}^{2}\left(M,dx\right)$ be the Hilbert space
of square integrable functions on $M$ with respect to the
translationally invariant Lebesgue measure. A pair of one
parameter groups of unitary operators can be defined: \beqa
\left(\Uo\left(q\right)\psi\right)\left(x\right)&=&\psi\left(x+q\right)
\nonumber \\
\left(\Vo\left(p\right)\psi\right)\left(x\right)&=&e^{i\la
p,x\ra/\hbar}\psi\left(x\right) \eeqa
Here $q$ is a vector of
coordinates $q^{a}$, labelling a $x$ element of $M$, while $p$ is
a covector of coordinates $p_{b}$ labelling a $\lambda$ element of
$M^{*}$; $\la p,x\ra$ represents the action of the function $\lambda$ on
the element $x$. $\Uo\left(q\right)$ is a unitary faithful
representation of the abelian group of translations
$\left(M,+\right)$. Harmonic analysis for this group says that its
dual $M^{*}$ is isomorphic to $M$, and $\Vo\left(p\right)$ can be
considered as a unitary faithful representation of
$\left(M^{*},+\right)$. These representations are not mutually
commuting: \beq
\Uo\left(q\right)\Vo\left(p\right)\,=\,e^{i\tilde{\omega}
\left(\left(q,0\right),\left(0,p\right)\right)/\hbar}
\,\Vo\left(p\right)\Uo\left(q\right) \label{weylexpcomm}\eeq In
this expression
$\tilde{\omega}\left(\left(q,0\right),\left(0,p\right)\right)=qp$
indicates the image of the 2-form on the pair of elements in $L$
whose components are $\left(q,0\right)$ and $\left(0,p\right)$.
Let: \beq
\Do\left(q,p\right)\,=\,\Uo^{\dagger}\left(q\right)\Vo\left(p\right)e^{i\la
p,q\ra/2\hbar} \label{wsys1} \eeq define a set of operators. They
are unitary operators, in correspondence with points of the space
$L=T^{*}M$, which satisfy the required properties (\ref{defws}).
These operators are a realization of a Weyl system for
$\left(L,\tilde{\omega}\right)$. Their image on a function $\psi$
in $\hil$, in the chosen realization, is: \beq
\left(\Do\left(q,p\right)\psi\right)\left(x\right)\,=\,e^{-\frac{i}{2\hbar}\la
p,q\ra} e^{\frac{i}{\hbar}\la p,x\ra}\psi\left(x-q\right)
\label{wmapexpl} \eeq Now, coming back to the pair of one
parameter groups of unitary operators $\Uo\left(q\right)$ and
$\Vo\left(p\right)$, one can consider the value of the coordinates $q$ and $p$
as infinitesimal parameters, to obtain an explicit form of their
generators: \beqa
\Uo\left(q\right)\,&=&\,e^{iq^{a}\Po_{a}/\hbar}\,\,\,\,\,\,\left(\Po_{a}\psi\right)\left(x\right)=-i\hbar\frac{d\psi}{dx^{a}}
\nonumber \\
\Vo\left(p\right)\,&=&\,e^{ip_{b}\Qo^{b}/\hbar}\,\,\,\,\,\,\left(\Qo^{b}\psi\right)\left(x\right)=x^{b}\psi\left(x\right)
\label{qpsugiu}
\eeqa
Commutation relations (\ref{CCRexact}) become: \beq
\left[\Qo^{a},\Po_{b}\right]\,=\,i\hbar\delta^{a}_{b}\idop \eeq
These are exactly the standard $\Qo$ and $\Po$ operators in the
space of square integrable functions on linear space, that are a
formal, well known solution to the problem of realizing the
canonical observables of position and momentum in point particle
quantum mechanics. The Displacement operators acquire the form:
\beq
\Do\left(q,p\right)\,=\,e^{i\left(p_{a}\Qo^{a}-q^{b}\Po_{b}\right)/\hbar}
\label{displacement} \eeq Weyl approach stresses the group
theoretical interpretation of the classical phase space as the
manifold that represents the translation group. Operators related
to the observables position and momentum are recovered as
generators of a representation of this action. This is the
solution to the initial problem of defining  quantum
displacements. Moreover, the phase factor that characterizes this
representation reproduces the commutation relations among its
generators.

This explicit realization of a Weyl system is called
Schr\"{o}dinger representation. It can be proved to be
irreducible. A fundamental result is known as Von Neumann theorem
\cite{reedsimon}. If $\Uo\left(q\right)$ and $\Vo\left(p\right)$
are strongly continuous one parameter group of unitary operators
on a separable Hilbert space $\mathcal{H}^{\prime}$ , satisfying,
in an irreducible representation, the Weyl form of the commutation
relations: \beq \Uo\left(q\right)\Vo\left(p\right)\,=\,e^{i\la
p,q\ra/\hbar}\Vo\left(p\right)\Uo\left(q\right) \eeq then there is
an isometry: $$
S\,:\,\mathcal{H}^{\prime}\,\mapsto\,\hil=\mathcal{L}^{2}\left(M,dx\right)
$$ such that, for $\psi\in\hil$: $$
\left(S\Uo\left(q\right)S^{-1}\psi\right)\left(x\right)=\psi\left(x+q\right)
$$ $$
\left(S\Vo\left(p\right)S^{-1}\psi\right)\left(x\right)=e^{i\la
p,q\ra/\hbar}\psi(\left(x\right) $$ This isometry says that there
is an equivalence among irreducible solutions of the Weyl
commutation relations\footnote{In the Weyl approach, selfadjoint
operators satisfying peculiar commutation relations are recovered
as generators of suitable strongly continuous one-parameter groups
of unitary operators. It is interesting to note that it is
possible to define \cite{reedsimon} the Hilbert space
$\mathcal{L}^{2}\left(Q,d\mu\right)$ where $Q$ is the Riemann
surface of the $\sqrt{z}$, $\left(z=x+iy\right)$and $d\mu$ is the
(local) Lebesgue measure. On this Hilbert space it can be proved
to exist a domain $\mathcal{D}$ and a pair of operators: $$
\Ao\,\equiv\,-i\,\frac{\del}{\del
x}\,\,\,\,\,\,\,\,\,\Bo\,\equiv\,x-i\,\frac{\del}{\del y} $$ which
satisfy:
\begin{itemize}
\item $$\Ao\,\,,\,\,\Bo\,\,:\,\mathcal{D}\,\mapsto\,\mathcal{D}$$
$\mathcal{D}$ is a common domain of essential self-adjointness for
the operators $\Ao,\Bo$
\item For all $\phi\,\in\,\mathcal{D}$
$$\Ao\Bo\phi-\Bo\Ao\phi\,=\,-i\phi$$
\end{itemize}
Nevertheless it can be proved that the unitary groups that these
two operators generate do not satisfy the Weyl relations
(\ref{weylexpcomm}).}.

This property is very important in the study of the problem of
quantizing a classical system. The notion of Weyl system is based,
on the vector space structure of $L$, and the invariance of the
symplectic form for the action of the translation group. It is
natural to study the covariance properties of this
map for transformations of $L$. In particular, such
transformations will respect, in a first approach, both the linear
structure of $L$ and the symplectic structure $\tilde{\omega}$:
they will be linear and symplectic. A transformation $T$ belongs
to the linear symplectic group $Sp\left(L,\tilde{\omega}\right)$
if:
$$\tilde{\omega}\left(Tz,Tu\right)=\tilde{\omega}\left(z,u\right)$$
Acting with such a transformation on $L$, one has: \beq
\Do\left(Tz+Tu\right)=\Do\left(Tz\right)\Do\left(Tu\right)
e^{\frac{i}{2\hbar}\tilde{\omega}\left(Tz,Tu\right)} \eeq by
linearity:\beq
\Do\left(T\left(z+u\right)\right)=\Do\left(Tz\right)\Do\left(Tu\right)
e^{-\frac{i}{2\hbar}\tilde{\omega}\left(z,u\right)} \eeq it is
possible to define: \beq
\Do\left(Tz\right)\equiv\Do_{T}\left(z\right) \label{weylsysT}\eeq
obtaining: \beq
\Do_{T}\left(z+u\right)=\Do_{T}\left(z\right)\Do_{T}\left(u\right)
e^{-\frac{i}{2\hbar}\tilde{\omega}\left(z,u\right)}\label{weylsysT2}
\eeq

This means that $\Do_{T}$ is a new Weyl system for
$\left(L,\tilde{\omega}\right)$, so it is unitarily equivalent to
$\Do\left(z\right)$. This equivalence enables to associate an
automorphism to this transformation $T$ $$
\nu_{T}\,:\ouh\,\mapsto\,\ouh $$ by putting:\beq
\Do\left(Tz\right)\equiv\nu_{T}\left(\Do\left(z\right)\right) \eeq
Since every automorphism for the group of unitary operators on a
Hilbert space can be written as a conjugation by a unitary
operator, this automorphism can be given the form: \beq
\nu_{T}\left(\Do\left(z\right)\right)=\Uo_{T}^{-1}\Do\left(z\right)\Uo_{T}
\label{unconjsws}\eeq where $\Uo_{T}$ is a unitary
transformation\footnote{It can be proved that the operator
$\Uo_{T}$ is determined by the transformation $T$ up to a phase.
These phases cannot be totally eliminated, but can be at best
reduced to a sign ambiguity. For such a pair of transformations
$T$ and $T^{\prime}$:
$$\Uo_{T}\Uo_{T^{\prime}}\,=\,\pm\Uo_{TT^{\prime}}$$ This
situation can be expressed by saying that one is dealing with a
representation of the $Mp\left(2\right)$, the metaplectyc group,
which is a double covering of $Sp\left(2\right)$.}.

This analysis also shows the reason why, in (\ref{qpsugiu}), the
operator $\Qo^{a}$ has a superscript index, while $\Po_{b}$ has a
subscript index. Given $O$, an automorphism of the linear space
$M$, represented as a map $q^{\prime\,a}=O^{a}_{b}q^{b}$, it can
be lift to a symplectic automorphism $\mathcal{O}$ of
$\left(L,\tilde{\omega}\right)$, defining a transformation on
fibers: $p^{\prime}_{a}\equiv \left(O^{-1}\right)^{b}_{a}p_{b}$.

Unitary equivalence expressed by (\ref{unconjsws}) is written as:
\beqa
\Uo^{\dagger}_{\mathcal{O}}\,e^{ip_{a}\Qo^{a}/\hbar}\,\Uo_{\mathcal{O}}&=&
e^{i\left(O^{-1}\right)^{a}_{b}p_{a}\Qo^{b}/\hbar} \nonumber \\
\Uo^{\dagger}_{\mathcal{O}}\,e^{iq^{a}\Po_{a}/\hbar}\,\Uo_{\mathcal{O}}&=&
e^{iO^{a}_{b}q^{b}\Po_{a}/\hbar} \eeqa and these two expressions
say that $\mathcal{O}$ symplectic transformation on $L$ acts on
generators of the Weyl system, by linearity via the linear
transformations $O$ and $O^{-1}$: \beqa
\Po_{\left(\mathcal{O}\right)\,a}&=&O^{b}_{a}\Po_{b} \nonumber \\
\Qo_{\left(\mathcal{O}\right)}^{a}&=&\left(O^{-1}\right)^{a}_{b}\Qo^{b}
\eeqa In particular, the transformation law for $\Po$'s operators
is contravariant with respect to that of $\Qo$'s operators.

This equivariance of Weyl systems by the linear symplectic group
is very interesting. One can consider a classical dynamics on $L$
that admits an Hamiltonian description in terms of the symplectic
structure in the canonical form and of a quadratic Hamiltonian
function (\ref{hamilton}). Its time evolution is written as a one
parameter group of linear symplectic map on $L$. Via this
formalism, it is possible to associate, with this classical
evolution, a one parameter group of unitary operators on a Hilbert
space, that is a quantum evolution. This is a quantization
procedure for certain classical dynamics, for example those of the
free particle, and of the harmonic oscillator.

\section{Weyl map}\label{secweylmap}

Using a realization of a Weyl system for a symplectic vector space
as a map into the set of unitary operators on a Hilbert space, it
is possible to define a map from a set of functions defined on
this vector space, to a larger class of operators on the same
Hilbert space. This application is called \emph{Weyl map}
\cite{weyl}: $$\Do\,:\left(L,\tilde{\omega}\right)\,\mapsto\,\ouh$$ $$
\omw\,:\,{\cal F}\left(L\right)\,\mapsto\,\oph $$

In his book, Weyl considered the form  of the unitary operators
(\ref{wsys1}) in terms of the Hermitian generators. The expression
becomes: \beq
\Do\left(q,p\right)\,=\,e^{i\left(p_{a}\Qo^{a}-q^{a}\Po_{a}\right)/\hbar}
\label{wsys2} \eeq His suggestion was to look at them as a sort of
formal plane wave basis in a suitable space of operators.
Following this idea, coefficients of an operator expansion are
given by Fourier coefficients of a function on the plane. Since
the geometrical structure underlying the whole construction is the
symplectic form on $L$, then the Fourier transform is defined
using this 2-form\footnote{In appendix \ref{digrSFT} there is a
description of the  symplectic Fourier transform.}: (the dimension
of $L$ is $2n$) \beq
\tilde{f}\left(w\right)\,=\,\int\,\frac{dz}{\left(2\pi\hbar\right)^{n}}\,f\left(z\right)\,e^{-i\tilde{\omega}\left(z,w\right)/\hbar}
\label{symft} \eeq The formal definition of the Weyl map is: \beq
\hat{\Omega}\left(f\right)\,=\,\hat{f}\,=\,\int\,\frac{dw}{\left(2\pi\hbar\right)^{n}}\,\tilde{f}
\left(w\right)\Do\left(w\right) \label{weylmap} \eeq This integral
should be understood in a distributional sense\footnote{This means that, for a pair of vectors $\psi$ and $\psi^{\prime}$
in the Hilbert space $\hil$, one properly has:
\beq
\la\psi\mid\hat{\Omega}\left(f\right)\mid\psi^{\prime}\ra\,\equiv\,
\int\,\frac{dw}{\left(2\pi\hbar\right)^{n}}\,\tilde{f}\left(w\right)\,\la\psi\mid\Do\left(w\right)\mid\psi^{\prime}\ra
\eeq
This definition will be implicitly used in the following, to evaluate the operators that the Weyl map associates to the
coordinate functions.}: in terms of a
generalized basis of the Hilbert space, it is possible to formally
estimate, the trace of the operators \beq
Tr\left[\Do\left(z\right)\right]\,=\,\left(2\pi\hbar\right)^{n}\delta\left(z\right)
\eeq from which one has: \beq
Tr\left[\hat{f}\right]\,=\,\frac{1}{\left(2\pi\hbar\right)^{n}}\,\int
dz\,f\left(z\right) \eeq The group properties of these
$\Do\left(z\right)$ operators, together with this distributional
trace estimate, give: \beq Tr\left[\Do\left(z\right)
\Do^{\dagger}\left(u\right)\right]\,=\,\left(2\pi\hbar\right)^{n}\delta\left(z-u\right)
\label{Dcomplet} \eeq

This means that the $\Do$ operators define a generalized
resolution of the identity in the space of operators: then Weyl
map can be inverted. The inverse is given by: $$
\tilde{f}\left(w\right)\,=\,Tr\left[\fo\Do^{\dagger}\left(w\right)\right]
$$ \beq
f\left(z\right)\,=\,\int\frac{dw}{\left(2\pi\hbar\right)^{n}}\,e^{-i\tilde{\omega}\left(w,z\right)/\hbar}
Tr\left[\fo\Do^{\dagger}\left(w\right)\right] \label{wignermap1}
\eeq This is usually called \emph{Wigner map}. Given an operator
$\Ao$, the function $A\left(z\right)$ to which it is associated by
the Wigner map is called \emph{Weyl symbol}. One can see that
Weyl-Wigner map takes the notion of Hermitian conjugation in the
space of operators into that of complex conjugation in the space
of symbols. If $A\left(z\right)$ is the symbol of $\Ao$, then:
\beq
\hat{\Omega}^{-1}\left(\Ao^{\dagger}\right)\,=\,A^{*}\left(z\right)
\eeq

To study the properties of this application it is useful to
consider a function $f$ which is square-integrable on the plane,
so that Plancherel theorem assures that Fourier transform is well
defined. The action of the operator $\hat{f}$ in the Schr\"odinger
representation (\ref{wmapexpl}) on
$\hil=\mathcal{L}^{2}\left(M\simeq\real^{n},ds\right)$ is: \beqa
\left(\hat{f}\psi\right)\left(s\right)\,&=&\,\int\frac{dz}{\left(2\pi\hbar\right)^{n}}\,\int\frac{dw}{\left(2\pi\hbar\right)^{n}}\,
f\left(z\right)e^{-i\tilde{\omega}\left(z,w\right)/\hbar}
\,e^{-\frac{i}{2\hbar}\la k,x\ra} e^{\frac{i}{\hbar}\la
k,s\ra}\psi\left(s-x\right) \nonumber \\
&=&\,\frac{1}{\left(\pi\hbar\right)^{n}}\int\,dz\,f\left(z\right)e^{2i\la\left(s-q\right),p\ra/\hbar}\psi\left(2q-s\right)
\eeqa In these equations it has been used the notation: $z=q\oplus
p$ and $w=x\oplus k$. Now the product: $$
\Wo\left(z\right)\,=\,2^{n}e^{2i\left(p_{a}\Qo^{a}-q^{a}\Po_{a}\right)/\hbar}\hat{\mathcal{P}}$$
with
$\left(\hat{\mathcal{P}}\psi\right)\left(s\right)=\psi\left(-s\right)$
the parity operator, defines a new set of Hermitian operators, a
new resolution of the identity: \beqa
\Wo\left(z\right)\,&=&\,2^{n}\,\Do\left(2z\right)\,\hat{\mathcal{P}}
\nonumber
\\ \Wo\left(z\right)\,&=&\,\Wo^{\dagger}\left(z\right) \nonumber \\
Tr\left[\Wo\left(z\right)\,\Wo^{\dagger}\left(u\right)\right]\,&=&\,\left(2\pi\hbar\right)^{n}
\delta\left(z-u\right)
\label{weylopprop}
\eeqa
These operators enable to write the Weyl
map (\ref{weylmap}) without directly using the concept of Fourier transform.
They are also called Moyal quantizers \cite{moquant}:
\beq
\hat{f}\,=\,\frac{1}{\left(2\pi\hbar\right)^{n}}\int
dz\,f\left(z\right)\Wo\left(z\right) \label{wmapw} \eeq
Properties (\ref{weylopprop}) of this system of operators clarify
how the Weyl map in this form can be inverted. Its inverse, the
Wigner map, is given by: \beq
f\left(z\right)\,=\,Tr\left[\hat{f}\Wo^{\dagger}\left(z\right)\right]
\label{wignermap2} \eeq
Moreover, one has:
\beq
Tr\left[\Ao^{\dagger}\Bo\right]\,=\,\frac{1}{\left(2\pi\hbar\right)^{n}}\int
dz\,A^{*}\left(z\right)B\left(z\right) \label{wwiso}
\eeq
This means that Weyl-Wigner map defines a
bijection between the set of square-integrable functions on the
plane, and the set of Hilbert-Schmidt operators in the Hilbert
space on which the Weyl system has been realized.

Although this bijection is well suited for square integrable functions,
it is of interest trying to calculate what are the operators Weyl
map associates to the coordinate functions.
Just to keep notation clear, in this example let $M=\real$. For
$f\left(q,p\right)=q$ one has, formally: $$
\left(\qo\psi\right)\left(s\right)\,=\,\int\frac{dqdp}{\pi\hbar}\,qe^{2i\left(s-q\right)p/\hbar}\psi\left(2q-s\right)=$$
integration in $dp$ gives a $\pi\hbar\delta\left(s-q\right)$
factor \beq=\int
dq\delta\left(s-q\right)\psi\left(2q-s\right)\,=\,s\psi\left(s\right)
\eeq This shows that $\Qo$ operator in the Schr\"{o}dinger representation, that is the multiplication by the
coordinate on the line, is the Weyl image of the coordinate
function $q$ on the phase space. For $f\left(q,p\right)=p$ one has: $$
\left(\po\psi\right)\left(s\right)\,=\,\int\frac{dqdp}{\pi\hbar}\,pe^{2i\left(s-q\right)p/\hbar}\psi\left(2q-s\right)=$$
the position $x=2q-s$ brings the integral in the form:
$$=\int\frac{dxdp}{2\pi\hbar}\,pe^{i\left(s-x\right)p/\hbar}\psi\left(x\right)=$$
The integration over $dx$ gives the Fourier transform of
$\psi\left(x\right)$, so
$$\left(\po\psi\right)\left(s\right)=\int\frac{dp}{\sqrt{2\pi\hbar}}\,p\tilde{\psi}\left(p\right)e^{isp/\hbar}$$
This expression is clearly equal to \beq
\left(\po\psi\right)\left(s\right)\,=\,-i\hbar\frac{d\psi}{ds}
\eeq So the coordinate function $p$ is mapped in the $\Po$
operator in the Schr\"{o}dinger representation.

The next step is the study of the operator it associates to a
generic monomial in the coordinate functions. It can be proved
that: \beq
\hat{\Omega}\left(q^{a}p^{b}\right)\,=\,\frac{1}{2^{a}}\sum_{k=0}^{a}\left(\begin{array}{c}
a \\ k \end{array}\right)\, \Qo^{k}\Po^{b}\Qo^{a-k} \label{standardmonomial}\eeq This
example shows what is the ordering that the Weyl map
introduces in the quantization of a sufficiently generic element
of the algebra of classical observables, depending by both coordinates $q$ and $p$, promoted to noncommuting variables.

\section{The Moyal product for the noncommutative plane}

The fact that the Weyl-Wigner map is invertible enables to define
a different product in the space of functions on this cartesian
phase space. This product is called the Moyal product: \beq
\hat{\Omega}\left(f*g\right)\,=\,\hat{\Omega}\left(f\right)\hat{\Omega}\left(g\right)
\eeq It is non commutative, being a realization, in the space of
functions, of the non commutative product among operators. Written
in terms of functions, its integral form is: \beq
\left(f*g\right)\left(z\right)=\int\frac{da}{\left(2\pi\hbar\right)^{n}}\,e^{-i\tilde{\omega}\left(a,z\right)/\hbar}
\,\int\frac{db}{\left(2\pi\hbar\right)^{n}}\,\tilde{f}\left(b\right)\tilde{g}\left(a-b\right)e^{-i\tilde{\omega}\left(a,b\right)/2\hbar}
\label{intmp} \eeq This product is nonlocal: this means that the
support of the product $f*g$ can be non void although the
intersection of the supports of the functions $f$ and $g$ is void.
It is seen to be related to a "commutative" convolution product
between symplectic Fourier transforms $\tilde{f}$ and $\tilde{g}$,
but now in some sense "deformed" by the integral kernel
$e^{-i\tilde{\omega}\left(a,b\right)/2\hbar}$, whose origin lies
in the symplectic structure on the phase space, that plays a
crucial role in the Weyl form of the commutation relations.

The Moyal product, written in the nonlocal form
(\ref{genmoyalprodint}), can be also cast as: \beq
\left(f*g\right)\left(z\right)\,=\,\frac{1}{\left(\pi\hbar\right)^{2n}}\int
dtdv\,f\left(t+z\right)g\left(v+z\right)\,e^{2i\tilde{\omega}\left(t,v\right)/\hbar}
\label{moyalprodrieffel} \eeq This integral form has been
exploited by M.Rieffel in a remarkable monograph \cite{rieffelRd},
as a starting point for a theory of general deformation of
$C^{*}-$ algebras\footnote{In appendix \ref{appendixcalg} there is an introduction to the algebraic
concepts mentioned in these pages}. The first space on which studying mathematical
properties of the Moyal product is the space of Schwartzian
functions $\mathcal{S}^{\infty}\left(\real^{2n}\right)$. It is
possible to prove \cite{marseticos} that Moyal product is
associative in the set of Schwartzian functions. The complex
conjugation defines an involution in this space, and in the limit
of $\hbar\,\rightarrow\,0$, one has
$\left(f*g\right)\left(z\right)=\left(fg\right)\left(z\right)$ for
every point $z$ in the vector space $L=\real^{2n}$. This can be summarised by
saying that
$\mathcal{A}_{\hbar}=\left(\mathcal{S}^{\infty}\left(\real^{2n}\right),*\right)$
is an associative, nonunital (because the function identically
equal to $1$ is not Schwartzian), involutive (involution is given
by the complex conjugation) algebra with a continuous product.

The Moyal product can be defined, by duality, on a set larger than
$\mathcal{S}^{\infty}$. The dual space of that of the Schwartzian
functions is the space of the tempered distributions
($\mathcal{S}^{\prime}$). For
$F\,\in\,\mathcal{S}^{\prime}\left(\real^{2n}\right)$, the
evaluation on $f\,\in\mathcal{S}\left(\real^{2n}\right)$ can be
written as a kind of scalar product $\langle
F,f\rangle\,\in\,\complex$. This notation is intended as a
shorthand for the integral of the kernel of the distribution $F$
times the function $f$. It is possible to define $F*f$ and $f*F$
as elements of $\mathcal{S}^{\prime}$ by: \beq \langle
F*f,g\rangle\,=\langle F,f*g\rangle\,\,\,\,\,\,\,\,\langle
f*F,g\rangle\,=\,\langle F,g*f\rangle \eeq while the involution is
extended to $\mathcal{S}^{\prime}$ by: \beq \langle
F^{\dagger},f\rangle\,=\,\overline{\langle F,f^{\dagger}\rangle}
\eeq It is possible to consider the left and right multiplier
algebras: \beqa \mathcal{M}_{\hbar}^{L}\,&=&\,\{
F\,\in\,\mathcal{S}^{\prime}\left(\real^{2n}\right)\,:\,F*g\,\in\,
\mathcal{S}\left(\real^{2n}\right)\,\forall\,g\,\in\,\mathcal{S}\left(\real^{2n}\right)\}\nonumber
\\
\mathcal{M}_{\hbar}^{R}\,&=&\,\{ F\,\in\,\mathcal{S}^{\prime}\left(\real^{2n}\right)\,:\,g*F\,\in\,
\mathcal{S}\left(\real^{2n}\right)\,\forall\,g\,\in\,\mathcal{S}\left(\real^{2n}\right)\}
\eeqa
The intersection of the two gives the multiplier algebra:
\beq
\mathcal{M}_{\hbar}\,=\,\mathcal{M}^{L}_{\hbar}\,\cap\,\mathcal{M}^{R}_{\hbar}
\eeq
This $\mathcal{M}_{\hbar}$ is a complete, unital $*-$algebra. It contains the identity, the constant function, the plane waves,
i.e. functions of the form $e^{izw}$, and even the Dirac $\delta$ function and the monomials
in the coordinates. So this algebra is a compactification of $\mathcal{S}^{\infty}$ defined by duality.

This algebra is the \emph{noncommutative Moyal plane}.

This analysis suggests to explicitly calculate the form of this
product in the case $f$ and $g$ are two Schwartzian functions. It
can be written as:
  \beq f*
g=f\left(z\right)e^{-\frac{i\hbar}{2}[\frac{\stackrel{\leftarrow}{\del}}{\del
q^{a}}\frac{\stackrel{\rightarrow}{\del}}{\del
p_{a}}-\frac{\stackrel{\leftarrow}{\del}}{\del
p_{a}}\frac{\stackrel{\rightarrow}{\del}}{\del
q^{a}}]}g\left(z\right) \label{diffmoyalstand}\eeq where the
arrows over the partial derivatives symbol indicate which is the
function on which they act. Equivalently: \beq
f*g=fe^{-\frac{i\hbar}{2}\tilde{\Lambda}_{ab}
\stackrel{\leftarrow}{\del_{a}}\wedge\stackrel{\rightarrow}{\del_{b}}}
g \eeq The fact that the Moyal product of $f$ times $g$ depends on
all the derivatives of $f$ and $g$ is a different way to look at
it as a nonlocal product. It is important to note that this notion
of nonlocality is not equivalent to that given before. The first
terms of this expansion are:
\begin{eqnarray} f*g&=&f{\cdot}
g+\frac{i\hbar}{2}\left(\frac{\del f}{\del p_{a}}\frac{\del g}{\del
q^{a}}- \frac{\del f}{\del q^{a}}\frac{\del g}{\del
p_{a}}\right)+o\left(\hbar^{2}\right)\\ {}
 \nonumber &=& f{\cdot}g-\frac{i\hbar}{2}\{f,g\}+o\left(\hbar^{2}\right)\end{eqnarray}
This expression shows that Moyal product can be seen as a
deformation of the usual pointwise among functions on the
plane\footnote{Wigner-Weyl-Moyal method, writing products and commutators of operators in phase space language,
has been instrumental in giving rise to the subject of deformation quantization \cite{defquant}.}.

The dependence on the parameter $\hbar$ is more intuitive in this
expression, as the meaning of the limit $\hbar\rightarrow 0$.

 Moreover, a careful analysis shows that formula (\ref{diffmoyalstand}) can be extended to evaluate the
product even between some functions not belonging to the space
$S^{\infty}\left(\real^{2}\right)$. For the generators: \beqa
q^{a}*p_{b}\,&=&\,q^{a}p_{b}-\frac{i\hbar}{2}\delta^{a}_{b} \nonumber \\
p_{b}*q^{a}\,&=&\,q^{a}p_{b}+\frac{i\hbar}{2}\delta^{a}_{b} \eeqa
The noncommutative plane algebra $\mathcal{M}_{\hbar}$ can be seen as the algebra formally generated by these noncommuting
coordinates.

An algebraic type analysis can be done even from a different
starting point. In the previous section it has been shown how
square integrable functions on the vector space $\real^{2n}$ are
mapped by the Weyl map onto Hilbert-Schmidt operators. So, it is
natural to study the Moyal product among these functions.
Properties of Hilbert-Schmidt operators show that, if $f$ and $g$
are in $\mathcal{L}^{2}\left(\real^{2n}\right)$, then the product
$f*g$ is still square integrable: moreover, the product is
continuous for $\hbar \rightarrow\,0$. This algebra can be
extended. One can define: \beq
A_{\hbar}\,=\,\{F\,\in\,\mathcal{S}^{\prime}\left(\real^{2n}\right)\,:\,F*g\,\in\,\mathcal{L}^{2}
\left(\real^{2n}\right)\,\forall\,g\,\in\,\mathcal{L}^{2}\left(\real^{2n}\right)\}
\eeq equipped with a norm, mutuated by the $\mathcal{L}^{2}$ norm:
\beq
\parallel F\parallel_{\hbar}\,=\,\sup\{\parallel F*g\parallel_{\mathcal{L}^{2}}/\parallel g\parallel
_{\mathcal{L}^{2}}\,:\,0\neq
g\,\in\,\mathcal{L}^{2}\left(\real^{2n}\right)\} \eeq This algebra
$\left(A_{\hbar},\parallel\cdot\parallel_{\hbar}\right)$ is proved
to be a unital $C^{*}-$algebra, isomorphic to the set of bounded
operators
$\mathcal{B}\left(\mathcal{L}^{2}\left(\real^{n}\right)\right)$.
It is interesting to note that the Weyl-Wigner isomorphism defines
a realization of the GNS construction for this algebra
\cite{pepevarillyGNS}. The Algebra $A_{\hbar}$ contains integrable
functions $\mathcal{L}^{1}\left(\real^{2n}\right)$ and plane
waves, but it does not contain non constant polynomials.

Since plane waves belong both to $A_{\hbar}$ and
$\mathcal{M}_{\hbar}$, the Moyal product for two plane waves of
covectors $u$ and $w$ is given by \beq
e^{iuz}\,*\,e^{iwz}\,=\,e^{i\tilde{\Lambda}\left(u,w\right)/2\hbar}\,e^{i\left(u+w\right)z}
\eeq So plane waves close an algebra, the Weyl algebra, that
represents an action of the translation group on the vector space
$\real^{2n}$. This action is put in a more intuitive form if a
"symplectic" plane wave basis is used: \beq
e^{i\tilde{\omega}\left(u,z\right)/\hbar}\,*\,e^{i\tilde{\omega}\left(w,z\right)/\hbar}\,=\,
e^{i\tilde{\omega}\left(u,w\right)/2\hbar}\,e^{i\tilde{\omega}\left(u+w,z\right)/\hbar}\eeq

\section{The classical limit of quantum mechanics in the
Weyl-Wigner formalism}

In the previous section it has been analysed the nature of the
Moyal product, and of the noncommutative algebra structure it
gives to the set of functions on the cartesian phase space.

The skewsymmetrised form of this product gives:
\begin{eqnarray}
\{f,g\}_{{\scriptscriptstyle M}}&\equiv&\frac{i}{\hbar}\left(f*
g-g*f\right) \nonumber \\ {}&=& \{f,g\}+o\left(\hbar\right)
\label{moyalbracket}\end{eqnarray} and this is called Moyal
bracket. It is the bilinear map that translates, in the set of
functions, the notion of commutator in the set of operators. This
is the reason why this map is bilinear, satisfies the Jacobi
identity, and satisfies a Leibniz rule with respect to the Moyal
product: \beq \{f*g,h\}_{{\scriptscriptstyle M}}\,=\,
f*\{g,h\}_{{\scriptscriptstyle
M}}\,+\,\{f,h\}_{{\scriptscriptstyle M}}*g \eeq

Relations (\ref{moyalbracket}) define a deformation of the Poisson
structure in the set of functions on the phase space.

In the space of functions on the plane, the introduction of Moyal
bracket enables to write:
\begin{eqnarray}
\{q^{a},p_{b}\}_{{\scriptscriptstyle M}}&=&\delta^{a}_{b}
\nonumber\\ {} \{q^{a},H\}_{{\scriptscriptstyle M}}&=&\frac{\del
H}{\del p_{a}} \nonumber
\\ {} \{p_{a},H\}_{{\scriptscriptstyle M}}&=&-\frac{\del H}{\del q^{a}}
\end{eqnarray}
Whatever the function $H$ were, the Moyal bracket with coordinate
functions gives the same result the Poisson bracket would give. Derivations 
associated with the coordinate functions are the same both in the classical 
algebra and in the ``deformed'' algebra.
So, what is the meaning of the expression for a generic $f$?

In classical formalism a theorem already mentioned says that smooth
derivations for the abelian algebra of functions on the phase
space are represented by vector fields, and vector fields are
infinitesimal generators of classical dynamics. So ``classical'' 
derivations are
related to the classical evolution of the observables.

If the same set of functions is given a non abelian algebraic
structure via the Moyal product, then the Moyal bracket provides a
class of derivations of this algebra. A function on the phase
space represents, via the Moyal bracket, a derivation of the
quantum algebra. These derivations are related to quantum
dynamics.

 If one considers the quantum evolution in the Heisenberg
picture, then operators are evolved: \beq
\Ao\left(t\right)=\Uod\left(t\right)\Ao\Uo\left(t\right) \eeq and,
applying the Wigner map to both sides, it can be written in terms
of an evolution of the symbols: \beq
A\left(t\right)=U^{*}\left(t\right)*A*U\left(t\right) \eeq If one
considers the infinitesimal form of this relation, with a quantum
evolution operator in the form $\Uo\left(t\right)=e^{-i\Ho
t/\hbar}$, this equation is written: \beqa
\frac{d}{dt}A\left(t\right)&=&\{H,A\left(t\right)\}_{\scriptscriptstyle
M} \nonumber \\ &=&\{H,A\left(t\right)\}+o\left(\hbar\right)
\label{moyalpoisson}\eeqa Derivations given by an element of the
algebra via the Moyal bracket represent the infinitesimal
form of a quantum dynamics in the Heisenberg picture.

Moreover, since the Moyal bracket is a deformation of the Poisson
bracket, this equation shows that the Weyl-Wigner formalism
enables to write the quantum evolution in terms of equations
involving functions on the phase space, carrying a classical
dynamics, in such a way to recover, in the limit
$\hbar\,\rightarrow\,0$, the classical evolution for the Weyl
symbols in the Poisson formalism.

This is the meaning of the classical limit procedure in the Weyl-Wigner formalism.

\section{Generalizing Weyl systems}

In the previous sections, a Weyl system has been defined
(\ref{defws}) in terms of a generic, though translationally
invariant, symplectic form on the vector phase space $L$.
Nevertheless the explicit realization of the Schr\"{o}dinger
representation and of the Weyl-Wigner maps have been studied in
the case of the symplectic form $\omega$ being in the canonical
Darboux form $\tilde{\omega}$. The aim of this section is to
generalize this explicit realization of a Weyl system. The first
generalization will be introduced to cover the case of a
symplectic structure $\omega$ which is no more in the canonical
form. The second generalization will be the study of a Weyl
system, defined as a unitary projective representation of the
abelian group of translations, with phase factors no longer simply
corresponding to a symplectic form.

\subsection{Weyl systems for translationally invariant symplectic structures}

To consider this first generalization of the Weyl system's notion,
it is useful to review the topics covered in the first two
subsections (\ref{schrorep}) and (\ref{staweylsys}). Let
$\Do\left(z\right)$ be a Weyl system (\ref{wsys1}) for
$\left(L,\tilde{\omega}\right)$. Let $T$ be an automorphism (an
invertible and linear map) in the vector space $L$. It does not
need to be symplectic: $T\in\mathcal{A}ut\left(L\right)$. Acting
with such a transformation on $L$, one has: \beq
\Do\left(Tz+Tu\right)=\Do\left(Tz\right)\Do\left(Tu\right)
e^{\frac{i}{2\hbar}\tilde{\omega}\left(Tz,Tu\right)} \eeq by
linearity:\beq
\Do\left(T\left(z+u\right)\right)=\Do\left(Tz\right)\Do\left(Tu\right)
e^{\frac{i}{2\hbar}\omega\left(z,u\right)} \eeq In this relations
it has been considered that the transformation of vectors of $L$
by such a $T$, can be dually read as a transformation of the
symplectic structure: \beq
\omega\left(z,u\right)\,\equiv\tilde{\omega}\left(Tz,Tu\right)\,\,\,\rightarrow\,\,\omega=\,T^{t}\,\tilde{\omega}\,T
\label{symplcov}\eeq Now it is possible to define: \beq
\Do\left(Tz\right)\equiv\Do_{T}\left(z\right)
\label{genweylsysT}\eeq obtaining: \beq
\Do_{T}\left(z+u\right)=\Do_{T}\left(z\right)\Do_{T}\left(u\right)
e^{\frac{i}{2\hbar}\omega\left(z,u\right)}\label{genweylsysT2}
\eeq In this approach, $\Do_{T}\left(z\right)$ is a Weyl system
for $\left(L,\omega\right)$ with $\omega$ obtained as
(\ref{symplcov}). Of course, $\Do_{T}\left(z\right)$ is recovered
as a standard Weyl system if the automorphism $T$ is also
symplectic. Moreover, the general analysis developed in the
previous sections clarifies that, if $T^{\prime}=T_{S}T$
($T^{\prime}$ is the composition of the automorphism $T$ with the
symplectic automorphism $T_{S}$) then the Weyl system
$\Do_{T^{\prime}}\left(z\right)$ is unitarily equivalent to
$\Do_{T}\left(z\right)$.

Properties of invertible skewsymmetric matrices whose coefficient
are constant, which represents translationally invariant
symplectic forms on the vector space $L$, show that the problem of
realizing a Weyl system for $\left(L,\omega\right)$ is solved by
$\Do_{T}\left(z\right)=\Do\left(Tz\right)$ where
$\Do\left(z\right)$ is a Weyl system for
$\left(L,\tilde{\omega}\right)$ (\ref{wsys1}), and $T$ is the
automorphism that solves the equation
$T^{t}\tilde{\omega}T=\omega$. Such a $T$ is defined by this
requirement and recovered up to the composition with an arbitrary
symplectic automorphism.

The problem of quantizing the dynamics of a particle in a constant
background magnetic field can be studied with these tools.
Classical evolution is formalized on $T^{*}\real^{3}$ in terms of
a symplectic structure with constant coefficients depending by the
components of the $\vec{B}$ fields, and a quadratic hamiltonian
function. Equations of motions are: \beqa
\dot{q}^{i}&=&p^{i}\nonumber \\ {}
\dot{p}_{i}&=&\epsilon_{i}^{jk}p_{j}B_{k} \eeqa for position
coordinates $q^{i}$ and gauge invariant momenta coordinates
$p_{i}$. These equations define an Hamiltonian vector field with
respect to the Poisson structure defined by:
 \beqa \{q^{i},q^{j}\}&=&0 \nonumber \\ {}
\{q^{i},p_{j}\}&=&\delta^{i}_{j} \nonumber \\ {}
\{p_{i},p_{j}\}&=&\epsilon_{ij}^{k}B_{k} \eeqa whose symplectic
counterpart is:\beq \omega\,=\,-\epsilon_{ijk}B^{k}dq^{i}\wedge
dq^{j}+dq^{i}\wedge dp_{i}\eeq The realization of a Weyl system
for this symplectic structure gives rise to a quantum Hamiltonian
operator defined without any reference to the vector potential
$\nabla\times\vec{A}=\vec{B}$, as it is the case in the standard
approach to quantization in terms of the so called minimal
coupling procedure. Even an ambiguity in the definition of this
Hamiltonian operator, due to the gauge transformation properties
of the vector potential, can be recovered inside this formalism:
it is related to the invariance of solution for a Weyl system by
symplectic transformations.

This generalization of a Weyl system can be naturally brought into
a generalization of the Weyl map, and the Moyal product. Formula
(\ref{weylmap}) can be extended to the case just shown: \beq
\hat{\Omega}\left(f\right)\,=\,\int\frac{dw}{\left(2\pi\hbar\right)^{n}}\left|T\right|
\tilde{f}\left(w\right)\Do_{T}\left(w\right) \label{symweylmap}
\eeq $f$ is a function on $\left(L,\omega\right)$ The symplectic
Fourier transform (appendix \ref{digrSFT})$\tilde{f}$ is obtained
via an automorphism $T$ that brings the symplectic form $\omega$
in the Darboux form $\tilde{\omega}$: $T^{t}\tilde{\omega}T$. This
transformation $T$ is the same that can be used to define the Weyl
system for $\left(L,\omega\right)$: $\left|T\right|$ is the
determinant of the matrix representing $T$.

This formula for the Weyl map simply shows that the role of the
$T$ transformation is to cast the problem in the form of defining
the Weyl map for functions on a vector space in a coordinate chart
in which the symplectic structure has the Darboux form.

 Even
this Weyl map is invertible, and the Wigner map is given by: \beq
\tilde{f}\left(w\right)\,=\,Tr\left[\fo\Do^{\dagger}\left(w\right)\right]
\label{symwignermap} \eeq This bijection now can be used to define
a Moyal product in a space of functions on the space $S$. The
relation \beq
\hat{\Omega}\left(f*g\right)\,=\,\hat{\Omega}\left(f\right)\hat{\Omega}\left(g\right)
\eeq acquires the nonlocal form: \beqa
\left(f*g\right)\left(z\right)&=&
\frac{\left|T\right|^{4}}{\left(2\pi\hbar\right)^{4n}}\int
dxdy\int d\beta d\alpha\,
f\left(x\right)g\left(y\right)\,e^{-i\omega\left(x,\beta\right)/\hbar}
e^{-i\omega\left(y,\alpha-\beta\right)/\hbar}e^{-i\omega\left(\alpha,z\right)/\hbar}
e^{i\omega\left(\beta,\alpha\right)/2\hbar} \nonumber \\ &=&
\frac{\left|T\right|^{2}}{\left(2\pi\hbar\right)^{2n}}\int\,d\alpha\,e^{-i\omega\left(\alpha,z\right)/\hbar}
\,\int\,d\beta\,
\tilde{f}\left(\beta\right)\tilde{g}\left(\alpha-\beta\right)\,
e^{i\omega\left(\beta,\alpha\right)/2\hbar}
\label{genmoyalprodint}\eeqa This product generalizes the product
(\ref{intmp}) for a generic translationally invariant symplectic
2-form. This generalization can be seen also in the asymptotic
expansion. On a suitable domain of functions, this product can be
written in a differential form, that is the generalization of
(\ref{diffmoyalstand}) \beq
f*g\,=\,f\,e^{-\frac{i}{2\hbar}\left(\overleftarrow{\del}_{a}\Lambda_{ab}\overrightarrow{\del}_{b}\right)}\,g
\label{expgenmoyal}\eeq Here the Poisson tensor is related to the
inverse of the 2-form represented by $\omega$. One of the reason
why it has been used the definition of symplectic Fourier
transform is that, if it were not, all this procedure would have
not been covariant for the whole symplectic group. Moreover, the
asymptotic expansion of the product (\ref{expgenmoyal}) would not
have been an exponentiation of the Poisson bivector, thus
eliminating the possibility to generalize the analysis culminated
in (\ref{moyalpoisson}) on the classical limit.

\subsection{Weighted Weyl systems}\label{onemorestep}

In the previous sections, the space $L$ has been looked at as a
real, even dimensional linear space. To proceed along the path of
studying a generalization of the concept of Weyl systems, $L$ can
be now considered as a realization of the abelian group of
translations $\left(\real^{2n},+\right)$, while $\Do$ as a
projective unitary representation of this group, where the phase
factors are given by the symplectic structure.

It is natural to consider now the definition of a more general
unitary representation for this group: \beq
\Do_{\Phi}\left(z+u\right)\,=\,e^{i\Phi\left(z,u\right)/2\hbar}\,\Do_{\Phi}\left(z\right)\,\Do_{\Phi}
\left(u\right) \eeq The obvious demand that this representation
preserves the associativity of group composition forces tha phase
factors to satisfy a peculiar condition: \beq
\Phi\left(z,u+v\right)+\Phi\left(u,v\right)\,=\,\Phi\left(z,u\right)+\Phi\left(z+u,v\right)
\eeq Without entering into a cohomological characterization of
this relation, it is enough to say that such a $\Phi$ is called a
cocycle. It is important to note that, if $\Phi\left(z,u\right)$
is linear in both entries, then it is necessarily a cocycle.
Following this analysis, it is clear that a standard Weyl system
(\ref{defws}) is such a representation, in which the group is even
dimensional, and the cocycle is a skewsymmetric nondegenerate
bilinear function.

A generalization of that construction is given by a choice of
$\Phi\left(z,u\right)$ with a nondegenerate skewsymmetric part,
and a nonvanishing symmetric part: $$
\Phi\left(z,u\right)\,=\,A\left(z,u\right)+S\left(z,u\right) $$
where: $$
A\left(z,u\right)\,=\,\frac{1}{2}\left[\Phi\left(z,u\right)-\Phi\left(u,z\right)\right]$$
$$
S\left(z,u\right)\,=\,\frac{1}{2}\left[\Phi\left(z,u\right)+\Phi\left(u,z\right)\right]$$
Now it is possible to follow the same path developed in the study
of standard Weyl systems. These definitions enable to write: \beq
\Do_{\Phi}\left(\alpha z\right)\Do_{\Phi}\left(\beta
z\right)\,=\,\Do_{\Phi}\left(\beta z\right)\Do_{\Phi}\left(\alpha
z\right) \eeq \beq
\Do_{\Phi}\left(\left(\alpha+\beta\right)z\right)\,=\,e^{i\alpha\beta
S\left(z,z\right)/2\hbar}\, \Do_{\Phi}\left(\alpha
z\right)\Do_{\Phi}\left(\beta z\right)\label{genwsequ} \eeq This
means that, restricted on a one dimensional subspace, $\Do_{\Phi}$
is no longer a faithful representation of the additive line.
$\Do_{\Phi}\left(\alpha z\right)$ is not anymore a one parameter
group of unitary operators. Stone's theorem cannot be invoked to
define generators to identify with physical observables.

Using as a guide the standard Weyl systems theory, it is natural
to define a set of hermitian operators $\Go\left(z\right)$
depending on an element of the group, and a real function of two
variables $w\left(\alpha,z\right)$ by the relation: \beq
\Do_{\Phi}\left(\alpha
z\right)\,=\,e^{i\left[\alpha\Go\left(z\right)+w\left(\alpha,z\right)\right]/\hbar}
\eeq Equation (\ref{genwsequ}) is satisfied if: \beq
w\left(\alpha+\beta,z\right)-w\left(\alpha,z\right)-w\left(\beta,z\right)\,=\,\frac{1}{2}\,S\left(z,z\right)\,
\alpha\beta \label{weightequ}\eeq Moreover, it can be seen that
this function $w$ should satisfy a sort of homogeneity condition
in the $z$ variable: \beq w\left(\alpha+\beta,\gamma
z\right)-w\left(\alpha,\gamma z\right)-w\left(\beta,\gamma
z\right)\,=\,\frac{\gamma^{2}}{2}\,S\left(z,z\right)\,\alpha\beta \eeq If one
tries to obtain the commutation relations among so defined
hermitian "generators", from the definition of Weyl systems: $$
\Do_{\Phi}\left(\alpha z+\beta
u\right)\,=\,e^{i\alpha\beta\Phi\left(z,u\right)/2\hbar}\,
\Do_{\Phi}\left(\alpha z\right)\Do_{\Phi}\left(\beta u\right) $$
$$ \Do_{\Phi}\left(\alpha z+\beta
u\right)\,=\,e^{i\alpha\beta\Phi\left(u,z\right)/2\hbar}\,
\Do_{\Phi}\left(\beta u\right)\Do_{\Phi}\left(\alpha z\right) $$
one obtains: \beq
\left[\Go\left(z\right),\Go\left(u\right)\right]\,=\,i\hbar
A\left(z,u\right) \eeq This indicates that commutation rules among
generators depend only by the skewsymmetric part of the cocycle
factor. The solution of equation (\ref{weightequ}) is: \beq
w\left(\alpha,
z\right)\,=\,\frac{\alpha^{2}}{4}\,S\left(z,z\right)
\label{weightsol} \eeq Campbell-Baker-Hausdorff formula enables to
cast a Weyl system in the form: \beq
\Do_{\Phi}\left(z\right)\,=\,e^{i\left[z^{a}\Go\left(e_{a}\right)+w\left(z^{a},e_{a}\right)\right]/\hbar}
\eeq

The generalization of the notion of Weyl system to the case of a
generic bilinear cocycle for the translation group is then of the
form: \beq
\Do_{\Phi}\left(z\right)\,=\,\Do\left(z\right)\,e^{iS\left(z,z\right)/4\hbar}
\eeq Here $\Do\left(z\right)$ is a Weyl system for a symplectic
structure given by the skewsymmetric part of the cocycle $\Phi$.
The extra term can be seen as a weight, depending only on the
symmetric part of the cocycle.

\subsection{Weighted Weyl map}\label{weightedwmap}

The generalization studied in the previous subsection is very
important, because it enables to define a different Weyl map,
which means a different ordering in going from commutative
variables to noncommutative ones, for the algebra of functions on
the space $\real^{2n}$. In the following it will be considered the
case of a cocycle $\Phi$ whose skewsymmetric part is in the
canonical form.

A Weyl map is now generalised to be: \beqa
\hat{\Omega}_{\Phi}\left(f\right)\,&=&\,\int
\frac{du}{\left(2\pi\hbar\right)^{n}}\,\tilde{f}\left(u\right)\,\Do_{\Phi}\left(u\right)
\nonumber
\\
&=&\,\int
\frac{du}{\left(2\pi\hbar\right)^{n}}\,\tilde{f}\left(u\right)\,e^{iS\left(u,u\right)/4\hbar}\,\Do\left(u\right)
\label{weighwm} \eeqa Also this map can be inverted: \beq
\tilde{f}\left(u\right)\,=\,e^{-iS\left(u,u\right)/4\hbar}\,
Tr\left[\hat{\Omega}_{\Phi}\left(f\right)\Do_{\Phi}^{\dagger}\left(u\right)\right]
\eeq This relation is based on the fact that operators
$\Do_{\Phi}$, being the product of standard $\Do$ times a phase,
close a relation of the same kind of (\ref{Dcomplet}). Two
examples are interesting: just to symplify notations, the vector
space considered will be of dimension 2, and vector $z$ will have
components $\left(q,p\right)$

The first is the case when the symmetric matrix $S$ is given by:
$$S\,=\,\left(\begin{array}{cc} 0 & 2 \\ 2 & 0\end{array}\right)$$
It can be seen that the coordinate function $q$ is mapped into the
generator $\Qo$ of a standard Weyl system, and the coordinate
function $p$ is mapped into the standard generator $\Po$. But this
new quantizing map defines a peculiar ordering for images of
monomials (cfr(\ref{standardmonomial})): \beq
\hat{\Omega}_{\Phi}\left(q^{a}p^{b}\right)\,=\,\Po^{b}\Qo^{a} \eeq
The noncommutative product one obtains in the space of functions
on the plane, via this \emph{weighted} Weyl map, is, in the formal
expansion valid on a suitable domain: \beqa
\left(f*_{\scriptstyle S}g\right)\left(q,p\right)\,&=&\,\sum_{k=0}^{\infty}\frac{\left(-i\hbar\right)^{k}}{k!}
\,\frac{\del^{k}f}{\del q^{k}}\,\frac{\del^{k}g}{\del
p^{k}}\nonumber
\\
\left(f*_{\scriptstyle S}g\right)\left(q,p\right)\,&=&\,f\,
e^{-i\hbar\left(\overleftarrow{\del}_{q}\overrightarrow{\del}_{p}\right)}\,g
\eeqa

The second is the case when the symmetric matrix $S^{\prime}$ is
given by: $$S^{\prime}\,=\,\left(\begin{array}{cc} 0 & -2 \\ -2 &
0\end{array}\right)$$ Even in this case, it can be seen that the
coordinate function $q$ is mapped into the generator $\Qo$ of a
standard Weyl system, and the coordinate function $p$ is mapped
into the standard generator $\Po$. Now the ordering this
quantizing map defines is, on images of monomials
(\ref{standardmonomial}): \beq
\hat{\Omega}_{\Phi}\left(q^{a}p^{b}\right)\,=\,\Qo^{a}\Po^{b} \eeq
The noncommutative product one obtains in the space of functions
on the plane, via this \emph{weighted} Weyl map, is, in the formal
expansion valid on a suitable domain: \beqa \left(f*_{\scriptstyle
S^{\prime}}g\right)\left(q,p\right)\,&=&\,\sum_{k=0}^{\infty}\frac{\left(i\hbar\right)^{k}}{k!}
\,\frac{\del^{k}f}{\del p^{k}}\,\frac{\del^{k}g}{\del
q^{k}}\nonumber
\\
\left(f*_{\scriptstyle S^{\prime}}g\right)\left(q,p\right)\,&=&\,f\,
e^{-i\hbar\left(\overleftarrow{\del}_{p}\overrightarrow{\del}_{q}\right)}\,g
\eeqa

It is very interesting to note that this two deformed products are
equivalent to the standard Moyal product,when the equivalence is
defined by the realization of an operator such that: \beqa
T^{\left(S\right)}\,:\,\left(\mathcal{F}\left(\real^{2}\right),*_{\scriptstyle
S}\right)\,\mapsto\,
\left(\mathcal{F}\left(\real^{2}\right),*\right) \nonumber \\
T^{\left(S\right)}\left(f*_{\scriptstyle S}g\right)
\,=\,\left(T^{\left(S\right)}f\right)*
\left(T^{\left(S\right)}g\right) \eeqa The operators, for these
two cases, can be proved to be \cite{zachos}: \beqa
T^{\left(S\right)}\,=\,\sum_{n=0}^{\infty}
\left(\frac{i\hbar}{2}\right)^{n}\,\frac{1}{n!}\,\left(\frac{\del}{\del
p}\right)^{n} \left(\frac{\del}{\del q}\right)^{n} \nonumber
\\
T^{\left(S^{\prime}\right)}\,=\,\sum_{n=0}^{\infty}
\left(-\frac{i\hbar}{2}\right)^{n}\,\frac{1}{n!}\,\left(\frac{\del}{\del
p}\right)^{n} \left(\frac{\del}{\del q}\right)^{n} \eeqa

\section{Weyl map from coherent states for the Heisenberg-Weyl-Wigner group}\label{WeylmapplaneCS}

Throughout this chapter, it has been pointed out how the
Weyl-Wigner formalism can be studied stressing the accent on a
group theoretical approach. A standard Weyl system has been
realized by Displacement operators, as a unitary projective
representation of the translation group, where phase factors are
related to the symplectic structure of the linear space on which
it acts. This clarifies some aspects of the deep contact between
geometrical foundations in the formulation of classical and
quantum dynamics. Even the way the first generalization of section
\ref{onemorestep} has been presented, goes towards an analysis
of a more general class of representations for the same group.

In section \ref{secweylmap} Weyl map has been written also
using a set of Weyl operators $\Wo\left(z\right)$, whose
properties are summarized in (\ref{weylopprop}). Displacement
operators have been defined via their composition properties. What
is the composition rule for this system of Weyl operators? It can
be checked: \beqa
\Wo\left(z\right)\,\Wo\left(z^{\prime}\right)\,&=&\,4^{n}\,
e^{2i\tilde{\omega}\left(z,z^{\prime}\right)/\hbar}\Do\left(z-z^{\prime}\right)
\nonumber \\ &=& 2^{n}\,
e^{2i\tilde{\omega}\left(z,z^{\prime}\right)/\hbar}\Wo\left(z-z^{\prime}\right)\hat{\mathcal{P}}
\eeqa This relation says that Weyl operators do not define a
group. But the introduction of the system of $\Wo\left(z\right)$
operators acquires an interesting geometrical meaning if it is
seen in the perspective of a representation of the so called
Heisenberg-Weyl-Wigner (HWW) group.

In the previous sections of this chapter, to stress that the
formalism was born to study the problem of quantization for a
classical dynamics, and classical limit for a quantum dynamics,
the role of $\hbar$ has been kept explicitly. From an algebraic
point of view, which is the one noncommutative geometry starts
from, $\hbar$ is just a parameter. It is the parameter that
represents a non commutativity in the quantum relations for
canonical observables, and it has been considered as a deformation
parameter in a formalism developed to unify both the classical and
the quantum ones.

In this section, since the stress will be just on geometrical
aspects of the formalism, an identification of the quantities with
physical observables will be abandoned. The deformation parameter
will be a constant $\theta$, and the space $L$ will be the space
$\real^{2}$ with coordinates $q$ and $p$ without a dimension of
position or momentum. And this is at the light of future themes of
this dissertation, where this Weyl formalism will be used to study
some specific non commutative spaces.

Canonical commutation relations define the Lie algebra of the
Heisenberg-Weyl (HW) group. The group manifold is $\real^{3}$, and
elements of the group are labelled as triples
$\left(q,p,\lambda\right)$. The composition rule is: \beq
\left(q,p,\lambda\right)\cdot
\left(q^{\prime},p^{\prime},\lambda^{\prime}\right)=
\left(q+q^{\prime},p+p^{\prime},\lambda+\lambda^{\prime}+\frac{1}{2}\left(qp^{\prime}-q^{\prime}p\right)\right)
\eeq Now the idea is to define a new group, obtained as a
semi-direct product of HW with the group $\zed_{2}$.

Among triples of the form
$\left(\xi,\lambda,\alpha\right)$ (where $\xi$ is a complex
number, and represents a point in a complex plane, $\lambda$ is a
real number, and $\alpha$ can take the discrete values $\pm 1$) it is possible to define
a composition by:
\beq
\left(\xi,\lambda,\alpha\right)\cdot\left(\xi^{\prime},\lambda^{\prime},\alpha^{\prime}\right)
\,\equiv\,\left(\xi+\alpha\xi^{\prime},\lambda+\lambda^{\prime}+\frac{i}{2\theta}\alpha
\left(\bar{\xi}\xi^{\prime}-\bar{\xi}^{\prime}\xi\right),\alpha\alpha^{\prime}\right)
\eeq
Then the set acquires the structure of a group, that is
topologically equivalent to 2 copies of $\real^{3}$. This is
called Heisenberg-Weyl-Wigner (HWW) group. The
identity element is
\be
\idop_{W^{\prime}}=\left(0,0,1\right)
\ee
and the inverse of a generic element is:
\be
\left(\xi,\lambda,\alpha\right)^{-1}=\left(-\alpha\,\xi,-\lambda,\alpha\right)
\ee

The next step is to define a system of coherent states for this
group. To this extent, a unitary irreducible representation of it
on a Hilbert space should be considered. Following Bargmann and
Fock \cite{perelomov}, one can introduce a space of functions
which are complex analytical in a $w$ variable, endowed with a
scalar product: \beq \la f\mid
g\ra\,\equiv\,\int\,\frac{d^{2}w}{\pi\theta}\,e^{-\bar{w}w/\theta}\,\bar{f}\left(w\right)g\left(w\right)
\label{fockspace}\eeq The Fock space $\fhil$ will be defined as
the set of those functions whose norm, resulting from this scalar
product, is finite. This can be proved to be a Hilbert space. On
this space a set of operators is defined:
$\hat{W^{\prime}}\left(\xi,\lambda,\alpha\right)$ whose action on
$f\,\in\,\fhil$ is given by: \beq
\left(\hat{W^{\prime}}f\right)\left(w\right)=
e^{i\lambda}e^{-\bar{\xi}\xi/2\theta}e^{\xi w/\theta}
f\left(\alpha\left(w-\bar{\xi}\right)\right) \label{unirephhw}
\eeq In these definitions a parameter $\theta$ has been
introduced. It can be thought to have the dimension of the square
of $w$ and $\xi$ variables, while $\lambda$ is considered
adimensional. In this space the most natural orthonormal basis is
\beq \psi_{n}\left(w\right)\,=\,\frac{w^{n}}{\sqrt{\theta^{n}n!}}
\eeq To define a system of coherent states, a \emph {fiducial
vector} in $\fhil$ must be chosen. The easiest choice is, of
course, $\psi_{0}\left(w\right)$. The action of $\hat{W^{\prime}}$
operators on this vector gives a new set of vectors in $\fhil$.
Among elements of the group, there are some whose action via the
representation gives just the $\psi_{0}$ multiplied by a phase:
these elements constitute the so-called isotropy subgroup of the
HWW group for the chosen fiducial state. But vectors which differ
by a phase can be identified,as physical states, from a quantum
mechanical point of view. The quotient of these set of states by
this relation gives a set of equivalence classes, the coherent
states. What can be proved is that each equivalence class can be
labelled by a complex number, so that the quotient space can be
seen as a complex plane. This means that there is a coherent state
for each point on a plane, whose explicit form is \beq
\mid\xi\ra\,\,\,\rightarrow\,\,\,\psi_{\xi}\left(w\right)=
e^{-\bar{\xi}\xi/2\theta}e^{\xi w/\theta} \label{cshww} \eeq The
coherent state labelled by the point $\xi$ in the complex plane
is
 an element of $\fhil$, so it is represented as a analytical function of $w$
 whose form is  exactly $\psi_{\xi}\left(w\right)$.
 It is possible to prove that this system of coherent states is
overcomplete:
\be
\idop\,=\,\int\frac{d^{2}\xi}{\pi\theta}\,\mid\xi\ra\la\xi\mid \label{HWWcomplete}\ee
and also, with $\psi_{n}$ an element of the basis already
considered:
\be
\la\xi\mid\psi_{n}\ra\,=\,e^{-\bar{\xi}\xi/2\theta}\frac{\bar{\xi}^{n}}{\sqrt{n!\theta^{n}}}
\ee
\be
\la\xi\mid f\ra\,=f\left(\bar{\xi}\right)e^{-\bar{\xi}\xi/2\theta}
\ee
where $f$ is an element in $\fhil$.

Now an action of the HWW group on the complex plane can be
defined. It is given by:
\be
\left(\xi,\lambda,\alpha\right)\cdot w\,=\,\xi\,+\alpha\,w \ee It
can be seen that the element $\left(0,\lambda,-1\right)$ defines a
reflection of the point $w$ with respect to the origin of the
plane, while the element $\left(2\xi^{\prime},\lambda,-1\right)$
defines a reflection with respect to the point $\xi^{\prime}$ of
the complex plane. The image of these reflection operators, in the
Fock representation, is the function:
\be
\hat{W^{\prime}}\left(2\xi^{\prime},\lambda,-1\right)\mid \xi\ra\,\,
\rightarrow\,\,e^{i\lambda}e^{-2\bar{\xi}^{\prime}\xi/\theta}e^{-2\xi^{\prime}w/\theta}
e^{-\bar{\xi}\xi/2\theta}e^{\xi\left(2\bar{\xi}^{\prime}-w\right)/\theta}
\ee

In the Fock space $\fhil$ the ladder operators:
\be
\hat{Z}^{\dagger}\mid\psi_{n}\ra\,=\,\sqrt{n}\mid\psi_{n-1}\ra \ee
\be
\hat{Z}\mid\psi_{n}\ra\,=\,\sqrt{n+1}\mid\psi_{n+1}\ra \ee are
just creation-annihilation operators. In the representation of
elements in $\fhil$ as analytic functions, they have the form:
\beqa
\left(\hat{Z}^{\dagger}f\right)\left(w\right)\,=\,\sqrt{\theta}\frac{df}{dw}
\nonumber \\
\left(\hat{Z}f\right)\left(w\right)\,=\,\frac{1}{\sqrt{\theta}}wf\left(w\right)
\eeqa

Now it is possible to write everything in the standard Hilbert space
of the Schr\"{o}edinger-Von Neumann representation, that of the
square integrable functions on the line with respect to the
translationally invariant Lebesgue measure $\hil={\cal
L}^{2}\left(\real\,,dx\right)$

The equation (\ref{unirephhw}) clarifies that the operators defining
the unitary representation can be written as:
\be
\hat{W}^{\prime}\left(\xi,\lambda,\alpha\right)\,=\,e^{i\lambda}e^{-\bar{\xi}\xi/2\theta}
e^{\xi\hat{Z}/\sqrt{\theta}}e^{-\bar{\xi}\hat{Z}^{\dagger}/\sqrt{\theta}}\hat{\Pi}_{\alpha}
\ee In this expression,$\hat{\Pi}_{\alpha}$ is the identity
operator if $\alpha=1$, or the parity operator if $\alpha=-1$.

On $\hil$, one can consider the operator
$\hat{W}^{\prime}\left(\xi,0,\alpha\right)$ realized in terms of
standard creation-annihilation operators:
\be
\hat{W}^{\prime}\left(\xi,0,\alpha\right)\,=\,
e^{\left(\xi\hat{a}^{\dag}-\bar{\xi}\hat{a}\right)/\sqrt{\theta}}\hat{\Pi}_{\alpha}
\ee Now, identifying:
\beqa
\hat{a}=\frac{1}{\sqrt{2\theta}}\left(\hat{Q}+i\hat{P}\right)\nonumber \\
\hat{a}^{\dag}=\frac{1}{\sqrt{2\theta}}\left(\hat{Q}-i\hat{P}\right)
\eeqa
\beqa
\xi=\frac{1}{\sqrt{2}}\left(q+ip\right)\nonumber \\
\bar{\xi}=\frac{1}{\sqrt{2}}\left(q-ip\right) \eeqa we have an
explicit realization of a Weyl system, in the form of a Displacement operator, composed with a parity
operator:
\be
\hat{W}^{\prime}\left(\xi,0,\alpha\right)\,=\,
e^{\frac{i}{\theta}\left(p\hat{Q}-q\hat{P}\right)}\hat{\Pi}_{\alpha}
\ee

This means that a Weyl operator is related to the representation of the elements of the Heisenberg-Weyl-Wigner
group which, acting on a plane, define a reflection \cite{moreno}:

\beq
\Wo\left(q,p\right)\,=\,2\Wo^{\prime}\left(2\xi,0,-1\right)
\eeq

\chapter{Weyl-Wigner formalism for compact Lie groups}\label{chapsecond}

In the first chapter an introduction to the Weyl-Wigner formalism
has been presented. It has been analysed the case where the
classical phase space is a cartesian vector space equipped with a
translationally invariant symplectic structure. In particular, the
phase space has been seen as the manifold representing the abelian
group of translations, and the noncommutativity, parametrized by
$\hbar$, has been introduced by an explicit use of the symplectic
structure. In section {\ref{onemorestep} the noncommutativity
of the quantum observables has been formalized by the
skewsymmetric term of the cocycle factor of the representation of
this abelian group. The aim of this chapter is to study a
generalization of the Weyl-Wigner formalism to the case where the
classical configuration space is no more a vector space, thus
identifiable with the noncompact abelian group of translations,
but a generic compact simple Lie group.

The chapter begins with a description of Wigner distributions in
the cartesian case: they are introduced using the machinery
previously developed. This section could also be considered as the end
of the first chapter. It is here because the notion of Wigner
distribution will be used as a guide in constructing the
Weyl-Wigner isomorphism in this case.

This isomorphism should take place between a set of operators on a
Hilbert space, and a set of functions on the classical phase
space, which is the cotangent bundle $T^{*}G$  of a compact simple
Lie group $G$. The path followed in the first chapter would
suggest, to consider, first of all, a notion of Fourier transform
for functions on this space. The second step would be the
definition of a kind of Weyl system for the dual of the classical
phase space, and then to define a Weyl map, and a Wigner map, by
the well known procedure. Section \ref{trialcylinder} shows what
are the results of following this path in the easiest case of
$G=U\left(1\right)\approx S^{1}$. It shows that a Weyl map for
functions on a cylinder (which is the manifold $T^{*}S^{1}$)
cannot be obtained in this way.

Nevertheless it shows that a kind of Wigner map can be defined,
introducing a set of operators that generalizes the properties of
Moyal quantizers (\ref{weylopprop}). These operators are used to
define a map from the space of operators to the space of functions
(symbols) defined on $S^{1}\times\zed$. The symbols of density
matrix operators are functions whose marginals distributions
reproduce the expected probability distribution for the quantum
mechanical system usually referred to as a particle constrained to
move on a circle. These symbols are then called Wigner
distributions, the map is called Wigner map, and Weyl map is
obtained as its inverse. The novelty of this approach
\cite{mukundi} is that this Weyl-Wigner isomorphism is defined
between the set of operators on a Hilbert space, and the set of
functions on the space $S^{1}\times\zed$. This space can be seen
as a \emph{quantum cotangent bundle} of the circle $S^{1}$.

In section \ref{wigdistsection} this approach is developed in
detail for the case where the configuration space of a classical
system is a compact simple Lie group. First of all the classical
kinematics is analysed, to set up its quantum version, which
generalizes the canonical commutation relations introduced by
Dirac, and already studied in detail. The noncommutativity of
quantum observables can be traced back to the non abelianess of
the group $G$, not related to a noncommutativity constant
parametrized by $\hbar$. It will not have any role in the
following studies.  Harmonic analysis on the group $G$ suggests
what is the space on which Wigner distributions are defined, and
what are the quantizer operators. Then the complete Weyl-Wigner
isomorphism is deduced. On the space of symbols (that are now
functions on the quantum cotangent space $G\times\Gamma$) a
noncommutative product can be set.

The last section shows how this
formalism can be used to define, when the group $G$ is nonabelian,
a Weyl-Wigner isomorphism between operators and functions on the
\emph{classical} cotangent space, thus solving the initial problem
of generalising the Weyl-Wigner isomorphism to a set of classical
systems larger than the cartesian ones.

\section{From Weyl map to Wigner functions}

In the standard quantum formalism an observable is formalized via
a self-adjoint operator on a separable Hilbert space, whose rays
represent the physical pure states. The set of measured values for
an observable, represented for example by the operator $\Ao$, if
the system is in a state represented by a normalized ket
$\left|\psi\rangle\right.$, is an experimental distribution whose
mean value is formalized as (in this analysis, it will be considered $L=\real^{2}=T^{*}\real$): 
\beq
\langle\Ao\ra\,=\,\la\psi\mid\Ao\mid\psi\ra 
\eeq The right hand
side of this equation can be written, on a suitable set of
operators, as: \beq
\la\Ao\ra\,=\,Tr\left[\Ao\mid\psi\ra\la\psi\mid\right] \eeq In the
Weyl formalism, the mapping between functions and operators is
such that relation (\ref{wwiso}) is valid, so that it is possible
to write the mean value of an observable as: \beq
\la\Ao\ra\,=\,Tr\left[\Ao\mid\psi\ra\la\psi\mid\right]\,=\int\,\frac{dqdp}{2\pi\hbar}\,A\left(q,p\right)
W_{\psi}\left(q,p\right) \eeq where $W_{\psi}\left(q,p\right)$ is
called Wigner distribution \cite{wignertd} function for the pure
state $\mid\psi\ra$. It is the Weyl symbol of the projector
$\mid\psi\ra\la\psi\mid$, while $A\left(q,p\right)$ is the Weyl
symbol for $\Ao$. In general, for a density operator $\hat{\rho}$:
\beq
W_{\hat{\rho}}\left(q,p\right)\,=\,\int\,\frac{dxdk}{2\pi\hbar}\,e^{-i\left(xp-kq\right)/\hbar}\,
Tr\left[\hat{\rho}\Do^{\dagger}\left(x,k\right)\right] \eeq It can
be written also using the system of Weyl operators, defined in
(\ref{weylopprop}), by equation (\ref{wignermap2}): \beq
W_{\hat{\rho}}\left(q,p\right)\,=\,Tr\left[\hat{\rho}\Wo^{\dagger}\left(q,p\right)\right]
\label{wigdistrweyl} \eeq

In the Schr\"{o}dinger realization, where $\psi\left(s\right)$ is
the wave function representing the ket state $\mid\psi\ra$, they
are given by: \beq W_{\psi}\left(q,p\right)\,=\,\int ds
\,e^{-ips/\hbar}\,\psi^{*}\left(q-s/2\right)\psi\left(q+s/2\right)
\eeq while for the density operator
$\hat{\rho}=\mid\phi\ra\la\psi\mid$: (both $\mid\phi\ra$ and
$\mid\psi\ra$ are normalised) \beq
W_{\phi\overline{\psi}}\left(q,p\right)\,=\,\int ds
\,e^{-ips/\hbar}\,\psi^{*}\left(q-s/2\right)\phi\left(q+s/2\right)
\eeq This Wigner distribution function has been introduced via the
standard Weyl formalism \cite{wignerreview}. It is thus natural to
wonder what is the behaviour with respect to the action of the
symplectic group. The symplectic group, in this case $Sp\left(2\right)$, acts on the classical phase space
(\ref{unconjsws}). For $T\in Sp\left(2\right)$, the Wigner function gets transformed as: \beq
W_{\psi}\left(T\left(q,p\right)\right)\,=\,W_{\Uo_{T}\left(\psi\right)}\left(q,p\right)
\eeq The values of the Wigner function along symplectic orbits is
related to the action of the unitary representation of the
symplectic (properly metaplectic) group on the Hilbert space of
states. This unitary representation is dictated by the Von Neumann
theorem, developed in the study of the covariance properties of a
standard Weyl system.

A very important aspect of this construction is that Wigner
functions can assume negative values in some regions of the
classical phase space: this is the reason why they are actually
called \emph{quasi-probabilities distributions}. Nevertheless
their marginal distributions do reproduce true probability
densities: \beqa \int
dp\,W_{\psi}\left(q,p\right)\,=\,2\pi\left|\psi\left(q\right)\right|^{2}
\nonumber
\\
\int dq\,W_{\psi}\left(q,p\right)\,=\,2\pi\left|\tilde{\psi}\left(p\right)\right|^{2}
\eeqa
Here $\tilde{\psi}$ is the Fourier transform of the wave function
$\psi$ in the usual Hilbert space of square integrable functions
on the line. In the dynamics of one dimensional point particle,
the modulus square of $\psi$, and $\tilde{\psi}$, represent, in
the Schr\"{o}dinger realization, the probability distributions
in the spectral representation of position and momentum.

In this picture, the Wigner functions can also be written as: \beq
W_{\hat{\rho}}\left(q,p\right)\,=\,\int\,dxdy\,
\delta\left(q-\frac{x+y}{2}\right)\,e^{ip\left(x-y\right)/\hbar}
\psi^{*}\left(x\right)\psi\left(y\right)\label{wigfundelta}\eeq
that is: \beq W_{\hat{\rho}}\left(q,p\right)\,=\,\int\,dxdy\,
\delta\left(q-\frac{x+y}{2}\right)\,e^{ip\left(x-y\right)/\hbar}
\la y\mid\hat{\rho}\mid x\ra \label{wigfundelgenq}\eeq while, in the momentum representation:
\beq
W_{\hat{\rho}}\left(q,p\right)\,=\,\int\,dkdl\,
\delta\left(p-\frac{k+l}{2}\right)\,e^{-iq\left(l-k\right)/\hbar}
\tilde{\psi}^{*}\left(l\right)\tilde{\psi}\left(k\right)
\label{wigfundeltap}
\eeq
or, equivalently:
\beq
W_{\hat{\rho}}\left(q,p\right)\,=\,\int\,dkdl\,
\delta\left(p-\frac{l+k}{2}\right)\,e^{-iq\left(l-k\right)/\hbar}
\la k\mid\hat{\rho}\mid l\ra
\eeq

In these relations, $\mid x\ra$ and $\mid y\ra$ are generalized
eigenstates of the position observable, while $\mid k\ra$ and
$\mid l\ra$ are generalised eigenstates of the momentum
observable.

\section{From Wigner functions to Weyl map}\label{trialcylinder}

In the last section, the theory of Wigner "quasi-probabilities"
functions has been developed as a nice example
in the Weyl formalism. This is not the historical order these concepts
were introduced. The method of Wigner distributions as a descriptions
of states of quantum mechanical systems appeared in 1932 \cite{wignertd}, quite early
in the history of quantum mechanics. As it has been outlined, for
systems whose kinematics is based upon the Heisenberg canonical
commutation relations, it gives a way of describing both pure and
mixed states in a classical phase space setting, at the level of
density operators. Moreover, they make it possible to write quantum
expectation values in terms of statistical averages on the classical
phase space of the system, in a formal analogy with the classical
statistical approach. In this perspective, the important aspect is
that these quantum distributions are not necessarily
positive, thus preventing from the complete identification with a classical distribution.
It was later appreciated that Wigner distribution approach to
quantum states is naturally dual to the Weyl approach to quantum
observables (described at lenght in the last chapter) \cite{weyl}:
together with the work of Moyal \cite{moyal}, who introduced the
concept of non abelian products in the space of classical functions,
these constitute a complete and coherent formulation of quantum
mechanics in terms of c-number dynamical phase space variables,
well suited for the comparison with classical mechanics.

\subsection{A Weyl-Wigner map for functions on a cylinder?}

The formalism developed up to now is perfectly fitting for the analysis
of quantum systems whose classical counterparts can be considered as
point particles moving in a cartesian configuration space, whose
cotangent bundle is a vector space.

Is it possible to define a Weyl-Wigner formalism for classical systems
whose configuration space is a compact simple Lie group?

In the last chapter, the path followed to define a Weyl map for functions on a plane,
i.e. the cotangent bundle of a line, has gone through the definition of a Weyl system
for the translation group, in terms of the so called Displacement operators.
These operators have been considered as a sort of generalized basis for a space of
operators: the coefficients of an expansion are given by the Fourier coefficients of
functions on the plane.
So a Weyl map seems to be traced to the study of harmonic analysis
for the group of translations, and unitary representations of abelian
groups defined on the dual space of the classical phase space.

As a first example of what are the features, and the problems, of
a generalization along these lines, one can consider the case where $\mathcal{Q}$,
the configuration space for a classical system, is a circle $S^{1}$, that is the group manifold
of $U\left(1\right)$. The cotangent space of a circle is a cylinder
$T^{*}S^{1}\,\approx\,S^{1}\times\real$. Considering $\real$ as the
real additive group in one dimension, the dual of this classical phase
space is the product $\zed\times\real$.

A first naive approach would be
that of defining a system of Displacement operators
for this space.

On the Hilbert space of square integrable functions on a circle
$\mathcal{H}\,=\,\mathcal{L}^{2}\left(S^{1}, d\theta/2\pi\right)$
an orthonormal basis is given by
$\phi_{n}\left(\theta\right)\,=\,e^{in\theta}$, with
$n\,\in\,\zed$. Using a ket notation, one can introduce two sets
of unitary operators: \beqa
\Uo\,:\,\zed\,\mapsto\,\mathcal{U}\left(\hil\right)\,\,&\,&\,\,
\Uo\left(m\right)\mid\phi_{n}\ra\,=\,\mid\phi_{n+m}\ra \nonumber
\\
\Vo\,:\,\real\,\mapsto\,\mathcal{U}\left(\hil\right)\,\,&\,&\,\,
\Vo\left(k\right)\mid\phi_{n}\ra\,=\,e^{-ink}\,\mid\phi_{n}\ra
\eeqa satisfying a sort of canonical commutation rules: \beq
\Uo\left(m\right)\Vo\left(k\right)\,=\,e^{imk}\,\Vo\left(k\right)\Uo\left(m\right)
\eeq A Displacement operator can be defined: \beq
\Do\left(m,k\right)\,=\,e^{imk/2}\,\Vo\left(k\right)\Uo\left(m\right)\,=\,
e^{-imk/2}\,\Uo\left(m\right)\Vo\left(k\right) \eeq This set of
operators do form a complete (trace orthonormal) basis for all
operators on $\mathcal{H}$.

A natural idea would be mapping an operator to a function via:
 \beq
\tilde{f}\left(m,k\right)\,=\,Tr\left[\fo\Do^{\dagger}\left(m,k\right)\right]
\eeq  At a first sight, this $\tilde{f}$ seems to be a
function defined on the dual $\zed\times\real$, eventually to be
identified with the Fourier transform of a function on the
cylinder. But a deeper inspection says that, for $k-k^{\prime}=4\pi$:
$$\tilde{f}\left(m,k\right)=\tilde{f}\left(m,k^{\prime}\right)$$
This means that $\tilde{f}$ is actually a function on
$S^{1}\times\zed$. This procedure does not enable to write a Weyl
symbol on the "right" classical phase space.

It is however possible to define a different system of operators,
the use of which does give a version of a Wigner formalism for this
class of quantum systems. The goal will be the introduction of a
set of Wigner quasi-probability densities, requiring that their
marginals are the correct probability distribution for these systems.
Instead of defining a set of Displacement operators, the line will
be the definition of a set of Weyl operators, generalizing those in (\ref{weylopprop}).

On the same Hilbert space, that of square integrable functions on
the circle, with the normalized Haar measure, one can define different pairs
of operators: \beqa
\Uo\,:\,\zed\,\mapsto\,\mathcal{U}\left(\hil\right)\,\,&\,&\,\,
\left(\Uo\left(m\right)\psi\right)\left(\theta\right)\,=\,e^{im\theta}\,\psi\left(\theta\right)
\nonumber
\\
\Vo\,:\,S^{1}\,\mapsto\,\mathcal{U}\left(\hil\right)\,\,&\,&\,\,
\left(\Vo\left(\theta^{\prime}\right)\psi\right)\left(\theta\right)\,=\,
\psi\left(\left[\theta-\theta^{\prime}\right]\right) \label{cylgen}\eeqa here
$\left[\theta-\theta^{\prime}\right]$ means
$\left(\theta-\theta^{\prime}\right)/mod 2\pi$.
$\Uo\left(\cdot\right)$ is a unitary representation of
$\left(\zed,+\right)$, while $\Vo\left(\cdot\right)$ is a unitary
representation of $U\left(1\right)$.

Their action on the separable orthonormal basis of ket vectors
$\mid\phi_{n}\ra$ is: \beqa
\Uo\left(m\right)\mid\phi_{n}\ra\,=\,\mid\phi_{n+m}\ra\nonumber \\
\Vo\left(\theta\right)\mid\phi_{n}\ra\,=\,e^{-in\theta}\mid\phi_{n}\ra
\label{UVopcircle}\eeqa and they satisfy an "Heisenberg"-type commutation relation:
\beq
\Vo\left(\theta\right)\Uo\left(m\right)\,=\,e^{-im\theta}\,\Uo\left(m\right)\Vo\left(\theta\right)
\eeq 
Here the phase factor can be properly formalised as a character 
of the representation $\Vo\left(\cdot\right)$ of $U\left(1\right)$: this notion generalizes the action of
covector on a vector in the case of the group of translations.
The composition of $\Uo$ operators and $\Vo$ operators with a
suitable phase factors would give the analogue of displacement
operators, one for each point of the space $S^{1}\times\zed$:
even in this case they do form a complete trace
orthonormal system: \beq
Tr\left[\Uo\left(m\right)\Vo\left(\theta\right)
\Vo^{\dagger}\left(\theta^{\prime}\right)\Uo^{\dagger}\left(m^{\prime}\right)
\right]\,=\,\delta_{mm^{\prime}}\delta\left(\left[\theta-\theta^{\prime}\right]\right)
\eeq In this relation the continuous
$\delta\left(\left[\theta-\theta^{\prime}\right]\right)$ is
referred to the measure $d\theta/2\pi$.

In the Hilbert
space on which these operators have been realized, besides the
separable basis $\phi_{n}\left(\theta\right)$ already mentioned,
it is possible to introduce a sort of generalized overcomplete
continuum basis of ket states $\mid\theta\ra$. They are introduced
in analogy to eigenstates of the position observable for a quantum point
particle in a cartesian space. They are defined by: \beq
\phi_{n}\left(\theta\right)\,=\,e^{in\theta}\,=\,\la\theta\mid\phi_{n}\ra
\label{thetabasis}\eeq Normalization and overcompleteness are written as: \beq
\la\theta\mid\theta^{\prime}\ra\,=\,\delta\left(\left[\theta-\theta^{\prime}\right]\right)
\eeq \beq
\idop\,=\,\int\,\frac{d\theta}{2\pi}\mid\theta\ra\la\theta\mid
\eeq In this basis these operators act as: \beqa
\Uo\left(m\right)\mid\theta\ra\,&=&\,e^{im\theta}\mid\theta\ra
\nonumber\\
\Vo\left(\theta^{\prime}\right)\mid\theta\ra\,&=&\,\mid\left[\theta+\theta^{\prime}\right]\ra
\eeqa

 Looking at this basis, that is related to a spectral decomposition
of a position observable for a particle whose configuration space
is a circle, one can see that  the discrete basis can be seen as made
of eigenstates for a momentum operator of that particle: as it is
well known, this observable has discrete spectrum. In a
group-theoretical approach, this can be seen as result in harmonic
analysis for the group $U\left(1\right)$. The dual of a compact
group is discrete \cite{simongroup}: it is possible to prove that the space of
UIRR's for a compact group is labelled by discrete indices.

Using these operators, it is possible to introduce a new set:
\beq \Wo\left(\theta,n\right)\,=\,\sum_{m}\int
\frac{d\sigma}{2\pi}\,\Uo^{\dagger}\left(m\right)\Vo\left(\sigma\right)\,
e^{in\sigma}e^{im\left(\theta+\sigma/2\right)} \label{circlequant}\eeq These
operators are hermitian (thus resembling one of
the properties of Weyl operators for the cartesian case
(\ref{weylopprop})), and form a complete trace orthonormal system.
They are a map from the space $S^{1}\times\zed$ to unitaries
$\mathcal{U}\left(\mathcal{H}\right)$: this means that the
standard procedure associates to an operator
$\hat{A}\,\in\,Op\left(\mathcal{H}\right)$ a symbol on that space:
\beq
A\left(\theta,n\right)\,=\,Tr\left[\Ao\Wo^{\dagger}\left(\theta,n\right)\right]
\eeq Which are the properties of these symbols? If one considers the
symbol of a projector $\rho\,=\,\mid\psi\ra\la\psi\mid$, then one
has: \beq
W_{\hat{\rho}}\left(\theta,n\right)\,=\,Tr\left[\hat{\rho}\Wo^{\dagger}\left(\theta,n\right)\right]\,=\,
\la\psi\mid\Wo\left(\theta,n\right)\mid\psi\ra \eeq

If one computes the marginal distribution for the symbol $W_{\hat{\rho}}\left(\theta,n\right)$, the result is:
\beqa
\int\frac{d\theta}{2\pi}\,W_{\hat{\rho}}\left(\theta,n\right)\,&=&
\,\la\psi\mid\phi_{n}\ra\la\phi_{n}\mid\psi\ra
\nonumber
\\
\sum_{n}W_{\hat{\rho}}\left(\theta,n\right)\,&=&
\,\la\psi\mid\theta\ra\la\theta\mid\psi\ra
\eeqa
This shows that marginals reproduce the probability density distribution
for the quantum dynamics of a point particle
on a circle, written in the dual basis of position and momentum.
By analogy with the planar case, this is the reason why the maps:
\beqa
\Ao\,&=&\,\sum_{n}\int\frac{d\theta}{2\pi}\,A\left(\theta,n\right)\Wo\left(\theta,n\right)
\nonumber
\\
A\left(\theta,n\right)\,&=&\,Tr\left[\Ao\Wo^{\dagger}\left(\theta,n\right)\right]
\eeqa that define an isomorphism between Hilbert-Schmidt operators
in the Hilbert space $\mathcal{H}$, and square modulus measurable
functions on the space $S^{1}\times\zed$ (considering an
integration over the continuous $\theta$ and a summation over the
discrete $n$), can be defined a Weyl-Wigner isomorphism for a
dynamics of the particle on a circle. The most important novelty
this formalism shows is that functions related to quantum
observables via this isomorphism are not on the classical phase
space, the classical cotangent space of the configuration space
$S^{1}$, but on a sort of quantum cotangent space of $S^{1}$, that
is the product of the configuration space itself (the group
manifold) with the dual space.

Moreover, a relation of the kind of (\ref{wwiso}) is valid:
\beq
Tr\left[\Ao^{\dagger}\Bo\right]\,=\,\sum_{n}\int \frac{d\theta}{2\pi}\,A^{*}\left(\theta,n\right)B\left(\theta,n\right)
\eeq

\section{The Wigner distributions in the Lie group case}\label{wigdistsection}

The path followed in the case of the cylinder, the cotangent bundle for
the group manifold of $U\left(1\right)$, can be generalized to the case
of a compact simple Lie group $G$, of order $n$.

 For a quantum system
whose classical counterpart has a configuration space which is
such a Lie group, setting the kinematics means defining the
commutation relations (the \emph{quantum conditions}, in Dirac's
approach) for a set of fundamental observables, and a realization
in terms of operators on a suitable Hilbert space. Quantum
kinematics for these system shows that noncommutativity among
fundamental observables are traced to the nonabelianess of the
group $G$, and not to the symplectic structure as in the cartesian
case.

The Wigner distributions are then introduced, attempting to
generalize the covariance properties of the Wigner distribution in
the cartesian case. The interesting aspect is the definition of a
kind of quantum cotangent bundle for a Lie group, related to the
space of unitary irreducible representations of the group itself.

Once the general isomorphism is set, it is possible to recover the
case of the classical cylinder, used as a guide in the last
section, as a particular case.

\subsection{Classical Kinematics}

The classical system  under analysis has a configuration space
$\mathcal{Q}$ which is a compact simple Lie group $G$. The
corresponding phase space is $T^{*}G$, the cotangent bundle of the
group. It can be described both in  intrinsic geometric terms, and
in a local coordinates system \cite{diffgeom}.

A Lie group is represented by a parallelizable and differentiable
manifold. Its Lie algebra $\underline{G}$ can be identified with
the tangent space of $G$ at the identity, and is isomorphic to the
dual, the cotangent space:
$$\underline{G}\,=\,T_{e}G\,=\,T_{e}^{*}G\,=\underline{G}^{*}$$
The Lie group brings with it the set of left translations $L_{g}$
and the set of right translations $R_{g}$: these are mutually
commuting actions of $G$ by mapping of $G$ on itself: \beqa
L_{g}\,:G\,&\mapsto&\,G\,\,\,\,L_{g}\left(g^{\prime}\right)=gg^{\prime}\nonumber
\\
R_{g}\,:G\,&\mapsto&\,G\,\,\,\,R_{g}\left(g^{\prime}\right)=g^{\prime}g^{-1}
\eeqa The corresponding tangent maps and pull backs act, as
nonsingular linear transformations, on the tangent and cotangent
spaces, respectively at general points of $G$, according to: \beqa
\left(L_{g}\right)_{*}\,:\,T_{g^{\prime}}G\,&\mapsto&\,T_{gg^{\prime}}G
\nonumber \\
\left(R_{g}\right)_{*}\,:\,T_{g^{\prime}}G\,&\mapsto&\,T_{g^{\prime}g^{-1}}G
\nonumber \\
\left(L_{g}\right)^{*}\,:\,T_{g^{\prime}}^{*}G\,&\mapsto&\,T_{g^{-1}g^{\prime}}^{*}G
\nonumber \\
\left(R_{g}\right)^{*}\,:\,T_{g^{\prime}}^{*}G\,&\mapsto&\,T_{g^{\prime}g}^{*}G
\eeqa If dual bases are introduced $\{e_{r}\}$ for $T_{e}G$ and
$\{e^{r}\}$ for $T^{*}_{e}G$, then the action of these right and
left tangent maps defines two bases for general vector fields on
$G$: \beqa
X_{r}\left(g\right)\,&=&\,\left(R_{g^{-1}}\right)_{*}\left(e_{r}\right)
\nonumber \\
\tilde{X}_{r}\left(g\right)\,&=&\,\left(L_{g}\right)_{*}\left(-e_{r}\right)
\eeqa Vector fields $\{X_{r}\}$ are right invariant, and are the
generators of the left translations $L_{g}$, while the vector
fields $\{\tilde{X}_{r}\}$ are left invariant and generate the
right translations $R_{g}$. They close a representation of the Lie
algebra $\underline{G}$: \beqa
\left[X_{r},X_{s}\right]\,&=&c_{rs}^{t}X_{t} \nonumber \\
\left[\tilde{X}_{r},\tilde{X}_{s}\right]\,&=&-c_{rs}^{t}\tilde{X}_{t}
\nonumber \\ \left[X_{r},\tilde{X}_{s}\right]\,&=&0 \eeqa In general
the elements of a Lie group $G$ cannot be described with the help
of coordinates in a globally smooth manner. In particular this is
so if $G$ is compact: one has to work with charts, with well
defined transition rules in overlaps.
 An element $g\,\in\,G$  can be locally labelled by $n$ real independent continuous coordinates $q^{r}$.
Conventionally it is set $q^{r}=0$ at the group identity $e$. The
bases elements for $T_{e}G$ and $T^{*}_{e}G$ are identified with:
(here the subscript $0$ means that these quantities are evaluated at the group identity)\beq e_{r}\,=\,\left(\frac{\del}{\del
q^{r}}\right)_{0}\,\,\,\,\,\,\, e^{r}\,=\,\left(dq^{r}\right)_{0}
\eeq The product of two group elements $g\left(q\right)$ and
$g^{\prime}\left(q^{\prime}\right)$ is a function: \beq
g\left(q\right)\cdot
g^{\prime}\left(q^{\prime}\right)\,=\,\left(gg^{\prime}\right)\left(f\left(q,q^{\prime}\right)\right)
\eeq and, infinitesimally: \beqa
\eta^{r}_{s}\left(q\right)&=&\left(\frac{\del f^{r}}{\del
q^{\prime s}}\right)\left(q^{\prime},q\right)_{q^{\prime}=0}
\nonumber \\
\tilde{\eta}^{r}_{s}\left(q\right)&=&\left(\frac{\del
f^{r}}{\del q^{\prime
s}}\right)\left(q,q^{\prime}\right)_{q^{\prime}=0} \eeqa These
quantities are related to the coordinate expression of the left
and right invariant vector fields: \beqa
X_{r}&=&\eta^{s}_{r}\left(q\right)\frac{\del}{\del q^{s}}
\nonumber \\
\tilde{X}_{r}&=&-\tilde{\eta}^{s}_{r}\left(q\right)\frac{\del}{\del
q^{s}} \label{rivecf}\eeqa In the sense of classical canonical
mechanics, on $T^{*}G$ there are local canonically conjugate
momentum variables $p_{r}$, and the classical Poisson bracket
relations are: \beqa \{q^{r},q^{s}\}&=&0 \nonumber \\
\{q^{r},p_{s}\}&=&\delta^{r}_{s} \nonumber \\ \{p_{r},p_{s}\}&=&0
\eeqa The range of the $p_{r}$ variables is usually taken to be
the entire real line: so to define an identification
$T^{*}_{e}G\,\simeq\,\real^{n}$ at each $g\,\in\,G$. These local
coordinates
 can be used to define a system of global ones, introducing a system of generalised momenta:
\beqa
J_{s}&=&\eta^{r}_{s}\left(q\right)p_{r}
\nonumber \\
\tilde{J}_{s}&=&-\tilde{\eta}^{r}_{s}\left(q\right)p_{r}
\eeqa
At the end of this description, there is the explicit evaluation of the Poisson bracket relations this system satisfies:
\beqa
\{q^{r},q^{s}\}&=&0
\nonumber \\
\{q^{r},J_{s}\}&=&\eta^{r}_{s}\left(q\right)
\nonumber \\
\{J_{r},J_{s}\}&=&c_{rs}^{t}J_{t}
\eeqa
Similar relations do occur for coordinates suited for the right action.

\subsection{Quantum Kinematics}
The most natural Hilbert space on which trying quantizing this class
of classical system is the set of complex square integrable functions on
$G$, with respect to the normalized Haar measure $\int_{G}\,d\mu\,=\,1$:
$$
\mathcal{H}\,=\,\mathcal{L}^{2}\left(G,d\mu\right)\,=\,
\{\psi\left(g\right)\,\in\,\complex\,:\,\parallel\psi\parallel^{2}\,=\,\int_{G}d\mu\,
\left|\psi\left(g\right)\right|^{2}<\infty\}
$$
As in the previous example, the first thing to do is to define a set of operators
from points of the group $G$ to unitary operators in this Hilbert space. They
will encode the informations about position observable for the system. From a
noncommutative geometry point of view, the notion of position coordinates is
intrinsically captured by the commutative algebra of smooth functions
$\mathcal{F}\left(G\right)$. To a function $f\,\in\,\mathcal{F}\left(G\right)$
one can associate an operator on $\mathcal{H}$:
\beq
\left(\hat{f}\psi\right)\left(g\right)\,=\,f\left(g\right)\psi\left(g\right)
\eeq
The algebra of functions on the group is mapped into an abelian algebra of multiplicative operators.
The next step is to define the analogous of momentum observables.
They will be related to the generators of the dual space of the group $G$.

Both left and right actions are defined on this Hilbert space. In particular, the left action is written as: \beq
\left(\Vo\left(g^{\prime}\right)\psi\right)\left(g\right)\,=\,\psi\left(g^{\prime-1}g\right)
\eeq These operators define a unitary representation of the group
$G$: $$
\Vo\left(g^{\prime}\right)\Vo\left(g\right)\,=\,\Vo\left(g^{\prime}g\right)
$$ So it is possible to identify the Hermitian generators. Fixed
the basis $e_{a}$ in the Lie algebra $\underline{G}$, the exponential map
for a compact group is surjective, so every element can be written
as: \beq
g\,=\,\exp\left(\alpha_{r}e_{r}\right) \eeq and, once more, Stone's
theorem enables to define a set of Hermitian generators for the
action represented by: \beq
\Vo\left(\exp\left(\alpha_{r}e_{r}\right)\right)\,=\,\exp\left(-i\alpha_{r}\hat{J}_{r}\right)
\eeq These generators do represent the Lie algebra
$\underline{G}$ on $\hil$, and will be considered as generalized momenta:
\beq
\left[\hat{J}_{a};\hat{J}_{b}\right]\,=\,ic_{ab}^{s}\,\hat{J}_{s}
\eeq On wave functions $\psi\left(g\right)$ of this
Schr\"{o}dinger representation, each $\hat{J}_{r}$ acts a first
order partial differential operator, that is a vector field, thus
defining another representation of $\underline{G}$ as derivations ($X_{r}$ are the right invariant vector fields (\ref{rivecf})):
\beq
\left(\hat{J}_{r}\psi\right)\left(g\right)\,=\,iX_{r}\psi\left(g\right)
\eeq

 It is possible to express functions of position also via
unitary operators, to be as close as possible to the Weyl
approach. For a real function in $\mathcal{F}\left(G\right)$, a
unitary operator is given by: $$
\Uo\left(f\right)\,=\,e^{i\hat{f}}\,\,\,\,\,\,\,\,\,
\left(\Uo\left(f\right)\psi\right)\left(g\right)\,=\,e^{if\left(g\right)}\,\psi\left(g\right)
$$ It can be seen that they satisfy a relation: \beq
\left(\Uo\left(f\right)\Vo\left(g^{\prime}\right)\psi\right)\left(g\right)\,=
\,e^{i\left[f\left(g\right)-f\left(g^{\prime-1}g\right)\right]}
\left(\Vo\left(g^{\prime}\right)\Uo\left(f\right)\psi\right)\left(g\right)
\eeq which is in the spirit of (\ref{displcomm}), except that $f$
is not restricted to be linear in any coordinate variables.

As in the previous case, two bases for the Hilbert space at hand
can be introduced. The first is a "momentum" basis, and can be set
up using the Peter-Weyl theorem involving the unitary irreducible
representations of $G$. The various UIR's can be denoted by an
index $j$, in general a collection labelling the Casimir operators
eigenvalues for the group. Since $G$ is compact, every UIR is
finite dimensional, and its dimension is $N_{j}$. Rows and columns
within the $j^{th}$ representation are labelled by $m$ and $n$, a
sort of generalised magnetic quantum numbers. So a unitary matrix
representing the element $g\,\in\,G$ in the $j^{th}$ UIR is:
\beq
g\,\mapsto\,\left(D^{j}_{mn}\left(g\right)\right)
\label{Delements}
\eeq
Moreover,
there is a freedom of unitary changes in the choice of $m,n$.
Unitarity and associativity of the representations are written as:
\beqa
\sum_{n}\,D^{j}_{mn}\left(g\right)^{*}D^{j}_{m^{\prime}n}\left(g\right)&=&\delta_{mm^{\prime}}
\nonumber
\\
\sum_{n}\,D^{j}_{mn}\left(g^{\prime}\right)D^{j}_{nn^{\prime}}\left(g\right)&=&
D^{j}_{mn^{\prime}}\left(g^{\prime}g\right) \eeqa while
orthogonality and completeness as: \beqa \int_{G}d\mu\,
D^{j}_{mn}\left(g\right)D^{j^{\prime}}_{m^{\prime}n^{\prime}}\left(g\right)^{*}&=&\delta_{jj^{\prime}}
\delta_{mm^{\prime}}\delta_{nn^{\prime}}/N_{j} \nonumber
\\
\sum_{jmn}N_{j}D^{j}_{mn}\left(g\right)
D^{j}_{mn}\left(g^{\prime}\right)^{*}&=&\delta\left(g^{-1}g^{\prime}\right)
\eeqa This completeness property, expressed by the presence of a
distributional $\delta$, is the main result of the mentioned
Peter-Weyl theorem. It enables to perform an harmonic analysis,
generalizing the Fourier analysis on linear spaces: \beqa
f\left(g\right)&=&\sum_{jmn}\,f^{j}_{nm}D^{j}_{nm}\left(g\right)\sqrt{N_{j}}
\nonumber \\ f^{j}_{nm}&=&\,\int_{G}
d\mu\,\left(D^{j}_{nm}\left(g\right)\right)^{*}f\left(g\right)\sqrt{N_{j}}
\eeqa This analysis shows that a basis in $\hil$ is given by, in
ket notation, vectors $\mid jmn\ra$, for which: \beqa \la
j^{\prime}m^{\prime}n^{\prime}\mid
jmn\ra\,&=&\,\delta_{jj^{\prime}}\delta_{mm^{\prime}}\delta_{nn^{\prime}}
\nonumber \\ \Vo\left(g\right)\mid
jmn\ra\,&=&\,\sum_{m^{\prime}}D^{j}_{mm^{\prime}}\left(g^{-1}\right)\mid
jm^{\prime}n\ra \label{laG}\eeqa The action of the operator
$\Vo\left(g\right)$ on this basis shows a well known fact. Left
regular representation of a compact group is not irreducible, and
the multiplicity of occurrence of the $j^{th}$ UIR in its
reduction is equal to the dimension $N_{j}$ of the irreducibility
subspace. The index $n$ counts this multiplicity \cite{simongroup}.

The second basis is related to "position" observables. For each
element of the group $G$, represented by a point $g$ on the
manifold, it is possible to set: \beq \la jmn\mid
g\ra\,=\,\sqrt{N_{j}}\left(D^{j}_{mn}\left(g\right)\right)^{*}
\eeq These vectors define an orthonormal, in a generalized sense
with respect to the Haar measure, and overcomplete system in
$\mathcal{H}$:
\beq
\la g\mid\
g^{\prime}\ra\,=\,\delta\left(g^{-1}g^{\prime}\right)
\,\,\,\,\,\,\,\,\,\,\,\,\,\,\, \idop\,=\,\int_{G}d\mu\,\mid
g\ra\la g\mid
\eeq
$\Vo\left(\cdot\right)$, the left representation,
acts as:
\beq
\Vo\left(g^{\prime}\right)\mid g\ra\,=\,\mid
g^{\prime}g\ra
\eeq
so that it is the representation of the left
multiplication for element of the group.

In this context, the right action of
$G$ can be defined as:
\beqa
\Ro\left(g^{\prime}\right)\mid g\ra\,&=&\,\mid
gg^{\prime -1}\ra \nonumber \\ \Ro\left(g^{\prime}\right)\mid
jmn\ra\,&=&\,\sum_{n^{\prime}}D^{j}_{n^{\prime}n}\left(g^{\prime}\right)\mid
jmn^{\prime}\ra
\label{raG}
\eeqa
It is clear that, for right
regular representation of $G$ on this space, it is the index
$m$ that counts the multiplicity of the occurrence of the
$j^{th}$ UIR in its reduction.

Given a state $\mid\psi\ra$ in $\hil$, it can be expanded in these
two basis: \beqa \la g\mid\psi\ra&=&\psi\left(g\right) \nonumber
\\
\la jmn\mid\psi\ra&=&\psi_{jmn}\,=\,\sqrt{N_{j}}\,\int_{G}
d\mu\left(D^{j}_{mn}\left(g\right)\right)^{*}\,\psi\left(g\right)
\nonumber
\\
\parallel\psi\parallel^{2}&=&\sum_{jmn}\left|\psi_{jmn}\right|^{2}\,
=\,\int_{G}d\mu\,\left|\psi\left(g\right)\right|^{2} \eeqa

\subsection{A Wigner distribution}

The analysis of the cartesian case, and of the examples above,
might give rise to the hypothesis that, for a state $\mid\psi\ra$,
the corresponding Wigner function $W$ were a function of
arguments $g$, bilinear in $\psi$, (more precisely involving one
$\psi$ factor and one $\psi^{*}$ factor) and $JMN$ (quantised
momenta), such that integration over $g$ yields
$\left|\psi_{JMN}\right|^{2}$ while summation over $JMN$ yields
$\left|\psi\left(g\right)\right|^{2}$. This would be a natural way
in which the marginal distributions are reproduced.

A natural requirement of covariance of this distribution for both
the right and left actions of $G$ on state $\psi$ poses the
problem of choosing the arguments of $W\left(\ldots\right)$ in
such a way to allow for a natural linear transformation law under
each of the changes
$\psi\left(g\right)\,\rightarrow\,\psi\left(g^{\prime-1}g\right)$
(left action) and
$\psi\left(g\right)\,\rightarrow\,\psi\left(gg^{\prime}\right)$
(right action) on $\psi$. In particular, eqs.(\ref{laG}) and
(\ref{raG}) show that, in the discrete $JMN$ basis, one index
carries the transformation properties for the left action, while
the other carries for the right action. Since $W$ should involve a
bilinear expression of the kind $\psi\psi^{*}$, it becomes a
reasonable assumption that a Wigner distribution function for this
system is a function: $$
\psi\left(g\right)\,\,\mapsto\,\,\,\tilde{W}\left(g;JMN\,M^{\prime}N^{\prime}\right)
$$ This means that, from this point of view, $W$ is a complex
function defined on a space which is the product of $G$ itself times a lattice,
related to the set of UIR's for the
group $G$. Moreover, this analysis does not appear to be necessary
in the cartesian case and in the previous $G=U\left(1\right)$
case, as they are abelian. In that case left and right actions are
the same.

The properties this distribution should satisfy are\footnote{It
can be checked that these properties are compatible among
themselves.}:
\begin{itemize}
\item
complex conjugation exchanges primed with unprimed indices (this
assures the expected transformation property for hermitian
conjugation when the Wigner distribution represents a density
operator): \beq \left(\tW\left(g;JMN\,M^{\prime}N^{\prime}\right)\right)^{*}\,=\,
\tW\left(g;JM^{\prime}N^{\prime}\,MN\right); \eeq
\item
it reproduces the expected marginal distributions: \beqa
\int_{G}d\mu\,\tW\left(g;JMN\,MN\right)&=&\left|\psi_{JMN}\right|^{2}
\nonumber
\\
\sum_{JMN}\tW\left(g;JMN\,MN\right)&=&\left|\psi\left(g\right)\right|^{2}
\eeqa
\item
for a left action of $G$ the transformation of the state
$\psi^{\prime}\left(g\right)=\psi\left(g^{\prime-1}g\right)\,\rightarrow$
\beq
\tW^{\prime}\left(g;JMN\,M^{\prime}N^{\prime}\right)=\sum_{M_{1}M_{1}^{\prime}}
D^{J}_{MM_{1}}\left(g^{\prime}\right)D^{J}_{M^{\prime}M_{1}^{\prime}}
\left(g^{\prime}\right)^{*}\tW\left(g^{\prime-1}g;JM_{1}N\,M_{1}^{\prime}N^{\prime}\right)
\eeq
\item
for a right action of $G$ the transformation of the state
$\psi^{\prime}\left(g\right)=\psi\left(gg^{\prime}\right)\,\rightarrow$
\beq
\tW^{\prime\prime}\left(g;JMN\,M^{\prime}N^{\prime}\right)=\sum_{N_{1}N_{1}^{\prime}}
D^{J}_{N_{1}N}\left(g^{\prime-1}\right)D^{J}_{N_{1}^{\prime}N^{\prime}}
\left(g^{\prime-1}\right)^{*}\tW\left(gg^{\prime};JMN_{1}\,M^{\prime}N_{1}^{\prime}\right)
\eeq
\end{itemize}

Eq.(\ref{wigfundelta}) suggests that a form of Wigner function
is: \beq
\tW\left(g;JMN\,M^{\prime}N^{\prime}\right)\,=\,N_{J}\int_{G}d\mu^{\prime}\int_{G}d\mu^{\prime\prime}
\,\delta\left(g^{-1}s\left(g^{\prime},g^{\prime\prime}\right)\right)
\,D^{J}_{MN}\left(g^{\prime}\right)\psi^{*}\left(g^{\prime}\right)
D^{J}_{M^{\prime}N^{\prime}}\left(g^{\prime\prime}\right)^{*}\psi\left(g^{\prime\prime}\right)
\label{tildewig} \eeq

 The expression
$s\left(g^{\prime},g^{\prime\prime}\right)$ is a group element,
depending on the two variables $g^{\prime},g^{\prime\prime}$. It
is a generalization of the average element $\frac{x+y}{2}$ in the
cartesian case. Without entering into a characterization of this
function, it can be seen that if
$s\left(g^{\prime},g^{\prime\prime}\right)$ is the midpoint along
the geodesic curve (with respect to the invariant Cartan-Killing metric
on $G$) joining $g^{\prime}$ to $g^{\prime\prime}$ then
the eq.(\ref{tildewig}) defines an acceptable Wigner distribution
for a quantum mechanical system on a compact semisimple Lie group
\footnote{The geodesic curve is the solution to the variational
problem:
\beq\delta\,\int_{\sigma_{1}}^{\sigma{2}}d\sigma\,\left[g_{rs}\left(q\left(\sigma\right)\right)
\frac{dq^{r}}{d\sigma}\frac{dq^{s}}{d\sigma}\right]^{1/2}\,=\,0\label{geodesics}\eeq
where $g_{rs}$ are the components of the Riemannian metric tensor
whose value in the identity of $G$ is given in terms of the
structure constants of the Lie algebra $\underline{G}$,
$g_{rs}\left(e\right)=-c_{ru}^{v}c_{sv}^{u}$, and whose value in a
generic point of is obtained by shifting that of the origin acting
with left and right shift (translations). }\cite{mukunda1}.

\subsection{A Weyl-Wigner isomorphism}

The definition (\ref{tildewig}) can be immediately extended to
associate a function
$\tW_{\Ao}\left(g;jmn\,m^{\prime}n^{\prime}\right)$, a symbol,
to every linear operator $\Ao$ on $\hil$ of Hilbert-Schmidt
class. In terms of the integral kernel $\la
g^{\prime\prime}\mid\Ao\mid g^{\prime}\ra$ of $\Ao$, in a way
similar to eq.(\ref{wigfundelgenq}), one has: \beq
\tW_{\Ao}\left(g;jmn\,m^{\prime}n^{\prime}\right)\,=
\,N_{j}\int_{G}d\mu^{\prime}\int_{G}\,d\mu^{\prime\prime}\,
\delta\left(g^{-1}s\left(g^{\prime}g^{\prime\prime}\right)\right)\,
D^{j}_{m^{\prime}n^{\prime}}\left(g^{\prime\prime}\right)^{*}D^{j}_{mn}\left(g^{\prime}\right)\,
\la g^{\prime\prime}\mid\Ao\mid g^{\prime}\ra \eeq

It is the case that this expression determines $\Ao$ 
completely, however this happens in an overcomplete
manner: there are certain linear relations obeyed by $\tW_{\Ao}$ 
which have an $\Ao$ independent form. Properties of the function 
$s\left(g^{\prime},g^{\prime\prime}\right)$ make it possible to prove that, for this symbol:
\beqa
&~&\sum_{m^{\prime}n^{\prime}}D^{j}_{m^{\prime}m^{\prime\prime}}\left(g\right)
D^{j}_{n^{\prime\prime}n^{\prime}}\left(g\right)\tW_{\Ao}\left(g;jmn\,m^{\prime}n^{\prime}\right)=\nonumber\\
&=&\,N_{j}\int_{G}d\mu^{\prime}\int_{G}d\mu^{\prime\prime}\,
\delta\left(g^{-1}s\left(g^{\prime},g^{\prime\prime}\right)\right)\,\la gg^{\prime-1}g\mid\Ao\mid g^{\prime}\ra
\,D^{j}_{n^{\prime\prime}m^{\prime\prime}}\left(g^{\prime}\right)
D^{j}_{mn}\left(g^{\prime}\right)\,\,\,\,\,\,\,\,\,\,\,\,\,\,\,\,\,\,\,\,
\eeqa

The r.h.s. of this relation shows a symmetry under the simultaneous 
interchanges $m\,\leftrightarrow\,n^{\prime\prime}$ and $n\,\leftrightarrow\,m^{\prime\prime}$, 
and this is independent of
$\Ao$: so l.h.s. should have this symmetry too. This is the sense in which 
$\tW_{\Ao}\left(g;jmn\,m^{\prime}n^{\prime}\right)$ contains information about $\Ao$ 
in an overcomplete manner, and this happens when $G$ is non abelian. Taking advantage of this, 
one can associate a symbol to an operator
in a simpler way:
\beqa
A\left(g,jmm^{\prime}\right)&=&N_{j}^{-1}\,\sum_{n}\tW_{\Ao}\left(g,jmn\,m^{\prime}n\right)
\nonumber
\\
&=&
\int_{G}d\mu^{\prime}\int_{G}d\mu^{\prime\prime}\,
\delta\left(g^{-1}s\left(g^{\prime},g^{\prime\prime}\right)\right)\,\la g^{\prime\prime}\mid\Ao\mid g^{\prime}\ra
\,D^{j}_{mm^{\prime}}\left(g^{\prime}g^{\prime\prime -1}\right)\,\,\,\,\,\,\,\,\,\,\,\,\,\,\,\,\,\,\,\,\,\label{wigfungroup}
\eeqa
This relation can be compared to the cartesian case (\ref{wigfundelgenq}). 
In the $\delta$ factor the role of $\frac{x+y}{2}$ is played by the geodesic 
average $s\left(g^{\prime},g^{\prime\prime}\right)$; position eigenstates are now
$\mid g^{\prime}\ra$ and $\mid g^{\prime\prime}\ra$ and the plane wave factor 
is now a unitary irreducible representation of the group $G$. For example, if $\Ao$ is a multiplication operator:
\beq
\Ao\,=\,\int_{G}\,d\mu\,f\left(g\right)\,\mid g\ra\la g\mid
\eeq
then its symbol is:
\beq
A\left(g,jmn\right)\,=\,f\left(g\right)\delta_{mn}
\eeq
This is defined the Weyl symbol corresponding to the operator $\Ao$, and it is very close to expression
(\ref{wigfundelgenq}).
The passage $\Ao\,\rightarrow\,\Ao^{\dagger}$ results in:
\beq
A^{\dagger}\left(g,jmm^{\prime}\right)\,=\,A\left(g,jm^{\prime}m\right)^{*}
\eeq
The transformation properties of the Weyl symbol under a left or right regular representation of $G$:
\beqa
\Ao^{\prime}&=&\Vo\left(g^{\prime}\right)\Ao\Vo^{\dagger}\left(g^{\prime}\right)\,\rightarrow
\nonumber
\\
A^{\prime}\left(g;jmm^{\prime}\right)&=&\sum_{m_{1}m_{1}^{\prime}}
D^{j}_{mm_{1}}\left(g^{\prime}\right)D^{j}_{m^{\prime}m_{1}^{\prime}}\left(g^{\prime}\right)^{*}
\,A\left(g^{\prime -1}g;jm_{1}m_{1}^{\prime}\right)
\eeqa
and:
\beqa
\Ao^{\prime\prime}&=&\Ro\left(g^{\prime}\right)\Ao\Ro^{\dagger}\left(g^{\prime}\right)\,\rightarrow
\nonumber
\\
A^{\prime\prime}\left(g;jmm^{\prime}\right)&=&A\left(gg^{\prime};jmm^{\prime}\right)
\eeqa
Had one chosen, in (\ref{wigfungroup}), to sum over the other pair of indices, the symbol would have
had these covariance properties interchanged. Another very interesting relation these symbols satisfy is:
\beq
Tr\left[\Ao\Bo\right]\,=\,\sum_{jmn}N_{j}\int_{G}d\mu\,
A\left(g,jmn\right)B\left(g;jnm\right)
\eeq
The very important aspect of the formalism one has developed is that 
it can be cast in the form of a Weyl-Wigner isomorphism, in a way formally 
similar to (\ref{wmapw}), introducing a set of quantizer operators.

It can be seen that the symbol can be written as: \beq
A\left(g;jmn\right)\,=\,Tr\left[\Ao\Wo\left(g,jmn\right)\right]
\eeq where the quantizer operators are given by: \beq
\Wo\left(g,jmn\right)\,=\,\sum_{j^{\prime}m^{\prime}n^{\prime}}
N_{j^{\prime}}\int_{G}d\mu^{\prime}\,
\Uo\left(j^{\prime}n^{\prime}m^{\prime}\right)\Vo\left(g^{\prime}\right)\,
D^{j}_{mn}\left(g^{\prime}\right)D^{j^{\prime}}_{n^{\prime}m^{\prime}}
\left(g^{-1}s_{0}\left(g^{\prime -1}\right)\right)
\label{quantizers} \eeq In this expression, operator
$\Uo\left(\cdot\right)$ is defined by: \beq
\left(\Uo\left(jmn\right)\psi\right)\left(g\right)\,=\,D^{j}_{mn}\left(g\right)\psi\left(g\right)
\eeq They are analogous to the operators $\Uo\left(m\right)$
defined by eq.(\ref{cylgen}) for the case of the group
$U\left(1\right)$: the difference with those is that they are not
unitary. Nevertheless they satisfy a related condition:
\beq
\sum_{M}\Uo^{\dagger}\left(jMn\right)\Uo\left(jMn^{\prime}\right)\,=\,\sum_{M}
\Uo^{\dagger}\left(jnM\right)\Uo\left(jn^{\prime}M\right)\,=\,\delta_{n^{\prime}n}\idop
\eeq
The analogy of this operator $\Wo\left(g,jmn\right)$ (\ref{quantizers}) with the quantizer used in the
cartesian case is evident, as $s_{0}\left(g\right)$ is a shorthand
for $s\left(e,g\right)$, the midpoint in the geodesic joining the
point $g$ to the identity. Exponential map for compact groups
enables to recover: \beq
s_{0}\left(g\right)\,=\,s_{0}\left(\exp\left(
\alpha_{r}e_{r}\right)\right)\,
=\,s_{0}\left(\exp\left(\frac{1}{2}\alpha_{r}e_{r}\right)\right)
\eeq These quantizers satisfy: \beq
\Wo^{\dagger}\left(g;jmn\right)\,=\,\Wo\left(g;jnm\right) \eeq and
are a complete trace orthonormal system: \beq
Tr\left[\Wo^{\dagger}\left(g^{\prime};j^{\prime}m^{\prime}n^{\prime}\right)
\Wo\left(g;jmn\right)\right]\,=\,N_{j}^{-1}\delta_{jj^{\prime}}\delta_{mm^{\prime}}\delta_{nn^{\prime}}
\delta\left(g^{-1}g^{\prime}\right) \eeq so that a Weyl map, a
quantization map, can be defined as: \beq
\Ao\,=\,\sum_{jmn}N_{j}\int_{G}d\mu\,A\left(g;jmn\right)\Wo\left(g;jnm\right)
\eeq This relation establish one of the map in a Weyl-Wigner
isomorphism built starting from a quantum system on a compact
simple Lie group: This isomorphism maps Hilbert-Schmidt operators
into functions in the space
$\mathcal{F}\left(G\times\Gamma\right)$ where $\Gamma$ is a
lattice, a set of discrete indices related to unitary irreducible
representations of the group $G$. This space, in analogy with the
exemplum case of $G=U\left(1\right)$, is a sort of quantum
cotangent space.

\subsection{A noncommutative product among functions on a Quantum Cotangent Space}

The isomorphism just outlined enables in a natural way to define a non abelian product on the set of functions
on what has been called a quantum cotangent space:
\beq
\left(A*B\right)\left(g;\gamma\right)\,=\,Tr\left[\Ao\Bo\Wo\left(g,\gamma\right)\right]
\eeq
(here $\left(g;\gamma\right)$ is a short cut for $\left(g;jmn\right)$ while $\left(g;\underline{\gamma}\right)$
stands for $\left(g;jnm\right)$- as it has been seen, the ordering of the labels is important -).

Hence:
\beq
\left(A*B\right)\left(g,\gamma\right)=\sum_{\tilde{\gamma}\,\check{\gamma}}
N_{\tilde{j}}N_{\check{j}}
\int_{G}d\tilde{g}\int_{G}d\check{g}\,
A\left(\tilde{g},\tilde{\gamma}\right)B\left(\check{g},\check{\gamma}\right)\left[Tr\,
\hat{W}\left(g,\gamma\right)\hat{W}\left(\tilde{g},\tilde{\underline{\gamma}}\right)
\hat{W}\left(\check{g},\check{\underline{\gamma}}\right)\right]\label{Gstarprod}
\eeq The product is non local, and the integral kernel is given by
the trace term between square brackets. To analyse this term, the
first step is to study the possibility of a kind of inversion of
(\ref{quantizers}).

Eventually\footnote{In appendix there are the details of the calculation related to this section.}: \beq
\hat{U}\left(\tilde{j}\tilde{n}\tilde{m}\right)\hat{V}\left(\tilde{g}\right)=\sum_{\gamma}\int_{G}dg
\,\hat{W}\left(g,\gamma\right)D^{\tilde{j}}_{\tilde{n}\tilde{m}}\left(s_{0}\left(\tilde{g}\right)g\right)
\left(D^{j}_{mn}\left(\tilde{g}\right)\right)^{*}N_{j}\label{Wopinv}\eeq
This can be seen as a sort of antitransform of (\ref{quantizers}).

The second step of the analysis just gives the composition
properties of $\hat{U}$and $\hat{V}$ operators:
\beq
\hat{U}\left(j^{\prime}n^{\prime}m^{\prime}\right)
\hat{U}\left(j^{\prime\prime}n^{\prime\prime}m^{\prime\prime}\right)=\sum_{JNM,\lambda}
C^{j^{\prime},j^{\prime\prime},J\,\lambda}
_{n^{\prime}m^{\prime},n^{\prime\prime}m^{\prime\prime},NM}\hat{U}\left(JNM\right)
\eeq
\beq
\hat{U}\left(j^{\prime}n^{\prime}m^{\prime}\right)\hat{V}\left(g^{\prime}\right)
\hat{U}\left(j^{\prime\prime}n^{\prime\prime}m^{\prime\prime}\right)\hat{V}\left(g^{\prime\prime}\right)
=\sum_{k=1}^{N_{j^{\prime\prime}}}\sum_{JNM,\lambda}
D^{j^{\prime\prime}}_{n^{\prime\prime}k}\left(g^{\prime-1}\right)
C^{j^{\prime},j^{\prime\prime},J\,\lambda}
_{n^{\prime}m^{\prime},km^{\prime\prime},NM}\hat{U}\left(JNM\right)
\hat{V}\left(g^{\prime}g^{\prime\prime}\right)
\label{compUV}
\eeq

This notation writes in a compact form the product of
two UIR's in terms of direct sum of UIR's \cite{simongroup}. The index $\lambda$ keeps track of
multiple occurrences of a given $D^{j}$\footnote{ In the case of the group $SU\left(2\right)$, in terms of
the standard Clebsh-Gordan coefficients, this would be:
\beq
C^{j^{\prime},j^{\prime},J}_{n^{\prime}m^{\prime},n^{\prime\prime}m^{\prime\prime},NM}=
c^{JN}_{j^{\prime}n^{\prime},j^{\prime\prime}n^{\prime\prime}}
c^{JM}_{j^{\prime}m^{\prime},j^{\prime\prime}m^{\prime\prime}}
\eeq
}.

The third step is to study the composition properties in the set of $\hat{W}$. It can be seen that:
\beqa
\hat{W}\left(g,\gamma\right)\hat{W}\left(\tilde{g},\tilde{\gamma}\right)&=&
\sum_{\gamma^{\prime}\,\gamma^{\prime\prime}\,\gamma^{\prime\prime\prime}}
N_{j^{\prime}}N_{j^{\prime\prime}}N_{j^{\prime\prime\prime}}
\int_{G}dg^{\prime}
\int_{G}dg^{\prime\prime}\int_{G}dg^{\prime\prime\prime}
\sum_{k=1}^{N_{j^{\prime\prime}}}\sum_{JNM,\lambda}\cdot\nonumber
 D^{j^{\prime\prime}}_{n^{\prime\prime}k}\left(g^{\prime-1}\right)
\\
&\cdot&
C^{j^{\prime},j^{\prime\prime},J\,\lambda}
_{n^{\prime}m^{\prime},km^{\prime\prime},NM}
\hat{W}\left(g^{\prime\prime\prime},\gamma^{\prime\prime\prime}\right)
D^{J}_{NM}\left(s_{o}\left(g^{\prime}g^{\prime\prime}\right)g^{\prime\prime\prime}\right)
\left(D^{j^{\prime\prime\prime}}_{m^{\prime\prime\prime}n^{\prime\prime\prime}}
\left(g^{\prime}g^{\prime\prime}\right)\right)^{*}\cdot\nonumber
\\
&\cdot& D^{j}_{mn}\left(g^{\prime}\right)
D^{j^{\prime}}_{m^{\prime}n^{\prime}}\left(g^{-1}s_{o}\left(g^{\prime-1}\right)\right)
D^{\tilde{j}}_{\tilde{m}\tilde{n
}}\left(g^{\prime\prime}\right)
D^{j^{\prime\prime}}_{m^{\prime\prime}n^{\prime\prime}}\left(\tilde{g}^{-1}
s_{o}\left(g^{\prime\prime-1}\right) \right)
\label{proWW}
\eeqa
The definition (\ref{Gstarprod}) indicates that the problem is evaluating the trace of the product of three
$\hat{W}$ operators:
\beqa
Tr\left[\hat{W}\left(g,\gamma\right)\hat{W}\left(\tilde{g},\tilde{\gamma}\right)
\hat{W}\left(\check{g},\check{\gamma}\right)\right]&=&
\sum_{\gamma^{\prime}\,\gamma^{\prime\prime}\,\gamma^{\prime\prime\prime}}
N_{j^{\prime}}N_{j^{\prime\prime}}N_{j^{\prime\prime\prime}}
\int_{G}dg^{\prime}
\int_{G}dg^{\prime\prime}\int_{G}dg^{\prime\prime\prime}
\sum_{k=1}^{N_{j^{\prime\prime}}}\sum_{JNM,\lambda}\cdot\nonumber
\\
&\cdot& D^{j^{\prime\prime}}_{n^{\prime\prime}k}\left(g^{\prime-1}\right)
C^{j^{\prime},j^{\prime\prime},J\,\lambda}
_{n^{\prime}m^{\prime},km^{\prime\prime},NM}
D^{J}_{NM}\left(s_{o}\left(g^{\prime}g^{\prime\prime}\right)g^{\prime\prime\prime}\right)
\cdot\nonumber
\\
&\cdot&
\left(D^{j^{\prime\prime\prime}}_{m^{\prime\prime\prime}n^{\prime\prime\prime}}
\left(g^{\prime}g^{\prime\prime}\right)\right)^{*}
D^{j}_{mn}\left(g^{\prime}\right)
D^{j^{\prime}}_{m^{\prime}n^{\prime}}\left(g^{-1}s_{o}\left(g^{\prime-1}\right)\right)
\cdot\nonumber
\\
&\cdot& D^{\tilde{j}}_{\tilde{m}\tilde{n
}}\left(g^{\prime\prime}\right)
D^{j^{\prime\prime}}_{m^{\prime\prime}n^{\prime\prime}}\left(\tilde{g}^{-1}
s_{o}\left(g^{\prime\prime-1}\right)
\right)\cdot\nonumber
\\
&\cdot&
Tr\left[\hat{W}\left(g^{\prime\prime\prime},\gamma^{\prime\prime\prime}\right)
\hat{W}\left(\check{g},\check{\gamma}\right)\right]\label{kernel}
\eeqa
Now from trace evaluation of
 the composition of two quantizer operators, one has  for the integral kernel of the star product (\ref{Gstarprod}):
\beqa
Tr\left[\hat{W}\left(g,\gamma\right)\hat{W}\left(\tilde{g},\tilde{\underline{\gamma}}\right)
\hat{W}\left(\check{g},\check{\underline{\gamma}}\right)\right]&=&
\sum_{\gamma^{\prime}\,\gamma^{\prime\prime}}\sum_{\Gamma,\lambda}
\int_{G}dg^{\prime}\int_{G}dg^{\prime\prime}\sum_{k=1}^{N_{j^{\prime\prime}}}
N_{j^{\prime}}N_{j^{\prime\prime}}D^{j^{\prime\prime}}_{n^{\prime\prime}k}\left(g^{\prime-1}\right)
\cdot\nonumber
\\
&\cdot&
C^{j^{\prime},j^{\prime\prime},J\,\lambda}
_{n^{\prime}m^{\prime},km^{\prime\prime},NM}
D^{J}_{NM}\left(s_{o}\left(g^{\prime}g^{\prime\prime}\right)\check{g}\right)
\left(D^{\check{j}}_{\check{m}\check{n}}\left(g^{\prime}g^{\prime\prime}\right)\right)^{*}\cdot\nonumber
\\
&\cdot& D^{j}_{mn}\left(g^{\prime}\right)
D^{j^{\prime}}_{m^{\prime}n^{\prime}}\left(g^{-1}s_{o}\left(g^{\prime-1}\right)\right)
D^{\tilde{j}}_{\tilde{n}\tilde{m
}}\left(g^{\prime\prime}\right)\cdot\nonumber
\\
&\cdot&
D^{j^{\prime\prime}}_{m^{\prime\prime}n^{\prime\prime}}\left(\tilde{g}^{-1}
s_{o}\left(g^{\prime\prime-1}\right)
\right)
\eeqa
(here $\Gamma$ is a short for $\left(J,N,M\right)$)
The product will be explicitly obtained putting this last equation into (\ref{Gstarprod})

There is also another way to evaluate the expression (\ref{kernel}), and it is based on:
\beq
\langle g\mid\hat{W}\left(\tilde{g},\tilde{\gamma}\right)\mid g^{\prime}\rangle=
D^{\tilde{j}}_{\tilde{m}\tilde{n}}\left(gg^{\prime-1}\right)
\delta\left(\tilde{g}^{-1}s\left(gg^{\prime}\right)\right)
\eeq
So:
\beqa
Tr\left[\hat{W}\left(g,\gamma\right)\hat{W}\left(\tilde{g},\tilde{\gamma}\right)
\hat{W}\left(\check{g},\check{\gamma}\right)\right]&=&\int_{G}dg\int_{G}dg^{\prime}
\int_{G}dg^{\prime\prime}\cdot\nonumber
\\
&\cdot&
D^{j}_{mn}\left(g^{\prime}g^{\prime\prime-1}\right)
D^{\tilde{j}}_{\tilde{m}\tilde{n}}\left(g^{\prime\prime}g^{\prime\prime\prime-1}\right)
D^{\check{j}}_{\check{m}\check{n}}\left(g^{\prime\prime\prime}g^{-1}\right)\cdot\nonumber
\\
&\cdot&
\delta\left(g^{-1}s\left(g^{\prime}g^{\prime\prime}\right)\right)
\delta\left(\tilde{g}^{-1}s\left(g^{\prime\prime}g^{\prime\prime\prime}\right)\right)
\delta\left(\check{g}^{-1}s\left(g^{\prime\prime\prime}g\right)\right)
\nonumber\eeqa

\subsection{Recovering the case $G=U\left(1\right)$}
In this section it will be shown how these calculations look like in the
special case where $G$ is the group $U\left(1\right)$. As already said, the group
has the manifold structure of a circle $S^{1}$: $\theta$ is the
"coordinate" on this space. The Haar measure is chosen to be
normalized:
\beq
\int_{G}d\mu\,=\,\int_{S^{1}}\frac{d\theta}{2\pi}\,=\,1
\eeq
The Hilbert space is
$\mathcal{H}=\mathcal{L}^{2}\left(S^{1},\frac{d\theta}{2\pi}\right)$,
and an orthonormal basis is the set
$\phi_{n}\left(\theta\right)=e^{in\theta}$
($n\,\in\,\zed$). Discrete Fourier transform and anti-transform can
be summarized in the formula:
\beq
\sum_{n=-\infty}^{\infty}\,e^{-in\left(\theta-\theta^{\prime}\right)}=
\delta\left(\theta-\theta^{\prime}\right)
\label{discdelta}
\eeq
An overcomplete unnormalizable basis is given by
$\mid\theta\rangle$ such that (eq.\ref{thetabasis})
$$\langle\theta\mid\psi\rangle=\psi\left(\theta\right)$$
$$\langle\theta\mid\phi_{n}\rangle=e^{in\theta}$$
This group is abelian: unitary irreducible representations are
one-dimensional. They will be labelled by an integer $n$: the D
functions of the preceding section become simply number (the so
called character), so $\Uo\left(\cdot\right)$ and $\Vo\left(\cdot\right)$ are:
\beq
\hat{V}\left(\theta\right)\mid\phi_{n}\rangle=e^{-in\theta}\mid\phi_{n}\rangle
\eeq
\beq
\hat{U}\left(m\right)\mid\phi_{n}\rangle=\mid\phi_{n+m}\rangle
\eeq
that are exactly the operators ad hoc introduced in (\ref{UVopcircle}).
The quantizer (\ref{quantizers}) acquires the specific form:
\beq
\hat{W}\left(\theta,n\right)=
\sum_{m^{\prime}}\int_{S^{1}}\frac{d\theta^{\prime}}{2\pi}\hat{U}\left(m^{\prime}\right)\hat{V}\left(\theta^{\prime}\right)
e^{in\theta^{\prime}}e^{-im^{\prime}\left(\theta+\theta^{\prime}/2\right)}
\eeq
which coincides with operator (\ref{circlequant}).
Now, as before, the first problem is to "invert" this relation:
\beq
\int_{S^{1}}\frac{d\theta}{2\pi}\hat{W}\left(\theta,n\right)e^{ik\theta}=
\sum_{m^{\prime}}\int_{S^{1}}\frac{d\theta}{2\pi}\int_{S^{1}}\frac{d\theta^{\prime}}{2\pi}
\hat{U}\left(m^{\prime}\right)\hat{V}\left(\theta^{\prime}\right)
e^{in\theta^{\prime}}e^{-im^{\prime}\left(\theta+\theta^{\prime}/2\right)}
e^{ik\theta}
\eeq
Using (\ref{discdelta}) enables to simplify the r.h.s.:
\beq
\int_{S^{1}}\frac{d\theta}{2\pi}\hat{W}\left(\theta,n\right)e^{ik\theta}=
\int_{S^{1}}\frac{d\theta^{\prime}}{2\pi}
\hat{U}\left(k^{\prime}\right)\hat{V}\left(\theta^{\prime}\right)
e^{in\theta^{\prime}}e^{-ik\theta^{\prime}/2}
\eeq
Again:
\beq
\sum_{n}e^{-in\theta^{\prime\prime}}
\int_{S^{1}}\frac{d\theta}{2\pi}\hat{W}\left(\theta,n\right)e^{ik\theta}=
\hat{U}\left(k\right)\hat{V}\left(\theta^{\prime\prime}\right)e^{-ik\theta^{\prime\prime}/2}
\eeq
so that:
\beq
\hat{U}\left(n\right)\hat{V}\left(\theta\right)=e^{in\theta/2}\sum_{m}e^{-im\theta}\int_{S^{1}}
\frac{d\theta^{\prime}}{2\pi}\hat{W}\left(\theta^{\prime},m\right)e^{in\theta^{\prime}}
\label{Wopinvdisc}\eeq This is the actual form of (\ref{Wopinv})
for this example. The second step is the study of composition
properties of two of these operators $W$:
\beqa
\hat{W}\left(\theta,n\right)\hat{W}\left(\tilde{\theta},\tilde{n}\right)&=&
\sum_{n^{\prime}\,n^{\prime\prime}}\int_{S^{1}}\frac{d\theta^{\prime}}{2\pi}
\int_{S^{1}}\frac{d\theta^{\prime\prime}}{2\pi}
\hat{U}\left(n^{\prime}\right)\hat{V}\left(\theta^{\prime}\right)
\hat{U}\left(n^{\prime\prime}\right)\hat{V}\left(\theta^{\prime\prime}\right)\cdot\nonumber
\\
&\cdot&
e^{in\theta^{\prime}}e^{i\tilde{n}\theta^{\prime\prime}}e^{-in^{\prime}\left(\theta+\theta^{\prime}/2\right)}
e^{-in^{\prime\prime}\left(\tilde{\theta}+\theta^{\prime\prime}/2\right)}
\eeqa
One can finally computes:
\beqa
\hat{W}\left(\theta,n\right)\hat{W}\left(\tilde{\theta},\tilde{n}\right)&=&
\sum_{n^{\prime}\,n^{\prime\prime}\,n^{\prime\prime\prime}}
\int_{S^{1}}\frac{d\theta^{\prime}}{2\pi}
\int_{S^{1}}\frac{d\theta^{\prime\prime}}{2\pi}
\int_{S^{1}}\frac{d\theta^{\prime\prime\prime}}{2\pi}\hat{W}\left(\theta^{\prime\prime\prime},n^{\prime\prime\prime}\right)
e^{i\left(n^{\prime}\theta^{\prime\prime}-n^{\prime\prime}\theta^{\prime}\right)/2}\cdot\nonumber
\\
&\cdot&
e^{-i\left[n^{\prime\prime\prime}\left(\theta^{\prime}+\theta^{\prime\prime}\right)
-\left(n^{\prime}+n^{\prime\prime}\right)\theta^{\prime\prime\prime}\right]}
e^{i\left(n\theta^{\prime}+\tilde{n}\theta^{\prime\prime}\right)}
e^{-i\left(n^{\prime}\theta+n^{\prime\prime}\tilde{\theta}\right)}
\eeqa This relation clearly indicates that the origin of the
properties of this formalism should be addressed to the
specific form of the commutation relations. Every phase factor in the
integral can be written as a skewsymmetric combination of two
variables.

The third step is to evaluate the trace. Noting that: \beq Tr
\left[\hat{W}\left(\theta,n\right)\hat{W}\left(\tilde{\theta},\tilde{n}\right)\right]
=\delta_{n\tilde{n}}\delta\left(\theta-\tilde{\theta}\right) \eeq
one obtains: \beqa Tr\left[
\hat{W}\left(\theta,n\right)\hat{W}\left(\tilde{\theta},\tilde{n}\right)\hat{W}\left(\check{\theta},\check{n}\right)\right]
&=&\sum_{n^{\prime}\,n^{\prime\prime}}\int_{S^{1}}\frac{d\theta^{\prime}}{2\pi}
\int_{S^{1}}\frac{d\theta^{\prime\prime}}{2\pi}\,
e^{i\left(n^{\prime}\theta^{\prime\prime}-n^{\prime\prime}\theta^{\prime}\right)/2}\cdot\nonumber
e^{i\left(\tilde{n}\theta^{\prime\prime}-n^{\prime\prime}\check{\theta}\right)}
\\
&\cdot&
e^{-i\left[\check{n}\left(\theta^{\prime}+\theta^{\prime\prime}\right)-
\left(n^{\prime}+n^{\prime\prime}\right)\check{\theta}\right]}
e^{i\left(n\theta^{\prime}-n^{\prime}\theta\right)}
\label{trWWWdisc}\eeqa Given two symbols, the product induced by
this isomorphism is: \beq
\left(A*B\right)\left(\theta,n\right)=\sum_{\tilde{n}\,\check{n}}\int_{S^{1}}\frac{d\tilde{\theta}}{2\pi}
\int_{S^{1}}\frac{d\check{\theta}}{2\pi}A\left(\tilde{\theta},\tilde{n}\right)B\left(\check{\theta},\check{n}\right)
Tr\left[
\hat{W}\left(\theta,n\right)\hat{W}\left(\tilde{\theta},\tilde{n}\right)\hat{W}\left(\check{\theta},\check{n}\right)\right]
\label{discstarprod} \eeq

\section{A noncommutative product on the classical cotangent space}

As it has been stressed, the formalism outlined defines a Weyl symbol for a certain class of quantum systems
as a function $A\left(g;jmn\right)$, not on the classical phase space. The nature of these three indices can
be once more analysed via the Peter-Weyl theorem. Index $j$ labels the irreducible representations
 $D^{j}\left(\cdot\right)$: each of these is realised on a finite dimensional Hilbert space, whose elements are
 labelled by the index $m$. The left regular representation, which is one of the building blocks of this
 construction, is highly reducible on the Hilbert space of square integrable functions on the group $G$.
 Exactly, the degeneracy of its occurence is equal to $N_{j}$, the dimension of the space of representation
$\mathcal{H}_{j}$. This occurrence is taken into account by the
index $n$. The form of the symbol suggests that it can be
considered as a complex function defined on the product
$G\times\mathcal{H}_{0}$ of the group with a smaller Hilbert
space, namely that carrying each UIR $D^{j}$ just once.

If
$\mathcal{H}_{j}$ is the linear span $Sp\,\{\mid j,m\ra\}$ of
dimension $N_{j}$, with $\la j^{\prime}m^{\prime}\mid j
m\ra=\delta_{jj^{\prime}} \delta_{mm^{\prime}}$, then this new
space is: \beq \mathcal{H}_{0}\,=\,\sum_{j}\oplus\,\mathcal{H}_{j}
\eeq In other words, a Weyl symbol can be regarded as a function
of $G$ tensor a matrix on $\mathcal{H}_{0}$, with the crucial
property that this matrix is block diagonal with respect to the
decomposition of $\mathcal{H}_{0}$ in terms of $\mathcal{H}_{j}$.
Of course, this interpretation is valid only for non abelian
groups: the space on which symbols are defined can be seen as a
quantum cotangent space of $G$.

A symbol can be mapped into a block diagonal, $g$ dependent operator on $\mathcal{H}_{0}$:
\beq
\underline{\Ao}\,=\,\sum_{jmn}\sqrt{N_{j}}\,A\left(g;jmn\right)\mid jm\ra\la jn\mid
\eeq
satisfying:
\beq
Tr_{\mathcal{H}}\left[\Ao\Bo\right]\,=\,\int_{G}d\mu\,Tr_{\mathcal{H}_{0}}
\left[\underline{\Ao}\,\underline{\Bo}\right]
\eeq
This operator can thus be written as a sum:
\beqa
\underline{\Ao}\left(g\right)&=&\sum_{j}\oplus\underline{\Ao}_{j}\left(g\right)
\nonumber
\\
\underline{\Ao}_{j}\left(g\right)&=&\sum_{m,n}\sqrt{N_{j}}\,A\left(g;jmn\right)\mid
jm\ra\la jn\mid \eeqa of terms acting on each irreducibility
subspace. On each of these subspaces there is a set of
$\Jo_{r}^{\left(j\right)}$, Hermitian generators of the left
action of $G$. In this context irreducibility means that it is
possible to expand every operator acting on $\mathcal{H}_{j}$ as a
sum of symmetrised polynomials in these variables:
\beq
\underline{\Ao}_{j}\left(g\right)\,=\,\sum_{N=0}^{N\left(j\right)}\sum_{r_{1},r_{2},\ldots,r_{N}}
\,a_{r_{1},r_{2},\ldots,r_{N}}\left(g;j\right)
\,\{\Jo_{r_{1}}^{\left(j\right)}\Jo_{r_{2}}^{\left(j\right)}\ldots\Jo_{r_{N}}^{\left(j\right)}\}_{S}
\label{irreducibilityexp}\eeq where the symmetrised sum is: \beq
\{\Jo_{r_{1}}^{\left(j\right)}\Jo_{r_{2}}^{\left(j\right)}\ldots\Jo_{r_{N}}^{\left(j\right)}\}_{S}
\,=\, \frac{1}{N!}\sum_{P\in
S_{N}}\left(\Jo_{r_{P\left(1\right)}}^{j}\ldots\Jo_{r_{P\left(N\right)}}^{j}\right)
\eeq In these relations, the upper limit in the summation over $N$ is determined by
the specific UIR $D^{j}$; $S_{N}$ is the permutation group on $N$
elements, the superscript $j$ denotes the specific realization in
the subspace $\mathcal{H}_{j}$. The coefficients
$a_{r_{1},\ldots,r_{N}}\left(g;j\right)$ are c-numbers quantities
symmetric in $r_{1},\ldots,r_{N}$. Their dependence by $j$ can be
replaced by a dependence on the mutually commuting Casimir
operators $\hat{C}$, themselves symmetric homogeneous polynomials
in the generators $\Jo_{r}^{\left(j\right)}$. The operator
$\underline{\Ao}$ can be written as: \beq \underline{\Ao}\,=\,
\sum_{j}\sum_{N=0}^{N\left(j\right)}\sum_{r_{1}\ldots,r_{N}}\,
a_{r_{1},\ldots,r_{N}}\left(g;\hat{C}\left(j\right)\right)\,
\{\Jo_{r_{1}}^{\left(j\right)}\Jo_{r_{2}}^{\left(j\right)}\ldots\Jo_{r_{N}}^{\left(j\right)}\}_{S}
\eeq This operator can be mapped into a function: \beq
a\left(g,\vec{J}\right)\,=\,
\sum_{j}\sum_{N=0}^{N\left(j\right)}\sum_{r_{1}\ldots,r_{N}}\,
a_{r_{1},\ldots,r_{N}}\left(g;C\right)\, J_{r_{1}}J_{r_{2}}\ldots
J_{r_{N}} \eeq In this expression $\vec{J}$ is a collection of
$J_{r}$, which are the commuting classical variables associated to
the canonical momentum coordinates of the classical phase space
$T^{*}G$, while $C$ are invariant Casimir homogeneous polynomials
in them. So, there is a correspondence: $$
\Ao\,\in\,Op\left(\hil\right)\,\,\Longleftrightarrow\,\,\underline{\Ao}\left(g\right)
\,\in\,b.d.Op\left(\mathcal{H}_{0}\right)\,\,\longleftrightarrow\,\,a\left(g,\vec{J}\right)
\,\in\,\mathcal{F}\left(T^{*}G\right) $$ (here $b.d.$ means block-diagonal operators) Weyl-Wigner isomorphism
sketched along this chapter maps an operator on the whole Hilbert
space $\hil$ of square integrable functions on the group $G$ to a
symbol on what has been called "quantum cotangent space". In this
section there has been shown how such a symbol can be mapped into
a block diagonal operator on a simpler Hilbert space $\hil_{0}$,
where the left action of $G$ is reducible without any degeneracy,
and then how this operator can be mapped into a function on the
classical cotangent space for the Lie group $G$.

This chain of invertible maps can be seen as a Weyl-Wigner isomorphism 
between the space of operators on a Hilbert space, and the set of functions 
on the cotangent bundle of a compact simple Lie group. This isomorphism then 
enables to define a noncommutative product in the space of these functions, 
so to open the possibility to study a new class of noncommutative spaces.

\chapter{A fuzzy disc}\label{chapthree}

It is well known that, in the conventional formulation of quantum
field theory as the theory of formally quantized classical fields
on a classical Minkowski spacetime, ultraviolet divergences arise
when one attempts to measure the amplitude of field oscillations
at a precise given point in spacetime. These divergences seem to
be related to a quantization procedure based on a continuum
manifold structure for spacetime.

An analysis of the relations between geometry of spacetime and
quantum formalism was started by Dirac \cite{dirac26}. In his
effort to describe quantum physics on a classical phase space, he
was aware that uncertainty relations led to the impossibility of
an infinitely precise localization of points in phase space. This
was originated by the noncommutativity among operators
representing positions and momenta, whose spectra would classically
define the phase space. On a related side, von Neumann was led to
study the possibility to replace the continuum phase space
structure with a lattice, introducing the idea of smearing out
points to Planck cells of area $\sim\,2\pi\hbar$. This idea also
led to the hypothesis, to cope these ultraviolet divergences, to
replace the continuum spacetime with a fundamental lattice.
Nevertheless this hypothesis does not fit with the requirement of
a natural symmetry action of continuous groups on these
approximating spaces.

It was Heisenberg to suggest that one could use a noncommutative
structure for spacetime coordinates at very short lenght scale.
Noncommutativity would have introduced an ultraviolet cutoff.
Snyder \cite{snyder} was the first to formalize this idea. He
wrote that if one assumes that the spectra of spacetime coordinate
operators are invariant under Lorentz transformations, then of
course the usual spacetime satisfies this requirement. But it is
not the only solution: there exists a Lorentz invariant spacetime
in which there is a fundamental lenght. This space is related to a
set of operators having a Lorentz invariant spectrum. This line
was developed by Yang \cite{yang}, who studied a discrete version
of spacetime, on which a larger group (including some sort of
translations) properly acts as a symmetry. His work was largely
ignored, mostly because, at around the same time, a first
renormalization program of quantum field theory finally proved to
be successful at accurately predicting numerical values for
physical measured quantities in quantum electrodynamics.

Noncommutative geometry \cite{NCG}, considering the topology and
the geometry of the space of states as encoded in the algebraic
relations among quantum observables, provides a natural
formalization for a pointless geometry, and then for quantum field
theories \cite{szaboQFT}, and for the analysis of finite
approximations to them. Moreover, quantum gravity models, and
string theoretical models, suggest the possibility that classical
general relativity would break down at a very short lenght scale,
spacetime being no longer described by a differentiable manifold
\cite{DFR,witten}.

Since in this geometry points are ill-defined, spaces are often
thought as fuzzy. The first formalization of the idea of a fuzzy
space was introduced by Madore \cite{madorefs}, for the
sphere\footnote{There has also been studied the possibility of a
similar approximation for the torus, based on the noncommutative
torus algebra \cite{rieffelFT,portoroz}, and for complex projective spaces
\cite{balcp2}}. In his approach, a fuzzy sphere is a sequence of
nonabelian algebras, more specifically of finite rank matrix
algebras, so that at each step of this sequence there is no
manifold structure for the set of pure states. Among elements of
this sequence, Madore analysed how the fuzzy sphere can be seen as
a specific filtration of functions on a sphere. This filtration
comes from studying how the sequence of matrix algebras converges
towards the set of infinite dimensional diagonal matrices, that is
an abelian algebra. This naive notion of convergence was replaced
by a meaningful one. M.Rieffel proved that the fuzzy sphere
``converges to the sphere'' if both each step of the sequence of
finite rank matrix algebras and the algebra of functions on the
sphere, are seen as compact quantum metric spaces, and the
distance among them is the Gromov-Hausdorff distance \cite{rieffelGHdist}.

 A path integral formalism for quantum field theories on these
 spaces can be introduced. It is based on the substitution of the
functional action in an infinite dimensional space with a
functional action depending on a finite number of functional
degrees of freedom \cite{madorebook}. Quantum fields on these
spaces do not present ultraviolet divergences \cite{grossestrohm}.

The aim of this chapter is to describe a new fuzzy space, the
fuzzy disc \cite{LVZ}. It is the first example of a fuzzy
approximation of a space with a boundary, and it has been proved
to act as a regulator for ultraviolet divergences in the case of a
noninteracting field theory.

In the first chapter the Weyl-Wigner approach has been presented
as a bridge connecting the quantum to the classical formalism. It
can be thought as way to study relations  between noncommutative
geometry and commutative geometry. Since a fuzzy space can be
seen, from the dual point of view of states of those matrix
algebras, as a sequence of "nonabelian" spaces, converging to an
ordinary, continuum manifold, it is natural to think that
Weyl-Wigner formalism can be suited to study these specific models
in noncommutative geometry.

In the first section of this chapter the fuzzy sphere is reviewed
as the prototype of a fuzzy space. The original approach of Madore
is outlined, then it is presented in terms of a Weyl-Wigner
isomorphism. This isomorphism is defined used the concept of
coherent states for the group $SU\left(2\right)$, following
Berezin \cite{berezinmap}. It is a map from the finite rank matrix
algebras to a subset of the space of functions on a 2-sphere: the
rank of matrices is related to the dimension of the space on which
unitary irreducible representations of the group
$SU\left(2\right)$ are defined. Noncommutativity in the space of
functions on the sphere, via this isomorphism, is related to the
rank $N$ of the range matrices, disappearing, as required, in the
limit of $N\,\rightarrow\,\infty$.

 Moreover, this isomorphism is explicitly written in terms of the
 properties of fuzzy harmonics, that arise in the study of the spectral
properties of a fuzzy Laplacian operator, defined in such a way to
converge, in the commutative limit, to the ordinary Laplacian
operator
 on the algebra of functions on a sphere.

The end of the first section describes the setting introduced by
M.Rieffel to prove the convergence of the fuzzy sphere algebras to
the algebra of functions on the sphere.

The second part is devoted to the description of the fuzzy disc. A
Weyl-Wigner isomorphism is introduced in terms of functions on a
plane, where noncommutativity is represented by a parameter
$\theta$. In this formalization, there is no natural concept of a
sequence of finite dimensional Hilbert spaces, or finite rank
matrix algebras. So it is necessary to introduce a truncation in
the algebra of operators, with respect to a specific basis in the
Hilbert space. If the dimension of truncation $N$ is constrained
to the noncommutativity parameter $N\theta=R^{2}$, then one
obtains a sequence of finite rank matrices, converging towards an
abelian algebra of operators, that approximates functions whose
support is concentrated on a disc of radius $R$. On this sequence
of states, fuzzy derivatives and a fuzzy Laplacian can be defined,
and a system of fuzzy Bessels can be introduced. This set of fuzzy
Bessels will be used to define, as fuzzy harmonics in the case of
the sphere, a Weyl-Wigner isomorphism between the set of finite
rank matrices and the set of functions on a disc. This is the way
it is obtained a sequence of finite rank matrix algebras, that
converges, in a formal
 commutative
limit, to the algebra of functions on a disc. Moreover, the last
section shows how this approximation enables to study a first
field theory model, namely a noninteracting one, and how this
method works as an ultraviolet regulator.

\section{The fuzzy sphere as a prototype of a fuzzy space}

This section starts with the description of the fuzzy sphere,
following the original paper \cite{madorefs}. The aim is to give a
first idea of what a fuzzy sphere is, of what is the difference
between a fuzzy approximation and the lattice approximation, and
of what is a notion, though almost naive, of a limit of the fuzzy
sphere to the algebra of functions on the sphere. In the second
part of the section the fuzzy sphere will be described making use
of a Weyl-Wigner isomorphism, while the third part will explain
the exact meaning of that convergence.

The approach that J.Madore used in the introduction of the fuzzy
sphere starts from the analysis of the algebra of functions on the
sphere $S^{2}$. This algebra $\mathcal{C}\left(S^{2}\right)$ is
made of continuous functions on the sphere, and can be seen as the
quotient $\mathcal{C}\left(\real^{3}\right)/\mathcal{I}$, where
$\mathcal{I}$ is the two-sided ideal of continuous functions on
$\real^{3}$ whose value is zero on points whose coordinates
satisfy\footnote{The radius of the sphere imbedded in $\real^{3}$
has been fixed equal to $1$.}: \beq \delta_{ab}x^{a}x^{b}\,=\,1
\label{spherequotient} \eeq Functions in
$\mathcal{C}\left(\real^{3}\right)$ have a formal polynomial
expansion: \beqa
f\left(x\right)&=&f_{0}+\sum_{a=1}^{3}\,f_{a}x^{a}\,+\,\frac{1}{2}\sum_{a,b=1}^{3}\,f_{ab}x^{a}x^{b}\,+\ldots
\nonumber \\
&=&\sum_{l=0}^{\infty}\,\frac{1}{l!}\,\sum_{a_{1}\ldots
a_{l}=1}^{3}x^{a_{1}}\cdots x^{a_{l}} \eeqa Quotienting
$\mathcal{C}\left(\real^{3}\right)$ by the above ideal translates
into a set of constraints for the coefficients $f_{a_{1}\ldots
a_{l}}$:
\begin{itemize}
\item
$f_{a_{1}\ldots a_{l}}$ should be totally symmetric in the
exchange of indices $a_{1}\ldots a_{l}$ (this requirement actually
comes from commutativity of both algebras of functions);
\item
$f_{a_{1}\ldots a_{l}}$ should be traceless with respect to every pair of indices.
\end{itemize}
At each order of expansion, represented by $l$, the number of
independent coefficients is $2l+1$, and one has that:
$$\sum_{l=0}^{N}\,\left(2l+1\right)\,=\,\left(N+1\right)^{2}$$
This relation is important in this context, because one can
consider a truncated expansion of elements of this algebra: \beq
f^{\left(N\right)}\left(x\right)\,\equiv\,\sum_{l=0}^{N-1}\,\frac{1}{l!}\,\sum_{a_{1}\ldots
a_{l}=1}^{3}\,f_{a_{1}\ldots a_{l}} x^{a_{1}}\cdots x^{a_{l}}
\label{madoretruncation} \eeq The number of independent
coefficients in $f^{\left(N\right)}$ is $N^{2}$. This set of
functions is no longer an algebra, if they are multiplied via the
standard pointwise commutative product. To introduce an algebraic
structure into this vector space one could consider coefficients
as elements of $\complex^{N^{2}}$. The easiest choice would be to
define a commutative componentwise product. Via this definition,
the space becomes algebraically isomorphic to the abelian algebra
of functions defined on $N^{2}$ points. This approximation is that
of a lattice.

Nevertheless $\complex^{N^{2}}$ can be seen as the space of
$N\times N$ matrices with complex coefficients. Then one can map
commutative coordinates into noncommutative coordinates, which are
the operators representing the Lie algebra of the group
$SU\left(2\right)$ on each space $\complex^{N}$: \beq
\left[\Lo^{\left(N\right)}_{a},\Lo^{\left(N\right)}_{b}\right]\,=\,i\epsilon_{abc}\Lo^{\left(N\right)}_{c}
\label{su2algebra} \eeq \beq
x^{a}\,\rightarrow\,\tilde{k}\Lo^{\left(N\right)}_{a}\,\equiv\,\hat{x}^{\left(N\right)}_{a}
\eeq The space of truncated functions is mapped into the space
$\mathbb{M}_{N}$: \beq
f^{\left(N\right)}\,\rightarrow\,\fo^{\left(N\right)}\,=\,
\sum_{l=0}^{N-1}\,\frac{1}{l!}\,\sum_{a_{1}\ldots
a_{l}=1}^{3}\,f_{a_{1}\ldots a_{l}}
\hat{x}^{\left(N\right)}_{a_{1}}\cdots\hat{x}^{\left(N\right)}_{a_{l}}
\label{madoremap} \eeq This map is well defined because the
quotienting relation (\ref{spherequotient}) is verified as a
Casimir relation for the group $SU\left(2\right)$. Fixing the
Casimir eigenvalue as the radius of the sphere: \beq
\left[\hat{x}^{\left(N\right)}_{1}\right]^{2}+
\left[\hat{x}^{\left(N\right)}_{2}\right]^{2}+
\left[\hat{x}^{\left(N\right)}_{3}\right]^{2}=1 \eeq fixes the
value of the constant $\tilde{k}$. This means that noncommuting
coordinates satisfy a relation of the form: \beq
\left[\hat{x}^{\left(N\right)}_{a},\hat{x}^{\left(N\right)}_{b}\right]\,=\,\frac{2i\epsilon_{abc}}{\sqrt{N^{2}-1}}
\,\hat{x}^{\left(N\right)}_{c} \label{madoregener} \eeq This
commutation relation says that, in the formal limit
$N\,\rightarrow\,0$, generators of this algebra are seen to
commute. Moreover, on each truncated subalgebra, isomorphic to the
algebra $\mathbb{M}_{N}$, there is a natural action of the group
$SU\left(2\right)$, that is the symmetry group acting on the
manifold $S^{2}$: this would have been impossible in the lattice
approximation.

The map (\ref{madoremap}) can be inverted\footnote{The
invertibility of this map can be proved using the fact that
generators $\Lo^{\left(N\right)}_{a}$ defines an irreducible
representation of the Lie algebra of $SU\left(2\right)$. This
property has also been used in (\ref{irreducibilityexp}).
Invertibility of this map will be clarified even in the following
subsection.}. In the notation of the original paper, this inverse
is represented by $\phi_{N}$, and its range is the set of
truncated (\ref{madoretruncation}) functions
$f^{\left(N\right)}\left(x\right)$, which can be seen as symbols
of the matrices $\fo^{\left(N\right)}$. To consider the limit of
this sequence of algebras for $N\rightarrow\infty$, Madore
stressed that the map $\phi_{N}$ is not an algebra morphism,
because the algebra of matrices is non abelian:
$\phi_{N}\left(\fo^{\left(N\right)}\go^{\left(N\right)}\right)$ is
the symbol of the matrix product $\fo^{\left(N\right)}
\go^{\left(N\right)}$, while
$\phi_{N}\left(\fo^{\left(N\right)}\right)\phi_{N}\left(\go^{\left(N\right)}\right)$
is the pointwise abelian product among symbols
$f^{\left(N\right)}\left(x\right)$ and
$g^{\left(N\right)}\left(x\right)$. The difference: \beq
\phi_{N}\left(\fo^{\left(N\right)}\go^{\left(N\right)}\right)\,-
\phi_{N}\left(\fo^{\left(N\right)}\right)\phi_{N}\left(\go^{\left(N\right)}\right)\,=\,o\left(l/N\right)
\eeq explicitly shows that the nonabelian product among matrices can be
written as a nonabelian product among truncated functions of order
$N$, and that noncommutativity can be estimated to be
 an infinitesimal term of the order $l/N$, where $l$ is the degree of the polynomials representing
$f^{\left(N\right)}\left(x\right)$ and
$g^{\left(N\right)}\left(x\right)$. In particular, it can be seen
that as the order of $l$ approaches $N$, the error involved in
considering $\phi_{N}$ a morphism becomes more and more important.

In the space of matrices a norm is introduced:
\beq
\parallel\fo^{\left(N\right)}\parallel^{2}\,
\equiv\,\frac{1}{N}\,Tr\,\left[\fo^{\left(N\right)\,\dagger}\fo^{\left(N\right)}\right]
\eeq
This norm can be seen as the integral norm on truncated functions on the sphere:
\beq
\parallel\fo^{\left(N\right)}\parallel^{2}\,=\,\frac{1}{4\pi}\,\int_{S^{2}}d\Omega\left|f^{\left(N\right)}\right|^{2}
\eeq It can be formally checked that: \beqa
\lim_{N\rightarrow\infty}\,
\frac{1}{N}\,Tr\,\left[\fo^{\left(N\right)}\right]&=&\frac{1}{4\pi}\,\int_{S^{2}}d\Omega\,f\,\,\,\,
\rightarrow \nonumber \\ \lim_{N\rightarrow\infty}\,\parallel
f^{\left(N\right)}\parallel&=&\parallel f \parallel_{S^{2}} \eeqa
One can estimate, in this approach, that a generic element
$\fo^{\left(N\right)}\,\in\,\mathbb{M}_{N}$ (whose elements do not
depend on $N$) has a norm satisfying: \beq
\lim_{N\rightarrow\infty}\,\parallel
f^{\left(N\right)}\parallel\,=\,\lim_{N\rightarrow\infty}\,\sqrt{N}
\eeq while a diagonal matrix will have a norm converging to
$\parallel f^{\left(N\right)}\parallel\rightarrow o\left(1\right)$
for increasing $N$. This definition of norm forces to consider
only those elements whose limit are diagonal matrices. This choice
is the realization of the specific filtration mentioned in the
introduction. In this perspective one can naively say that the limit of
the fuzzy sphere is a commutative algebra, that ``looks like'' the
algebra of functions on a sphere.

\subsection{The fuzzy sphere in the Weyl-Wigner formalism}

In the previous pages, it has been described how the fuzzy sphere
can be looked at as a peculiar sequence of finite rank matrix algebras. It
has been stressed how the truncation of the algebra of functions
on the sphere can be cast in a matrix form using the properties of
the generators of the Lie algebra of $SU\left(2\right)$. The main
tool to transform a set of functions into a noncommutative algebra
is the mapping (\ref{madoremap}), while to study the behaviour of
that sequence in the large $N$ limit the main role has been played
by its inverse $\phi_{N}$. These can be properly formalised as a
Weyl-Wigner isomorphism \cite{quantforsphere}.

The first step of this analysis is to set up an isomorphism
between a space of operators and a space of functions. Since a
sphere is the coadjoint orbit of the group $SU\left(2\right)$, the
basic tool to introduce this map is a system of coherent states,
specialising the general arguments of appendix \ref{CSappendix}.

To define a system of coherent states
the first problem is the study of unitary irreducible
representations for the group. It is well known which are these
UIRR's. On each finite dimensional Hilbert space $\complex^{N}$,
with $N=2L+1\,\,\,\left(L\,=\,0,1/2,1,3/2\ldots\right)$, where a
basis is given by vectors that, in ket notation, are represented
as $\mid L,M\ra$ with $M\,=\left(\,-L,-L+1\ldots L-1,L\right)$,
one has: \beq
u\,\in\,SU\left(2\right)\,\stackrel{\hat{R}^{\left(L\right)}}{\mapsto}\,
\mathcal{B}\left(\complex^{N}\right) \eeq The matrix elements of
this representation are given by: \beq \la
L,M\mid\hat{R}^{\left(L\right)}\left(u\right)\mid
L,M^{\prime}\ra\,=\,D_{MM^{\prime}}^{L}\left(u\right) \eeq
These\footnote{These Wigner functions are a specific example of
the general definition used in chapter \ref{chapsecond}, for
matrix elements of UIRR of a Lie group (\ref{Delements}).} are
called Wigner functions \cite{QTAngMom}.

The second step is to fix a fiducial state. One can choose the so
called highest weight in the representation:
$\mid\psi_{0}\ra\,=\,\mid L,L\ra$. If the group manifold is
parametrised by Euler angles, then $u$ represents a point whose
``coordinates'' range through
$\,\alpha\in\left[\left.0,4\pi\right)\right.\,$,$\,
\beta\in\left[\left.0,\pi\right)\right.\,$,
$\,\gamma\in\left[\left.0,2\pi\right)\right.\,$.
Fixed the fiducial vector, its stability subgroup $H_{\psi_{0}}$
by the $\hat{R}^{\left(L\right)}$ representation is made by elements for which
$\beta=0$ (this condition can be seen to be valid whatever the
dimension $N$ of the space of representation is).

Two elements $u$ and $u^{\prime}$ are equivalent if
$u^{\dagger}u^{\prime}\in H_{\psi_{0}}$. It is possible to prove
that \beq SU\left(2\right)/H_{\psi_{0}}\approx S^{2} \eeq
identifying $\theta=\beta$ and $\varphi=\alpha/mod\,2\pi$. Chosen
a representative $\tilde{u}$ element in each equivalence class of
the quotient, the set of coherent states is defined as: \beq
\mid\theta,\varphi,N\ra=\hat{R}^{\left(L\right)}\left(\tilde{u}\right)\mid L,L\ra
\eeq
The left hand side ket now explicitly depends on $N$, the
dimension of the space on which the representation takes place.
Projected on the basis elements, one has: \beqa &{}&\la L,M\mid
\theta,\varphi,N\ra=D_{ML}^{L}\left(\tilde{u}\right)\nonumber \\
&{}&\mid\theta,\varphi,N\ra=\sum_{M=-L}^{L}
\left[\frac{\left(2L\right)!}{\left(L+M\right)!\left(L-M\right)!}\right]^{1/2}
\left(\cos\theta/2\right)^{L+M}\left(\sin\theta/2\right)^{L-M}e^{-i\varphi
M}\,\mid L,M\ra\,\,\,\,\,\,\,\,\,\,\,\,\,\,\,\,\,\,\,\,\,\,\, \eeqa
 This set of states is
nonorthogonal, and overcomplete: \beqa
\la\theta^{\prime},\varphi^{\prime},N\mid\theta,\varphi,N\ra &=&
e^{-iL\left(\varphi^{\prime}-\varphi\right)}\,
\left[e^{i\left(\varphi^{\prime}-\varphi\right)}\cos \theta/2\cos
\theta^{\prime}/2 +\sin \theta/2\sin\theta^{\prime}/2\right]^{2L}
\nonumber
\\
\idop&=&\frac{2L+1}{4\pi}\int_{S^{2}}\,d\Omega\mid\theta,\varphi,N\ra\la\theta,\varphi,N\mid
\label{SU2complete}
\eeqa Using this set of vectors it is possible to define a map from
the space of operators on a finite dimensional Hilbert space to
the space of functions on the sphere $S^{2}$: \beqa
\Ao^{\left(N\right)}\,\in\,\mathcal{B}\left(\complex^{N}\right)\,\approx\,
\mathbb{M}_{N}\,\,\,&\mapsto&\,\,\,A^{\left(N\right)}
\,\in\,\mathcal{F}\left(S^{2}\right) \nonumber
\\
A^{\left(N\right)}\left(\theta,\varphi\right)\,&=&
\,\la\theta,\varphi,N\mid\Ao^{\left(N\right)}\mid\theta,\varphi,N\ra
\label{berezinsymsph}\eeqa So this is a way to map every finite rank matrix into a
function on a sphere, called Berezin symbol. Among these
operators, there are $\Yo^{\left(N\right)}_{JM}$ whose symbols are
the spherical harmonics, up to order $2L$ (here
$J=0,1,\ldots,2L$ and $M=-J,\ldots,+J$): \beq
\la\theta,\varphi,N\mid\Yo^{\left(N\right)}_{JM}\mid\theta,\varphi,N\ra\,=
\,Y_{JM}\left(\theta,\varphi\right) \eeq These operators are
called \emph{fuzzy harmonics}. The origin of this name must be
traced back to the definition, on each finite rank matrix algebra
$\mathbb{M}_{N}$, of an operator, in terms of the generators
(\ref{su2algebra}) $\Lo_{a}^{\left(N\right)}$ representing the Lie algebra of the
group $SU\left(2\right)$ on the space $\complex^{N}$: \beqa
\nabla^{2}\,&:&\,\mathbb{M}_{N}\,\mapsto\,\mathbb{M}_{N} \nonumber
\\
\nabla^{2}\Ao^{\left(N\right)}\,&=&\,\left[\Lo_{s}^{\left(N\right)},
\left[\Lo_{s}^{\left(N\right)},\Ao^{\left(N\right)}\right]\right]
\eeqa This operator is called \emph{fuzzy Laplacian}. It can be
seen that the spectrum of this fuzzy Laplacian is given by
eigenvalues $\mathcal{L}_{j}=j\left(j+1\right)$, where
$j\,=\,0,\ldots,2L$, and every eigenvalue has a multiplicity of
$2j+1$. The spectrum of this fuzzy Laplacian thus coincides with the
spectrum of the continuum Laplacian defined in the space of
functions on a sphere, up to order $2L$. The cut-off of this
spectrum is of course related to the dimension of the rank of the
matrix algebra under analysis. Fuzzy harmonics are the eigenstates
of this operator, or so to say, "eigenmatrices" of this fuzzy
Laplacian \cite{ramgoolam}.

Fuzzy harmonics are a basis in each space of matrices $\mathbb{M}_{N}$. They are
trace orthogonal with respect to the scalar product: \beq
Tr\left[\left(\Yo^{\left(N\right)}_{JM}\right)^{\dagger}\,\Yo^{\left(N\right)}_{J^{\prime}M^{\prime}}
\right]\,=\,\alpha_{NJ}\,\delta_{JJ^{\prime}}\delta_{MM^{\prime}}
\eeq
where 
\beq
\alpha_{NJ}\,=\,\left(\frac{J!}{\left(2LJ\right)^{J}}\right)^{2}\left[\frac{\left(2L+J+1\right)!}{4\pi\left(2L-J\right)!}\right]\eeq
In the space of finite rank matrices of
order $N=2L+1$, that is the set of operators $\mathbb{M}_{N}$, it
is possible to introduce a set of $N^{2}$ polarization operators
\cite{QTAngMom} $\To^{\left(N\right)}_{JM}$. They satisfy: \beqa
\left[\Lo_{\mu}^{\left(N\right)},\To^{\left(N\right)}_{JM}\right]\,&=&\,
\sqrt{L\left(L+1\right)}\,C^{L\,M+\mu}_{LM\,1\mu}\,\To^{\left(N\right)}_{LM+\mu}
\nonumber
\\
Tr\left[\To^{\left(N\right)\,\dagger}_{JM}\,\To^{\left(N\right)}_{J^{\prime}M^{\prime}}\right]
&=&\delta_{JJ^{\prime}}\delta_{MM^{\prime}} \nonumber
\\
\To^{\left(N\right)\,\dagger}_{JM}&=&\left(-1\right)^{M}\To^{\left(N\right)}_{J-M}
\eeqa These three conditions completely determine the polarization
operators: fuzzy harmonics are proportional to polarization
operators: \beq
\Yo^{\left(N\right)}_{JM}\,=\,\sqrt{\alpha_{NJ}}\,\To^{\left(N\right)}_{JM}\eeq

 Using this basis, an element $\Fo^{\left(N\right)}$ belonging
to $\mathbb{M}_{N}$ can be expanded as: \beq
\Fo^{\left(N\right)}\,=\,\sum_{J=0}^{2L}\sum_{M=-J}^{J}\,F^{\left(N\right)}_{JM}\,\Yo^{\left(N\right)}_{JM}
\eeq Coefficients of this expansion are given by: \beq
F^{\left(N\right)}_{JM}\,=\,Tr\left[\Yo^{\left(N\right)\dagger}_{JM}\,\Fo^{\left(N\right)}\right]/\lambda_{NJM}
\eeq A Weyl-Wigner map can be defined simply mapping spherical
harmonics into fuzzy harmonics: \beq
\Yo^{\left(N\right)}_{JM}\,\,\Leftrightarrow\,\,Y_{JM}\left(\theta,\varphi\right)
\eeq This map clearly depends on the dimension $N$ of the space on
which fuzzy harmonics are realized. It can be linearly extended
by: \beq \hat{F}^{\left(N\right)}=\sum_{J=0}^{2L}\sum_{M=-J}^{J}\,
F^{\left(N\right)}_{JM}\hat{Y}^{\left(N\right)}_{JM}\,\,\,\,\leftrightarrow\,\,\,\,
F^{\left(N\right)}\left(\theta,\phi\right)\,=\,\sum_{J=0}^{2L}\sum_{M=-J}^{+J}F_{JM}^
{\left(N\right)}Y_{JM}\left(\theta,\phi\right)
\label{WWmapsphere}\eeq

This is a Weyl-Wigner isomorphism. How can this formalization be used to define a fuzzy sphere? Given
a function on a sphere, if it is square integrable with respect to
the standard measure $d\Omega=d\varphi\,\sin\theta d\theta$, then it can be expanded in the basis of
spherical harmonics: \beq f\left(\theta,\varphi\right)\,=\,
\sum_{J=0}^{\infty}\sum_{M=-J}^{J}
\,f_{JM}\,Y_{JM}\left(\theta,\varphi\right) \eeq This expansion
can be truncated: \beq
f^{\left(N\right)}\left(\theta,\varphi\right)\,=\,
\sum_{J=0}^{2L}\sum_{M=-J}^{J}
\,f_{JM}\,Y_{JM}\left(\theta,\varphi\right) \eeq This is a set of
functions whose expansion in spherical harmonics is up to order
$2L=N-1$. It is a vector space, but it is no more an algebra, with
the standard definition of sum and pointwise product of two
functions, as the product of two spherical harmonics of order say
$2L$ has spherical components of order larger\footnote{ The
product of two spherical harmonics is: \beq
\left(Y_{J^{\prime}M^{\prime}}Y_{J^{\prime\prime}M^{\prime\prime}}\right)\left(\theta,\varphi\right)
=
\sum_{J=\left|J^{\prime}-J^{\prime\prime}\right|}^{J=\left|J^{\prime}+J^{\prime\prime}\right|}
\sum_{M=-J}^{J}\sqrt{\frac{\left(2J^{\prime}+1\right)\left(2J^{\prime\prime}+1\right)}{4\pi\left(2J+1\right)}}
\,C^{J0}_{J^{\prime}0\,J^{\prime\prime}0}\,
C^{JM}_{J^{\prime}M^{\prime}\,J^{\prime\prime}M^{\prime\prime}}\,Y_{JM}\left(\theta,\varphi\right)
\eeq in terms of Clebsh-Gordan coefficients for
$SU\left(2\right)$. } than $2L$. If these truncated functions are
mapped, via the Weyl-Wigner procedure, into matrices: \beq
f^{\left(N\right)}\left(\theta,\varphi\right)\,\mapsto\,\hat{f}^{\left(N\right)}
=\sum_{J=0}^{2L}\sum_{M=-J}^{+J}\,f_{JM}\hat{Y}^{\left(N\right)}_{JM}
\label{fsmap}\eeq then this set of truncated functions is given
the vector space structure of $\mathbb{M}_{N}$. Nevertheless
$\mathbb{M}_{N}$ is even an algebra, a non abelian algebra.
Invertibility of this association (\ref{fsmap}) enables to define,
in the set of the truncated functions on the sphere, a non abelian
product, isomorphic to that of matrices\footnote{ The product of
two fuzzy harmonics can be obtained by the product of two
polarization operators \cite{QTAngMom}:
\beq
\To^{\left(N\right)}_{J^{\prime}M^{\prime}}\To^{\left(N\right)}_{J^{\prime\prime}M^{\prime\prime}}\,=\,
\sum_{J=\left|J^{\prime}-J^{\prime\prime}\right|}
^{\left|J^{\prime}+J^{\prime\prime}\right|}\left(-1\right)^{2L+J}\sqrt{\left(2J^{\prime}+1\right)\left(2J^{\prime\prime}+1\right)}
\{
\begin{array}{ccc} J^{\prime} & J^{\prime\prime} & J \\ L & L & L \end{array}
\}
C^{JM}_{J^{\prime}M^{\prime}\,J^{\prime\prime}M^{\prime\prime}}\hat{T}^{\left(N\right)}_{JM}
 \eeq The sum over the index $M$ is made superfluous by the properties
of the Clebsh-Gordan coefficients, fixing $M=M^{\prime}+M^{\prime\prime}$.}: \beq
\left(f^{\left(N\right)}*g^{\left(N\right)}\right)\left(\theta,\varphi\right)\,=\,
\sum_{J=0}^{2L}\sum_{M=-J}^{J}\,Tr\left[\fo^{\left(N\right)}\go^{\left(N\right)}
\Yo_{JM}^{\left(N\right)\,\dagger}\right]\,Y_{JM}\left(\theta,\varphi\right)/\alpha_{NJ}
\eeq The Weyl-Wigner map (\ref{WWmapsphere}) has been used to make
each set of truncated functions a non abelian algebra
$\mathcal{A}^{\left(N\right)}\left(S^{2},*\right)$, isomorphic to
$\mathbb{M}_{N}$. These algebras can be seen as formally generated
by matrices which are the images of the norm $1$ vectors in
$\real^{3}$, that are points on a sphere. They are mapped into
multiples of the generators $\Lo_{a}^{\left(N\right)}$ of the Lie
algebra: \beq \frac{x_{a}}{\parallel
\vec{x}\parallel}\,\mapsto\,\hat{x}_{a}^{\left(N\right)}\,\,\,\,\,\,\,\,
\left[\hat{x}_{a}^{\left(N\right)},\hat{x}_{b}^{\left(N\right)}\right]\,=\,
\frac{2i\varepsilon_{abc}}{\sqrt{N^{2}-1}}
\,\hat{x}_{c}^{\left(N\right)} \eeq This relation perfectly fits
with (\ref{madoregener}): once more, the commutation rules
satisfied by generators of the algebras in the
 sequence
$\mathcal{A}^{\left(N\right)}\left(S^{2},*\right)$ make it
intuitively clear that the limit for $N\,\rightarrow\,\infty$ of
this sequence is an abelian algebra. This is the reason why this
sequence is called \emph{fuzzy sphere}\footnote{A coherent state approach to the study of the fuzzy sphere is in \cite{hammou}.}.

The formal proof of this convergence towards the algebra of
functions on a sphere has been given by Rieffel, in terms of the
so called Quantum Gromov-Hausdorff distance among quantum metric
spaces.

\subsection{An analysis of the convergence of matrix algebras to the sphere}\label{rieffelsection}

 In a series of papers, M.Rieffel studied the problem
of giving a precise meaning to the notion of the convergence of
the fuzzy sphere to the classical sphere, that is the convergence
of that sequence of finite rank matrix algebras to the algebra of
functions on a sphere. The aim of this section is to report some
aspects of this analysis. Specifically, it will be sketched how
these algebras can be considered as elements of a peculiar metric
space, where the convergence to the algebra of functions on the
sphere can be formalized.

In noncommutative geometry, the natural way to specify a metric is
by means of a suitable "Lipschitz seminorm". This idea was
developed by Connes \cite{connesbyrieffel}. He pointed out that
from a Lipschitz seminorm one obtains in a simple way an ordinary
metric on the state space of a $C^{*}$-algebra.

Given a compact metric space $\left(Z,\rho\right)$ ($\rho$ is an
ordinary metric), a Lipschitz seminorm is defined for functions on
$Z$ ($x,y$ are points of $Z$): \beq
L_{\rho}\left(f\right)\,\equiv\,
\sup\,\{\left|f\left(x\right)-f\left(y\right)\right|/\rho\left(x,y\right)\,:\,x\,\neq\,y\}
\eeq This is actually a seminorm as it is $0$ for constant
functions, and can take the value $+\infty$, unless the domain is
restricted to Lipschitz functions. From $L_{\rho}$ the metric
$\rho$ can be obtained: \beq
\rho\left(x,y\right)\,=\,\sup\,\{\left|f\left(x\right)-f\left(y\right)\right|\,:\,L_{\rho}\left(f\right)\,\leq\,
1\}\eeq This metric on pure states can be extended to the set of all
states (probability measures on $Z$): \beq
\rho\left(\mu,\nu\right)\,\equiv\,\sup\,\{\left|\mu\left(f\right)-\nu\left(f\right)\right|
\,:\,L_{\rho}\left(f\right)\,\leq\,1\} \eeq This metric induces,
in the set of states $\mathcal{S}\left(Z\right)$ of the algebra of
functions on $Z$, a topology that coincides with the weak
$*$-topology that $\mathcal{S}\left(Z\right)$ would have if it were
considered as the dual of the $C^{*}$-algebra $C\left(Z\right)$ of
continuous functions on $Z$\footnote{ If $\mathcal{A}^{*}$ is the
space of states of a suitable algebra $\mathcal{A}$, i.e. the
Banach space of norm $1$ continuous linear functionals
$\phi:\mathcal{A}\mapsto\complex$ then the weak $*$-topology is
given as the pointwise convergence on elements on $\mathcal{A}$.}.

These results are used to extend the notion of metric to states on
noncommutative algebras. If $\mathcal{A}$ is a unital
$C^{*}$-algebra, and $L$ is a Lipschitz seminorm on this algebra,
satisfying $L\left(a\right)=L\left(a^{\dagger}\right)$, then, on
the set of states $\mathcal{S}\left(\mathcal{A}\right)$, a metric
is defined by ($a$ is an element of the algebra, $\mu,\nu$ are
states of the algebra): \beq
\rho\left(\mu,\nu\right)\,\equiv\,\sup\,\{\left|\mu\left(a\right)-\nu\left(a\right)\right|
\,:\,L_{\rho}\left(a\right)\,\leq\,1\} \label{metricfromlipsm}\eeq
If this metric induces on $\mathcal{S}\left(\mathcal{A}\right)$
the weak $*$-topology, then $\left(\mathcal{A},L\right)$ is a
\emph{compact quantum metric space}. The word ``quantum'' is used
to stress that its origin lies in noncommutative geometry.

The notion of Gromov-Hausdorff distance for compact quantum metric
spaces is an evolution of the "classical" Hausdorff distance for compact metric spaces.

If $\left(Z,\rho\right)$ is again a compact metric space, and $X$
is a closed subset $X\,\subset\,Z$, an $r$-neighborood of $X$ of radius
$r$ is given by all the points of $Z$ whose "distance" from $X$ is
less than $r$: \beq
\mathcal{N}^{\rho}_{r}\left(X\right)\,\equiv\,\{z\,\in\,Z\,:\,\exists\,x\,\in\,X\,:\,\rho\left(x,z\right)\,<\,r\}
\eeq If $X\,\subset\,Z$ and $Y\,\subset\,Z$ are two such subsets,
the Hausdorff distance between them is: \beq
d_{\mathcal{H}}^{\rho}\left(X,Y\right)\,\equiv\,\inf\,\{r\,:\,X\,\subseteq\,\mathcal{N}^{\rho}_{r}\left(Y\right)\,
;\,Y\,\subseteq\,\mathcal{N}^{\rho}_{r}\left(X\right)\}
\label{hausdistclassic}\eeq

Now it is possible to consider $\left(X,\rho_{X}\right)$ and
$\left(Y,\rho_{Y}\right)$, a pair of independent compact metric
spaces. $\mathcal{U}\,\equiv\,X\,\underline{\cup}\,Y$ is the
disjoint union of the two.
$\mathcal{M}\left(\rho_{X},\rho_{Y}\right)$ is the set of metrics
on $\mathcal{U}$ such that $\rho_{X}$ is the quotient on $X$ and
$\rho_{Y}$ is the quotient on $Y$. The Gromov-Hausdorff distance
between $X$ and $Y$ is: \beq
d_{\mathcal{G}\mathcal{H}}\left(X,Y\right)\,\equiv\,\inf\,\{d_{\mathcal{H}}^{\rho}\left(X,Y\right)\,:
\,\rho\,\in\,\mathcal{M}\left(\rho_{X},\rho_{Y}\right)\}\eeq

This reasoning is extended to compact quantum metric spaces.

Given a pair of compact quantum metric spaces
$\left(\mathbb{A},L_{\mathbb{A}}\right)$ and
$\left(\mathbb{B},L_{\mathbb{B}}\right)$ (here $\mathbb{A}$ and
$\mathbb{B}$ can be considered as operator algebras, while
$L_{\mathbb{A}}$ and $L_{\mathbb{B}}$ are Lipschitz seminorm), one
can consider the "disjoint" union:
$$\left(\mathbb{A}\oplus\mathbb{B},\mathcal{M}\left(L_{\mathbb{A}},L_{\mathbb{B}}\right)\right)$$
where $\mathcal{M}\left(L_{\mathbb{A}},L_{\mathbb{B}}\right)$ is
the set of Lipschitz seminorm whose quotient is $L_{\mathbb{A}}$
on $\mathbb{A}$ and $L_{\mathbb{B}}$ on $\mathbb{B}$. On the set
of states $\mathcal{S}\left(\mathbb{A}\oplus\mathbb{B}\right)$ one
can induce a metric $\rho_{L}$ from a Lipschitz seminorm $L$ in
$\mathcal{M}\left(L_{\mathbb{A}},L_{\mathbb{B}}\right)$ via
(\ref{metricfromlipsm}), and then has an Hausdorff distance
\ref{hausdistclassic} between the states of $\mathbb{A}$, $\mathcal{S}\left(\mathbb{A}\right)$,
and the states of $\mathbb{B}$,  $\mathcal{S}\left(\mathbb{B}\right)$, seen as subsets of
$\mathcal{S}\left(\mathbb{A}\oplus\mathbb{B}\right)$. The
Gromov-Hausdorff distance between $\mathbb{A}$ and $\mathbb{B}$
is: \beq
d_{\mathcal{G}\mathcal{H}}\left(\mathbb{A},\mathbb{B}\right)\,\equiv\,\inf\,
\{d_{\mathcal{H}}^{\rho_{L}}\left(\mathcal{S}\left(\mathbb{A}\right),\mathcal{S}\left(\mathbb{B}\right)\right)\,:
\,L\,\in\,\mathcal{M}\left(L_{\mathbb{A}},L_{\mathbb{B}}\right)\}\eeq

The notion of Gromov-Hausdorff distance is then well adapted to
study problems of convergence is a space of suitable algebras. The
next step is to express the algebra of functions on a sphere in
this context.

Let $G$ be a compact Lie group, and $U\left(g\right)$ a unitary
irreducible representation on a finite dimensional Hilbert space
$\complex^{N}$. The set
$\mathcal{B}\left(\complex^{N}\right)\,=\,\mathcal{B}^{\left(N\right)}\,=\,\mathbb{M}_{N}$
of bounded operators on this space, which are finite rank
matrices, is a unital $C^{*}$-algebra. The group $G$ acts on this
space of operators. This action $\alpha$ is a conjugation, and maps the
space $\mathcal{B}\left(\complex^{N}\right)$ into itself: \beq
\alpha\left(g\right)\cdot
T\,\equiv\,U\left(g\right)\,\cdot\,T\,\cdot\,U\left(g^{-1}\right)
\eeq This action is ``ergodic'', in the sense that the only
invariant element is a multiple of the identity $T=c\idop$,
because the representation $U$ is irreducible.

It can be introduced a function on elements of $G$, measuring a ``lenght'' from the identity:
\begin{itemize}
\item $l\left(g\right)\,\geq\,0$. Moreover, $l\left(g\right)\,=\,0$ iff $g\,=\,e$
\item $l\left(g\right)\,=\,l\left(g^{-1}\right)$
\item $l\left(g\,g^{\prime}\right)\,\leq\,l\left(g\right)\,+\,l\left(g^{\prime}\right)$
\item $l\left(g\right)\,=\,l\left(g^{\prime}\,g\,g^{\prime\,-1}\right)$
\end{itemize}
On the space of matrices $\mathcal{B}^{\left(N\right)}$ a Lipschitz seminorm can be defined:
\beq
L_{\mathcal{B}}\left(T\right)\,\equiv\,
\sup\{\parallel\alpha\left(g\right)\cdot T\,-\,T\parallel/l\left(g\right)\,:\,l\left(g\right)
\,\neq\,0\} \label{lipBseminorm}
\eeq
The space $\left(\mathcal{B}\left(\complex^{N}\right),L_{\mathcal{B}}\right)$
is a compact quantum metric space.

In the set of operators $\mathcal{B}\left(\complex^{N}\right)$ one
can choose a rank one projector $P$, and use it to define the
Berezin covariant symbol of an operator $T$: \beqa
\sigma_{T}\left(g\right)&=&Tr\left[T\left(\alpha\left(g\right)\cdot
P\right)\right] \nonumber
\\
&=&Tr\left[TU\left(g\right)PU^{\dagger}\left(g\right)\right] \eeqa
It is evident that, if $H$ is the isotropy subgroup of $P$ for the
action of $\alpha$, then the symbol $\sigma_{T}\left(g\right)$ is
clearly a function on the quotient space $G/H$, whose points are
represented by $x$. The algebra of functions on this quotient is
$\mathcal{A}\left(G/H\right)$. This symbol is related to the
symbol introduced in (\ref{berezinsymsph}). If $P$ is considered
in the form $\mid\psi_{0}\ra\la\psi_{0}\mid$, then
$U\left(g\right)\mid\psi_{0}\ra$ is a system of coherent states
for the group $G$, and the ciclicity of the trace proves that the
symbol $\sigma_{T}\left(x\right)$ can be written as the mean value
of $T$ on the coherent states labelled by $x$. The identification
of this symbol with the one defined in (\ref{berezinsymsph}) for
the sphere is obtained by the requirement that this symbol were
zero only in correspondence to the matrix $T=0$ (usually this
condition is referred to as a faithfulness condition on the
mapping $\sigma_{T}$). This happens only if the compact group $G$
is semisimple, and the projector $P$ ranges over the highest
weight vector in the Cartan analysis\footnote{This condition means
that, if $\mid\psi_{0}\ra$ is the highest weight vector for the
considered representation $U\left(g\right)$, then
$\tilde{U}\left(E_{+}\right)\mid\psi_{0}\ra=0$, where $\tilde{U}$
is the representation of the Lie algebra $\underline{G}$ induced
by $U$, and $E_{+}$ are the positive Cartan's roots in
$\underline{G}$.}.

There is a natural left action of the group $G$ on this algebra:
\beq
\left(\lambda_{g}f\right)\left(x\right)\,\equiv\,f\left(g^{-1}\cdot
x\right) \eeq where $g\cdot x$ represents the left multiplication
action of the element $g\in G$ on the equivalence class in $G/H$
whose label is $x$. Also the lenght function can be quotiented to a
lenght function $\tilde{l}$ on $G/H$\footnote{ In the case of
compact semi-simple Lie group an admissible lenght function is
given by the geodesics lenght with respect to the Cartan-Killing
metric (\ref{geodesics}).}. Equipped with the Lipschitz seminorm: \beq
L_{\mathcal{A}}\left(f\right)\,\equiv\,\sup\,
\{\left|f\left(x\right)-
\left(\lambda_{g}f\right)\left(x\right)\right|/\tilde{l}\left(x\right)\,:\,x\,\neq\,e\}
\eeq one has that $\left(\mathcal{A},L_{\mathcal{A}}\right)$ is a
compact quantum metric space.

The fuzzy sphere is a sequence of finite rank matrix algebras,
obtained as the set of operators on \emph{each} finite dimensional
Hilbert space $\complex^{N}$, because a unitary irreducible
representation of the group $SU\left(2\right)$ is allowed on such
space for each $N$. Moreover, the symbol (\ref{berezinsymsph}) is
introduced using a set of coherent states whose defining fiducial
projector ranges over the highest weight vector. This means that
the fuzzy sphere can be seen as a sequence of compact quantum
metric spaces, with Lipschitz seminorm
$L_{\mathcal{B}^{\left(N\right)}}$.

It is now possible to consider the ``distance'' between $\mathcal{B}^{\left(N\right)}$ and $\mathcal{A}$ in the space
$\left(\mathcal{B}^{\left(N\right)}\oplus\mathcal{A},\mathcal{M}\left(L_{\mathcal{B}^{\left(N\right)}},L_{\mathcal{A}}\right)\right)$.
The main result is that the sequence of Gromov-Hausdorff distances:
\beq
\lim_{N\rightarrow\infty}\,d_{\mathcal{G}\mathcal{H}}^{\left(N\right)}
\left(\mathcal{B}^{\left(N\right)},\mathcal{A}\right)\,=\,0
\eeq
 This is the meaning of the sentence that the fuzzy sphere converges to the algebra of functions on a sphere.

\section{A fuzzy disc}

In the previous section, the fuzzy sphere has been introduced via
a kind of Weyl-Wigner formalism, which made great use of the fact
that the 2-dimensional sphere is the coadjoint orbit of the group
$SU\left(2\right)$, that is compact. Compactness of the group
brought to a natural definition of a sequence of finite
dimensional Hilbert spaces, that is to a natural identification of
the dimension of these spaces as a cut-off index for a suitable
expansion of a generic function on the sphere. Moreover,
compactness of the group played a fundamental role in the analysis
of the convergence performed by M.Rieffel.  Properties of the
sphere as an orbit of that group made it possible to formalize
every point of the sphere as a coherent state, defined on every of
those finite dimensional space carrying the UIRR's. So the map
between operators (finite rank matrices) and functions on the
sphere has been introduced via the natural Berezin procedure, and
has been refined stressing the role of the fuzzy Laplacian
operator in the definition of a fuzzy harmonics system as a basis
in the set $\mathbb{M}_{N}$.

It would be intuitively natural, to draw a path towards the
definition of a fuzzy disc, to analyse the possibility that a disc
were a coadjoint orbit for a Lie group. If this were the case, it would be possible to introduce a
system of coherent states labelled by its points, and some sort of
Berezin map \cite{berezinmap} between operators and a suitable set of functions on
the disc.

This, in some sense naive, approach meets some troubles. There is
a group, from which it is possible to define a system of coherent
states in correspondence with points of a disc. This group is
$SU\left(1,1\right)$, but it is non compact, so its UIRR's are not
realized on finite dimensional Hilbert spaces. There is no more an
intrinsic concept of a cut-off index, the dimension of the
fuzzyfication, in this context.

To pursue the task of defining a fuzzy version of the algebra
of functions on a disc, it is possible to follow, and extend, the
second line of the path sketched in sections above. This will circumvent the
problem of an intrinsic definition of fuzzyfication dimension. It
is well known that a basis for the space of functions on a disc is
related to a suitable system of Bessel functions. As the
Weyl-Wigner map for functions on a sphere has been introduced
(\ref{WWmapsphere}) mapping spherical harmonics into fuzzy
harmonics, is it possible to define a system of \emph{fuzzy
Bessels} in terms of finite rank matrices, and mapping again
continuum Bessels into them?

Since Bessel functions are defined on a plane, introduced to solve
a class of boundary values problems for a Laplacian operator, the
first step of this analysis will be the introduction of a specific
noncommutative plane. In this algebra it will be then introduced a
system of "fuzzy" derivations, and a "fuzzy" Laplacian, mimicing,
as much as possible, the properties of the continuum version. The
final answer to the problem will be given studying the spectral
resolution of this fuzzy Laplacian.

\subsection{A noncommutative plane as a matrix algebra}

In section \ref{WeylmapplaneCS} it has been shown how the Weyl
map defined in the first chapter can be obtained in terms of a set
of quantizer operators arising in the study of a specific unitary
representation of the Heisenberg-Weyl-Wigner group. In that
section a system of generalised coherent states for this group has
been introduced, and it has been seen that such coherent states
are in correspondence with points of a complex plane. But they
have been used just to define an action of the HWW group on the
plane, and to select which elements of the group act as a
reflections on the plane.

In this section a noncommutative plane will be defined. A
Weyl-Wigner map will be introduced following the general procedure
of Berezin: so the first step will be the definition of a set of
generalised coherent states for the Heisenberg-Weyl group, since
they are labelled by points of a plane.

An analysis of the Heisenberg-Weyl group has already been
performed in section \ref{WeylmapplaneCS}. Since in this chapter
the identification of the two dimensional plane with the phase
space carrying a classical dynamics of a one dimensional point particle will be
definitely abandoned, it will be assumed a system of coordinates
of the kind $\left(x,y\right)$ for a point of $\real^{2}$;
$\theta$ will be a parameter introducing an explicit
noncommutativity.

With this notation, Heisenberg-Weyl group is a manifold
$\real^{3}$, whose points are represented by a triple
$\left(x,y,\lambda\right)$, with the composition rule: \beq
\left(x,y,\lambda\right)\cdot
\left(x^{\prime},y^{\prime},\lambda^{\prime}\right)=\left(x+x^{\prime},y+y^{\prime},\lambda+\lambda^{\prime}-
\frac{1}{\theta}\left(xy^{\prime}-x^{\prime}y\right)\right) \eeq
The identity of this group is given by: \beq
id_{W}=\left(0,0,0\right) \eeq and the inverse of a generic
element is: \beq
\left(x,y,\lambda\right)^{-1}=\left(-x,-y,-\lambda\right) \eeq
Complexifying the plane via $z=x+iy$ and $\bz=x-iy$ the group law
acquires the form: \beq
\left(z,\lambda\right)\cdot\left(z^{\prime},\lambda^{\prime}\right)=
\left(z+z^{\prime},\lambda+\lambda^{\prime}+
\frac{i}{2\theta}\left(\bz z^{\prime}-\bz^{\prime}z\right)\right)
\eeq

The Hilbert space on which representing this group is again the
Fock space $\mathcal{F}$ (\ref{fockspace}) of finite norm complex
analytical functions in the $w$ variable where the norm is obtained
by the scalar product: \beq \langle f\mid
g\rangle\,\equiv\,\int\,\frac{d^{2}w}{\pi\theta}\,e^{-\baw w
/\theta}\,\bar{f}\left(w\right)g\left(w\right) \eeq and the
orthonormal basis chosen is given by:
 \beq \psi_{n}\left(w\right)\,=\,\frac{w^{n}}{\sqrt{\theta^{n}n!}}
\eeq The unitary representation of the Heisenberg-Weyl group is given by operators
$\To\left(z,\lambda\right)\,\in\,Op\left(\mathcal{F}\right)$: \beq
\left(\To f\right)\left(w\right)= e^{i\lambda}e^{-\baz z
/2\theta}e^{z w/\theta} f\left(w-\baz\right) \label{unirephw} \eeq

Chosen the first basis element $\psi_{0}\left(w\right)$ as
fiducial state, the procedure already outlined gives a set
of coherent states perfectly coincident with the system obtained
for the more general Heisenberg-Weyl-Wigner group (\ref{cshww}).
In the realization of the Fock space as complex analytical
functions, a coherent state is then: \beq \mid
z\rangle\,\,\,\rightarrow\,\,\,\psi_{\left(z\right)}\left(w\right)=
e^{-\baz z/2\theta}e^{zw/\theta} \eeq and, with $\psi_{n}$ an
element of the basis already considered: \beq \mid z\ra
\,=\,\sum_{n=0}^{\infty} e^{-\baz
z/2\theta}\frac{z^{n}}{\sqrt{n!\theta^{n}}}\mid\psi_{n}\ra \eeq
Such coherent states are non orthogonal, and overcomplete: \beqa
\la z\mid
z^{\prime}\ra&=&e^{-\left(\left|z\right|^{2}+\left|z^{\prime}\right|^{2}-2\baz
z^{\prime}\right)/\theta} \nonumber \\
\idop&=&\int\frac{d^{2}z}{\pi\theta}\,\mid z \rangle\langle
z\mid\label{HWcomplete} \eeqa

On this Hilbert space $\mathcal{F}$, it is possible to introduce a
pair of creation-annihilation operators: \beqa
\left(\hat{a}f\right)\left(w\right)&=&\theta\frac{df}{dw} \nonumber
\\
\left(\hat{a}^{\dagger}f\right)\left(w\right)&=&wf\left(w\right)
\eeqa such that
\be
\left[\hat{a},\hat{a}^{\dagger}\right]=\theta\idop \ee In the chosen
orthonormal basis, one has: \beqa \hat{a}\mid
\psi_{n}\rangle&=&\sqrt{n\theta}\mid \psi_{n-1}\rangle \nonumber
\\
\hat{a}^{\dagger}\mid\psi_{n}\rangle&=&\sqrt{\left(n+1\right)\theta}\mid\psi_{n+1}\rangle
\eeqa  and \beqa \langle z\mid \hat{a}\mid z\rangle =z \nonumber
\\
\langle z\mid \hat{a}^{\dagger}\mid z\rangle=\bar{z} \eeqa

These relations can be extended. A Berezin symbol can be
associated to an operator in the Fock space:
\be
f\left(\baz,z\right)\,=\,\langle z\mid \fo\mid z\rangle \ee This
can be seen as a Wigner map. It can be inverted:
\be
\fo\,=\,\int\,\frac{d^{2}\xi}{\pi\theta}\,\int\,\frac{d^{2}z}{\pi\theta}\,f\left(z,\bar{z}\right)\,
e^{-\left(\bar{z}\xi-\bar{\xi}z\right)/\theta}\,e^{\xi\hat{a}^{\dagger}/\theta}\,e^{-\bar{\xi}\hat{a}/\theta}
\label{discweylmap}\ee

This quantization map for functions on a plane can be given an
interesting form. A first analysis can be restricted to functions
which can be written as Taylor series in $\baz,z$: \beq
f\left(\baz,z\right)\,=\,\sum_{m,n=0}^{\infty}\,f_{mn}^{Tay}\baz^{m}z^{n}
\eeq An easy calculation says that this $f$ is the symbol of the
operator: \beq
\fo\,=\,\sum_{m,n=0}^{\infty}\,f_{mn}^{Tay}\ao^{\dagger m}\ao^{n}
\label{taylorexp}\eeq The second analysis starts from an operator
written in a density matrix notation: \beq
\fo\,=\,\sum_{m,n=0}^{\infty}\,f_{mn}\mid\psi_{m}\ra\la\psi_{n}\mid
\eeq The Berezin symbol of this operator is the function:
\beq f\left(\baz,z\right)\,=\,e^{-\left|z\right|^{2}/\theta}
\sum_{m,n=0}^{\infty}\,f_{mn}\frac{\baz^{m}z^{n}}{\sqrt{m!n!\theta^{m+n}}}\label{dmsymbol}
\eeq
Given the Taylor coefficients $f_{mn}^{Tay}$, one has: \beq
f_{lk}\,=\,\sum_{q=0}^{\min\left(l,k\right)}\,f_{l-q\,k-q}^{Tay}\,
\frac{\sqrt{k!l!\theta^{l+k}}}{q!\theta^{q}} \eeq while the
inverse relation is given by: \beq
f_{mn}^{Tay}\,=\,\sum_{p=0}^{\min\left(m,n\right)}\,
\frac{\left(-1\right)^{p}}{p!\sqrt{\left(m-p\right)!\left(n-p\right)!\theta^{m+n}}}\,f_{m-p\,n-p}\eeq

 Equation (\ref{taylorexp}) shows that the quantization of a
monomial in the variables $z,\baz$ is an operator in $\ao,\aod$,
formally a monomial in these two noncommuting variables, with all
terms in $\aod$ acting at the left side with respect to terms in
$\ao$. This means that this quantization brings a specific
ordering. In section (\ref{weightedwmap}) it has been analysed how
ordering is usually related to a weight in the Weyl map that
defines the quantization. The Weyl map (\ref{discweylmap}) can be
written, restoring real variables, as (here
$u=\left(a+ib\right)/2$): \beq
\hat{f}\,=\,\int\,\frac{dadb}{2\pi\theta}\,\int\,\frac{dxdy}{2\pi\theta}\,f\left(x,y\right)\,
e^{-i\left(bx-ay\right)/\theta}\,e^{u\hat{a}^{\dagger}/\theta}\,e^{-\bar{u}\hat{a}/\theta}
\eeq This expression is equal to: \beq
\fo\,=\,\int\,\frac{dadb}{2\pi\theta}\,\tilde{f}\left(b,a\right)\,
e^{\left(a^{2}+b^{2}\right)/8\theta}\,\Do\left(a/\sqrt{2},b/\sqrt{2}\right)
\label{weightdisc}\eeq The presence of the factor $1/\sqrt{2}$ in the argument of
the Displacement operator follows from the specific
complexification of the plane via $z=x+iy$. It can be compared to
equation(\ref{weighwm}). The factor
$e^{\left(a^{2}+b^{2}\right)/8\theta}$ is a weight to the standard
Weyl map, so the quantization via Berezin procedure is just an
example of a weighted quantization.

The invertibility of Weyl map (on a suitable domain of functions
on the plane) enables to define a noncommutative product in the
space of functions. It is known as Voros product: \beq
\left(f*g\right)\left(\baz,z\right)\,=\,\la z\mid\fo\go\mid z\ra
\label{vorospr}\eeq It is a non local product: \beq
\left(f*g\right)\left(\baz,z\right)\,=\, e^{-\baz
z/\theta}\int\,\frac{d^{2}\xi}{\pi\theta}\,f\left(\baz,\xi\right)g\left(\bar{\xi},z\right)
\,e^{-\bar{\xi}\xi/\theta}e^{\bar{\xi}z/\theta}e^{\baz\xi/\theta}
\eeq Its asymptotic expansion acquires the form: \beq
\left(f*g\right)\left(\baz,z\right)\,=\,f\,e^{\theta\overleftarrow{\del}_{\baz}\overrightarrow{\del}_{z}}\,g
\eeq and makes it clear that it is a deformation, in $\theta$, of
the pointwise commutative product. Since it is the translation, in
the space of functions, of the product in the space of operators,
given  symbols  expressed in the form (\ref{dmsymbol}), the
product acquires a matrix form: \beq
\left(f*g\right)_{mn}\,=\,\sum_{k=0}^{\infty}f_{mk}g_{kn} \label{matrixprod}\eeq

The space of functions on the plane, with the standard definition of sum, and the product given by
the Voros product (\ref{vorospr}), is a nonabelian algebra, a noncommutative plane.
This algebra $\mathcal{A}_{\theta}=\left(\mathcal{F}\left(\real^{2}\right),*\right)$ is isomorphic to an
algebra of operators, or, as equation (\ref{matrixprod}) suggests,
to an algebra of infinite dimensional matrices.

\subsection{A sequence of non abelian algebras}

A fuzzy space has been presented as a sequence of finite rank matrix algebras
converging, as compact quantum metric spaces, to an algebra of functions. In the case of the fuzzy
sphere the rank of the matrices involved is the dimension
of the Hilbert space on which UIRR's of the group $SU\left(2\right)$ are realised. In the approach sketched in
the last section, based on a definition of a noncommutative plane, there is no natural, say intrinsic,
definition of a set of finite dimensional matrix algebras. This, following the general comment already
expressed, can be thought of as a consequence of the fact that also this noncommutative plane has been
realised via a Berezin quantization based on coherent states originated by a group, the Heisenberg-Weyl, which
is noncompact.

In this context the strategy to obtain finite dimensional matrix algebras is different.
$\mathcal{A}_{\theta}$ can be considered, once a basis in the Hilbert space
has been chosen, as a matrix algebra, formally made up of,
infinite dimensional matrices. One can define a set of finite dimensional matrix algebras simply truncating
$\mathcal{A}_{\theta}$.
The notion of truncation is formalised via the introduction of a set of projectors in the space of operators.
Their symbols are projectors in the algebra $\mathcal{A}_{\theta}$ of the noncommutative plane, in the sense that they are
idempotent functions of order $2$ with respect to the Voros product (here $z=re^{i\varphi}$):
\beqa
P_{\theta}^{\left(N\right)}\left(r,\varphi\right)&=&
\sum_{n=0}^{N}\la z\mid\psi_{n}\ra\la\psi_{n}\mid z\ra\,=\,e^{-r^{2}/\theta}
\sum_{n=0}^{N}\,\frac{r^{2n}}{n!\theta^{n}}
\nonumber \\
P_{\theta}^{\left(N\right)}*P_{\theta}^{\left(N\right)}&=&P_{\theta}^{\left(N\right)}
\eeqa
This finite sum can be performed yielding a rotationally symmetric function:
\beq
P_{\theta}^{\left(N\right)}\left(r,\varphi\right)\,=
\,\frac{\Gamma\left(N+1,r^{2}/\theta\right)}{\Gamma\left(N+1\right)}
\eeq
in terms of the ratio of an incomplete gamma function by a gamma function \cite{prudnikov}.
If $\theta$ is kept fixed, and nonzero, in the limit for $N\,\rightarrow\,\infty$ the symbol
$P_{\theta}^{\left(N\right)}\left(r,\varphi\right)$ converges, pointwise, to the constant function
$P_{\theta}^{\left(N\right)}\left(r,\varphi\right)=1$, which can be formally considered as the symbol of
the identity operator: in this limit one recovers the "whole" noncommutative plane.

This situation changes if the limit for $N\,\rightarrow\,\infty$
is performed keeping the product $N\theta$ equals to a constant,
say $R^{2}$. In a pointwise convergence, chosen $R^{2}=1$: \beq
P^{\left(N\right)}_{\theta}\,\,\rightarrow\,\,\,\left[\begin{array}{cc}
1 & r<1 \\ 1/2 & r=1 \\ 0 & r> 1
\end{array}\right]\,=\,Id\left(r\right) \eeq This sequence of
projectors converges to a sort of step function in the radial
coordinate $r$, or a characteristic function of a disc on the
plane. The shape of this function is plotted in
figure~\ref{disc3d}.
\begin{figure}[htbp]
\epsfxsize=2.5 in
\bigskip
\centerline{\epsffile{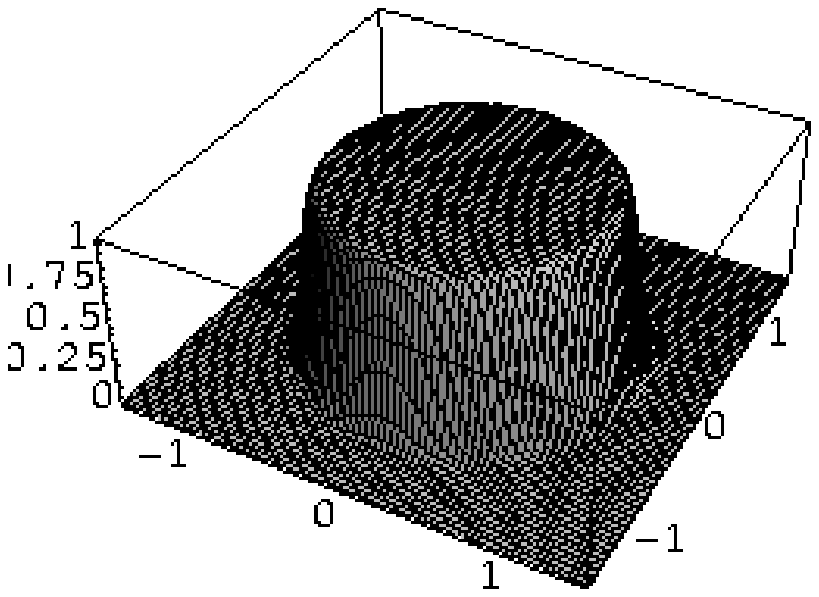}} \caption{\baselineskip=12pt
{\it The function $P^{\left(N\right)}_\theta$ for $N=10^2$.}}
\bigskip
\label{disc3d}
\end{figure}
Already for $N= 10^3$ it is well approximated (see
figure~\ref{identity}) by a step function.
\begin{figure}[htbp]
\epsfxsize=2.5 in
\bigskip
\centerline{\epsffile{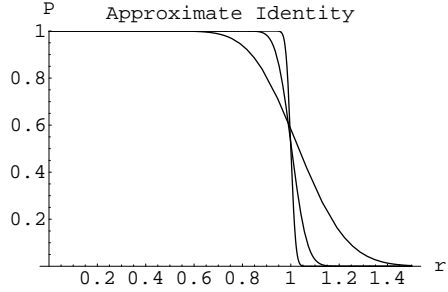}} \caption{\baselineskip=12pt
{\it Profile of the spherically symmetric function $P^N_\theta$ for the
choice $R^2=N\theta=1$ and $N=10, 10^2, 10^3$. As $N$ increases the step
becomes sharper.}}
\bigskip
\label{identity}
\end{figure}
Thus a sequence of subalgebras
$\mathcal{A}_{\theta}^{\left(N\right)}$ can be defined by: \beq
\mathcal{A}_{\theta}^{\left(N\right)}\,=\,P_{\theta}^{\left(N\right)}*\mathcal{A}_{\theta}
*P_{\theta}^{\left(N\right)} \eeq As it has been said, the full
algebra $\mathcal{A}_{\theta}$ is isomorphic to an algebra of
operators. What the previous relation says is that
$\mathcal{A}_{\theta}^{\left(N\right)}$ is isomorphic to
$\mathbb{M}_{N+1}$, the algebra of
$\left(N+1\right)\times\left(N+1\right)$ rank matrices: the
important thing is that this isomorphism (this truncation) is
obtained via a specific choice of a basis in the Fock space
$\mathcal{F}$ on which coherent states for the Heisenberg-Weyl
group are realised. The explicit effect of this projection on a
generic function is: \beq
\Pi^{\left(N\right)}_{\theta}\left(f\right)\,=\,f^{\left(N\right)}_{\theta}\,=\,P^{\left(N\right)}
_{\theta}*f*P^{\left(N\right)}_{\theta}\,=\,
e^{-\left|z\right|^{2}/\theta}\sum_{m,n=0}^{N}f_{mn}\frac{\baz^{m}z^{n}}{\sqrt{m!n!\theta^{m+n}}}
\eeq On every subalgebra $\mathcal{A}^{\left(N\right)}_{\theta}$,
the symbol $P^{\left(N\right)}_{\theta}\left(r,\varphi\right)$ is
then an identity, because it is the symbol of the projector
$\Po^{\left(N\right)}=\sum_{n=0}^{N}\mid\psi_{n}\ra\la\psi_{n}\mid$,
which is the identity operator in
$\mathcal{A}_{\theta}^{\left(N\right)}$, or, equivalently, the
identity matrix in every $\mathbb{M}_{N+1}$.

It is important to note that the rotation group on the plane,
$SO\left(2\right)$, acts in a natural way on these subalgebras.
Its generator is the number operator
$\No\,=\,\sum_{n=0}^{N}\,n\theta\mid\psi_{n}\ra\la\psi_{n}\mid$,
that is diagonal in each $\mathcal{A}_{\theta}^{\left(N\right)}$.

Cutting at a finite $N$ the expansion provides an infrared cutoff.
This cutoff is "fuzzy" in the sense that functions in the
subalgebra are still defined outside the disc of radius $R$, but
are exponentially damped. To analyse the nature of this cutoff, it
is interesting to study how this truncation works in a test case
of a gaussian function, cylindrically symmetric and centred at the
origin of the plane: \beq
\Phi\left(r\right)\,=\,\frac{1}{\pi\alpha}\,e^{-r^{2}/\alpha} \eeq
The width of this gaussian is proportional to the parameter
$\alpha$. This function can be expanded in Taylor series: \beq
\Phi\left(r\right)\,=\,\frac{1}{\pi\alpha}\,e^{-r^{2}/\alpha}\,=\,
\frac{1}{\pi\alpha}\,\sum_{s=0}^{\infty}\left(-\frac{1}{\alpha}\right)^{s}\frac{1}{s!}\,\baz^{s}z^{s}
\eeq Formulas (\ref{discweylmap}) and (\ref{taylorexp}) show that
this function is mapped into the operator: \beq \hat{\Phi}\,=\,
\frac{1}{\pi\alpha}\,\sum_{s=0}^{\infty}\left(-\frac{1}{\alpha}\right)^{s}\frac{1}{s!}\,
\ao^{\dagger\,s}\ao^{s} \eeq This operator is a diagonal operator,
whose form is given by: \beq
\hat{\Phi}\,=\,\frac{1}{\pi\alpha}\sum_{n=0}^{\infty}\,
\left(1-\frac{\theta}{\alpha}\right)^{n}\,\mid\psi_{n}\ra\la\psi_{n}\mid
\eeq The symbol of the truncated version is then: \beq
\Pi^{N}_{\theta}\left(\Phi\right)\,=\,e^{-r^{2}/\theta}\,\sum_{n=0}^{N}\,\frac{1}{\pi\alpha}\,
\left(1-\frac{\theta}{\alpha}\right)^{n}\,\frac{r^{2n}}{\theta^{n}n!}\,=\,
e^{-r^{2}/\alpha}\,
\frac{\Gamma\left(N+1,r^{2}\left(1/\theta\,-1/\alpha\right)\right)}
{\pi\alpha\Gamma\left(N+1\right)} \eeq This formula clearly shows
that the behaviour of the projected function, for increasing $N$,
is related to a comparison between the values of $\theta=1/N$ and
$\alpha$. A first interesting analysis on the nature of the
projection is given by a direct computation: \beqa
\hat{\Phi}^{\dagger}\hat{\Phi}&=&\left(\frac{1}{\alpha\pi}\right)^{2}\,\sum_{n=0}^{\infty}\,
\left(1-\frac{\theta}{\alpha}\right)^{2n}\,
\mid\psi_{n}\ra\la\psi_{n}\mid\nonumber
\\
Tr\left[\hat{\Phi}^{\dagger}\hat{\Phi}\right]&=&
\left(\frac{1}{\alpha\pi}\right)^{2}\,\sum_{n=0}^{\infty}\,
\left(1-\frac{\theta}{\alpha}\right)^{2n}\,=\,
\frac{1}{\left(\pi\alpha\right)^{2}}\lim_{n\rightarrow\infty}\,\frac{1-\left(1-\theta/\alpha\right)^
{2\left(n+1\right)}}{1-\left(1-\theta/\alpha\right)^{2}}\,\,\,\,\,\,\,\,\,\,\,\,\,\,\,\,\,\,\,\,\,\,\,\,
\label{gaussianexp}\eeqa
The limit in the RHS of this relation is finite if $\left|1-\theta/\alpha\right|^{2}\,<\,1$:
\begin{itemize}
\item
for $\alpha\,>\,\theta$, $\hat{\Phi}$ is an Hilbert-Schmidt
operator, so the projection gives a sequence, for
$N\,\rightarrow\,\infty$, of finite rank matrices
$\hat{\Phi}^{\left(N\right)}_{\theta}$ converging in the strong
operator topology
 to
$\hat{\Phi}$. This means that the sequence of symbols
$\Phi^{\left(N\right)}_{\theta}\left(r\right)=\Pi^{\left(N\right)}_{\theta}\left(\Phi\right)$
converges (pointwise) to a function which is equal to $
\Phi\left(r\right)$ inside the disc of radius $1$, and equal to
zero outside the disc. The plot is in figure~\ref{figgaussorigin}.
\item
for $\alpha=\theta$ the projection is trivial, as in this case the
operator $\hat{\Phi}$ is just a multiple of the projector: $$
\hat{\Phi}^{\left(N\right)}_{\theta}\,=\,\frac{1}{\pi\theta}\,\mid\psi_{0}\ra\la\psi_{0}\mid
$$
\item
for $\alpha\,<\,\theta\,<\,2\alpha$ the operator $\hat{\Phi}$ is
still Hilbert-Schmidt, so the sequence of truncated operators
converges to the operator $\hat{\Phi}$ in the strong operator
topology, and the sequence of projected symbols converges again to
the symbol $\Phi\left(r\right)$ inside the disc, and to zero
outside.
\end{itemize}

\begin{figure}[htbp]
\epsfxsize=2.3 in \centerline{\epsfxsize=2.2
in\epsffile{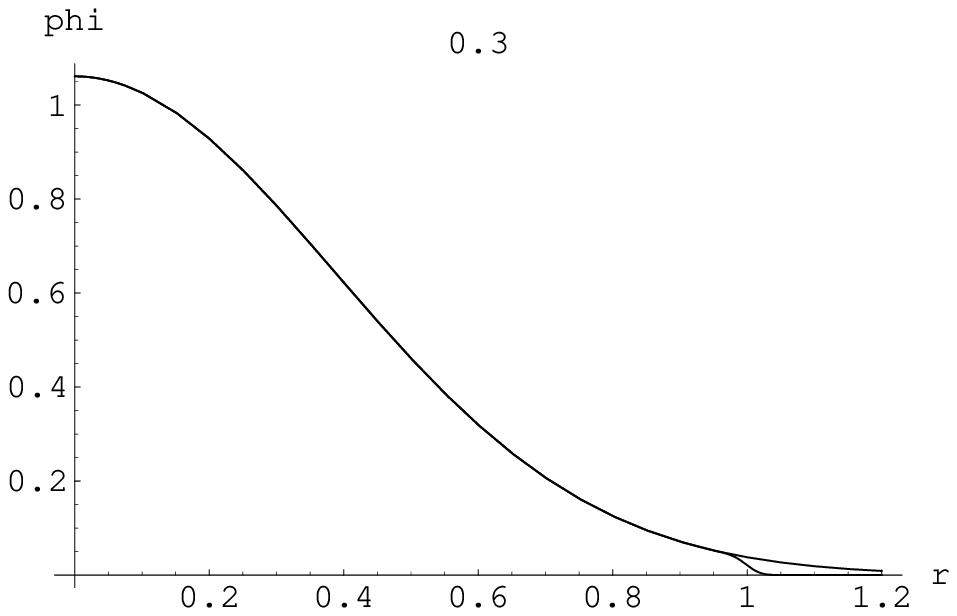}\epsfxsize=2.2
in\epsffile{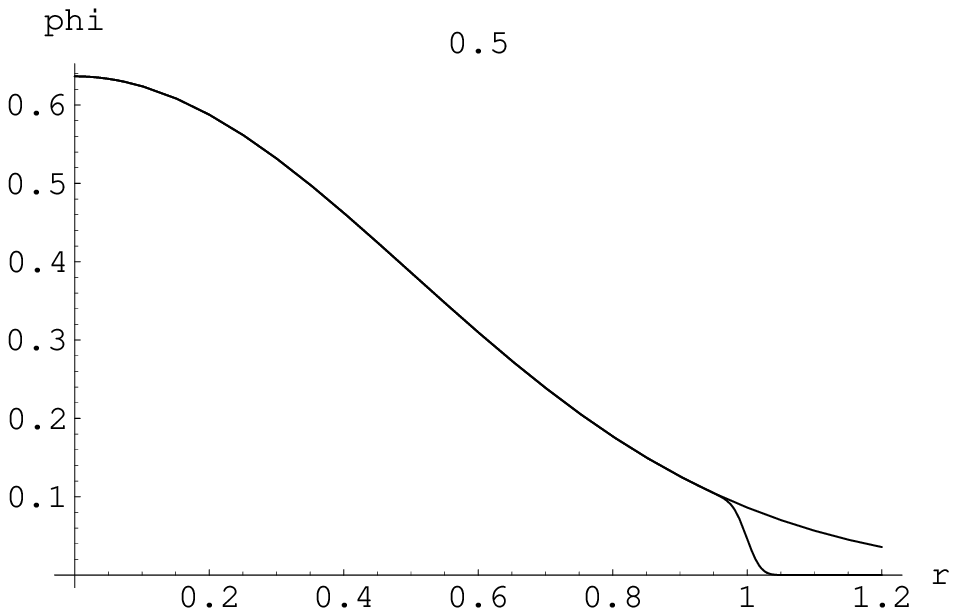}\epsfxsize=2.2 in\epsffile{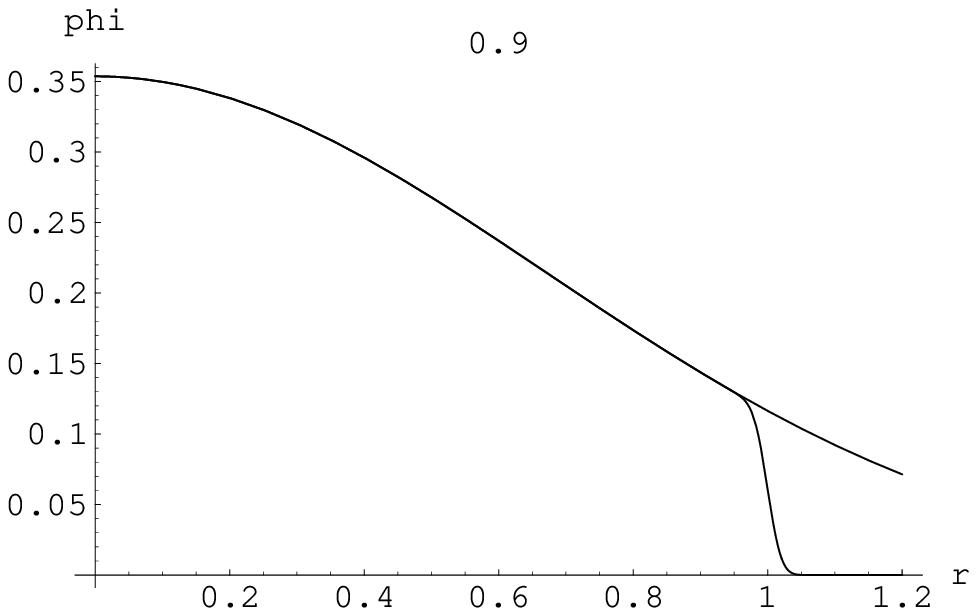}}
\caption{\baselineskip=12pt {\it Profile of the spherically
symmetric function
$\Pi^N_\theta\left(\frac{1}{\pi\alpha}e^{-\frac{r^2}{\alpha}}\right)$
for the choice $R^2=N\theta=1$. Here $N=10^3$, so $\theta=0.001$,
and $\alpha$ has been chosen to be the value on the top of each
plot. For these three choices $\alpha$ is much larger than
$\theta$.  Both the projected and the unprojected functions are
plotted, although inside the disc they are practically
indistinguishable. The unprojected function is always the larger
one.}}
\bigskip
\label{figgaussorigin}
\end{figure}

The operator $\hat{\Phi}$ is no more of the Hilbert-Schmidt class
starting from $\theta=2\alpha$, and for $\theta\,>\,2\alpha$ it is
no more compact. That something is happening to the sequence of
projected symbols is evident by figure~\ref{figorigingauss}

With $\alpha=.5\theta$ a small "bump" at $r=1$ appears. Already
for $\alpha=.49\theta$ the "bump" has become a large gaussian
sitting at the infrared cutoff; the part close to the origin is
still present, but it is quickly dwarfed by this bump, growing
very fast: in a numerical approximation, for $\alpha\sim .4\theta$
it is already of the order of $10^{17}$. In this case the series
expansion (\ref{gaussianexp}) is alternating, and individual terms
are divergent. A very interesting point is to stress: keeping
$\alpha$ fixed, and increasing $N$ with the introduced constraint
$N\theta=1$, the bump disappears. This limit forces in fact
$\theta$ to go to $0$, thus recovering the "nice" behaviour.

\begin{figure}[htbp]
\epsfxsize=2.3 in \centerline{\epsfxsize=2.2
in\epsffile{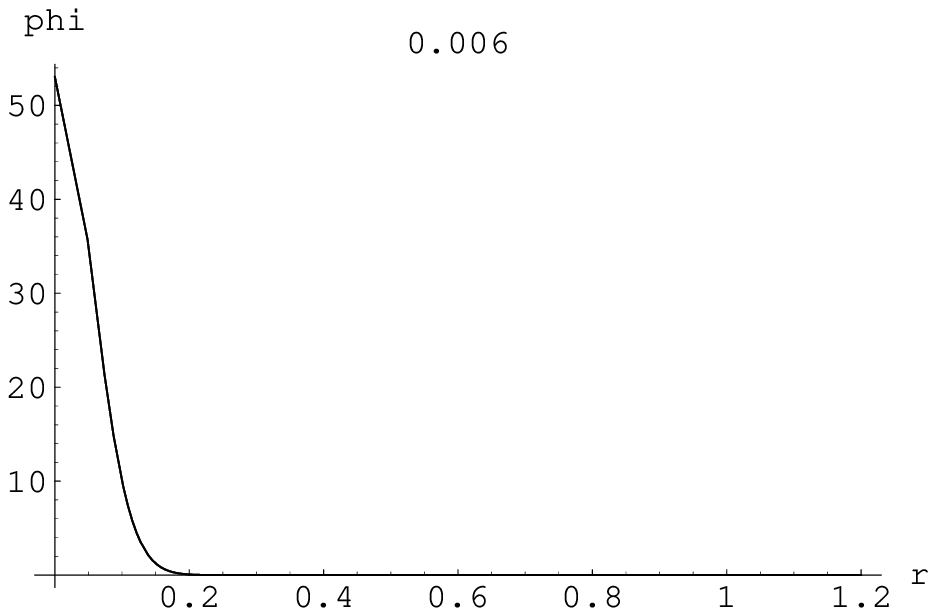}\epsfxsize=2.2
in\epsffile{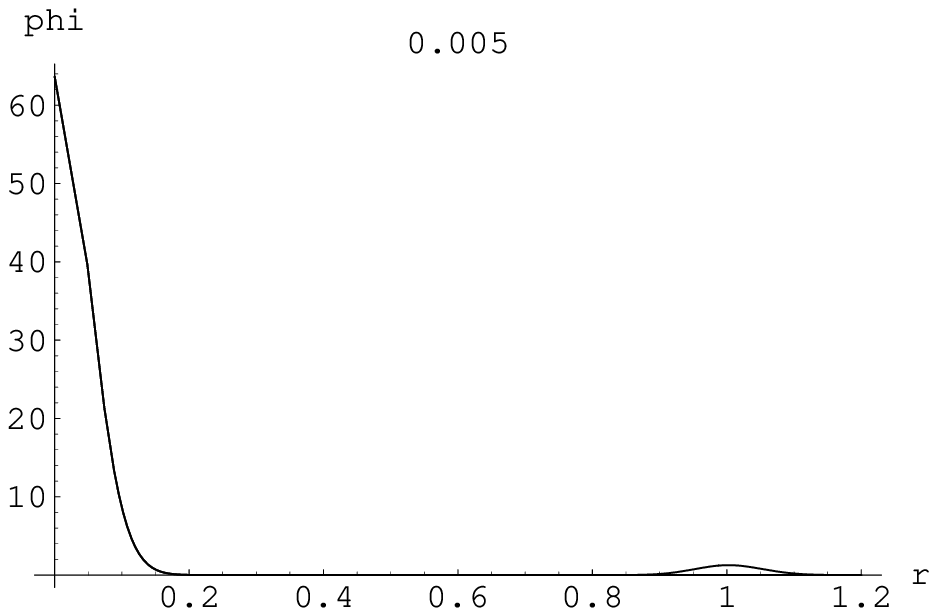}\epsfxsize=2.2
in\epsffile{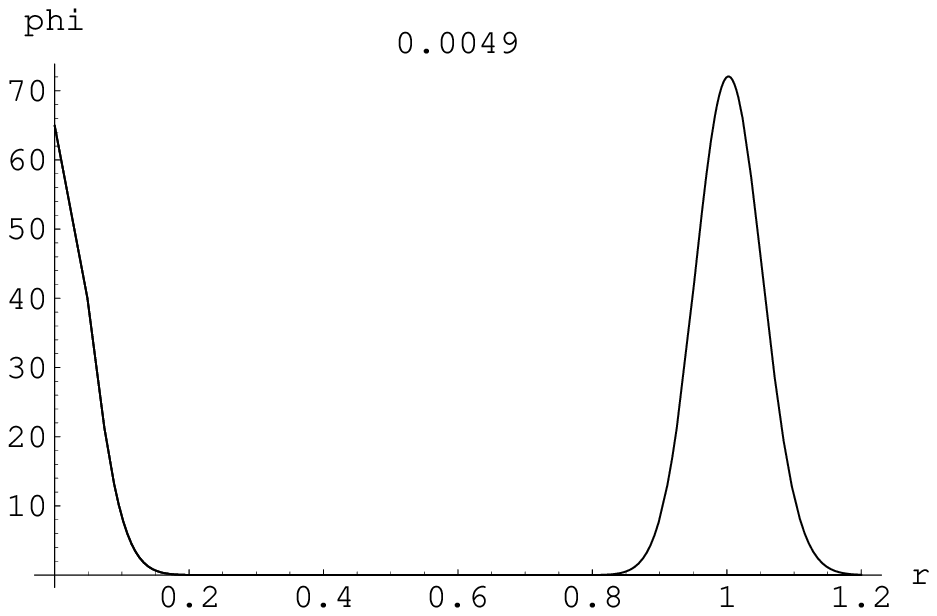}} \caption{\baselineskip=12pt {\it
Profile of the spherically symmetric function
$\Pi^N_\theta\left(\frac{1}{\pi\alpha}e^{-\frac{r^2}{\alpha}}\right)$
for the choice $R^2=N\theta=1$, $N=10^2$. Here $\theta=0.01$ and
the value of $\alpha$ is chosen to be around the ``turning point''
$\alpha=\theta/2$.}}
\bigskip
\label{figorigingauss}
\end{figure}

This detailed analysis shows that the Weyl map used to define this
quantization (\ref{discweylmap}) is very different from the
standard Weyl map described in the first chapter (\ref{weylmap}).
In the standard case it has been proved that it is an isomorphism
between square integrable functions on the plane, and
Hilbert-Schmidt class operators. In this case, it is evident that
the role of the weight factor in (\ref{weightdisc}) is to change
the set of applicability of this correspondence.

In general the function $\Phi$ is close to its projected version
$\Phi^{\left(N\right)}_{\theta}$ if it is mostly supported on a
disc of radius $1$ (of radius $R=\sqrt{N\theta}$, in general),
otherwise it is simply exponentially cut, and if it has not
oscillations of too small wavelenght (compared to $\theta$). In
this case the projected function becomes very large on the boudary
of the disc. This can be seen as a compact example of the
\emph{ultraviolet-infrared} mixing, which is one of the most
interesting characteristic of noncommutative geometry \cite{UVIR}.
If one tries to localise the function too much, unavoidably an
infrared divergence on the boundary of the disc appears.

There are however functions which are localised sharply near the
edge of the disc. These can be seen as edge states \cite{edges},
which play an important role in Chern-Simons theory. These edge
states are given, in each finite rank approximation, by the
symbols of the highest one dimensional projectors: \beq
\phi^{edge}\,\equiv\,\frac{1}{\theta}\la
z\mid\psi_{N}\ra\la\psi_{N}\mid z\ra\,=\,
e^{-r^{2}/\theta}\,\frac{r^{2N}}{N!\theta^{N+1}}
\label{edgestate}\eeq They are plotted in figure~\ref{edge}.
\begin{figure}[htbp]
\epsfxsize=2.5 in \centerline{\epsfxsize=2.5
in\epsffile{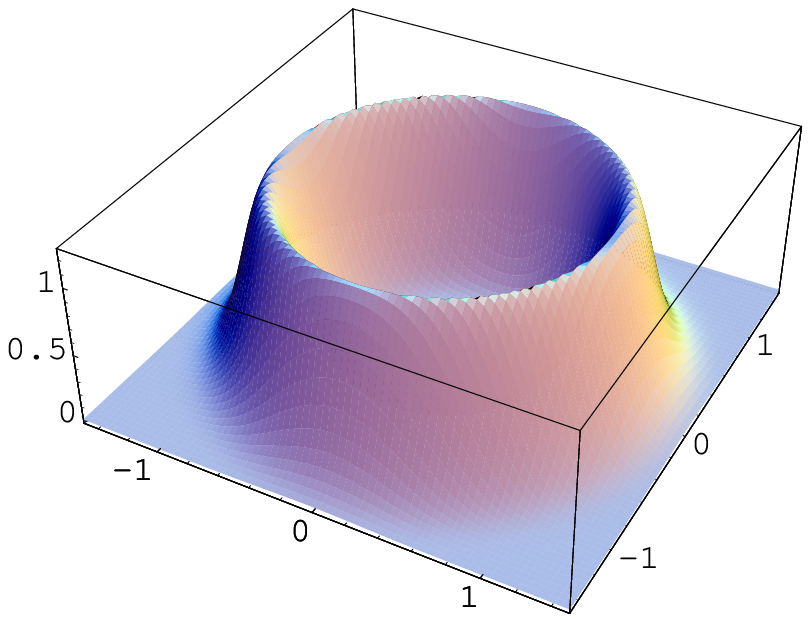} \epsfxsize=2.5
in\epsffile{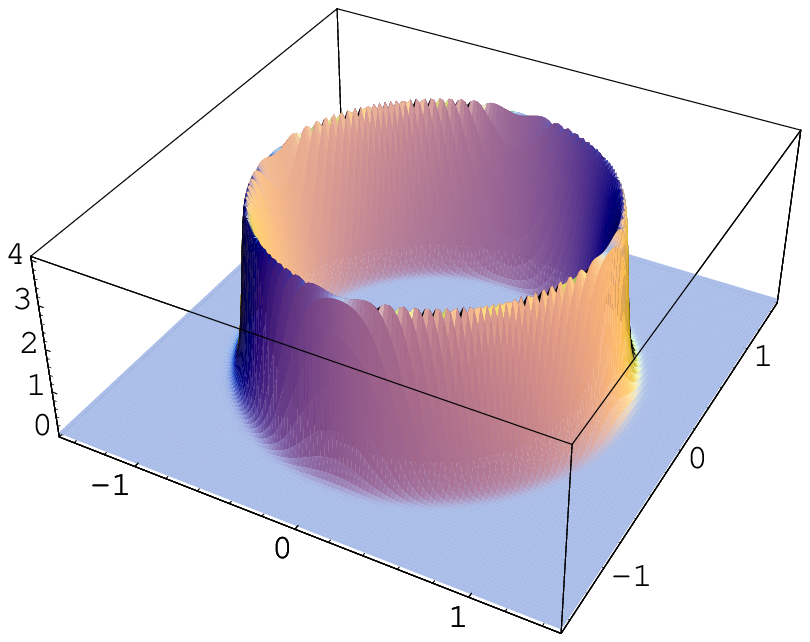}} \caption{\baselineskip=12pt {\it
The edge states $\varphi^{\mbox{\rm edge}}$ for $N=10$ and
$N=100$.  }}
\bigskip
\label{edge}
\end{figure}

\subsection{Fuzzy derivatives}

So far the Weyl-Wigner formalism, and the projection procedure,
have provided a way to associate to functions on the plane a
sequence of finite dimension
$\left(N+1\right)\times\left(N+1\right)$ matrices. An appropriate
choice of $\theta$, the noncommutativity parameter introduced by
the quantization map, and $N$, showed that it is possible to
obtain a good approximation of a certain class of functions
supported on a disc.

The next step is the analysis of the geometry this algebras can
formalize. To pursue this task, it is important to define, in this sequence of algebras, 
both derivations and a Laplacian. The
starting point to define the matrix equivalent of the derivations
is:
 \bea
\del_{z}f\,&=&\,\frac{1}{\theta}\langle z\mid
\left[\hat{f},\hat{a}^{\dagger}\right]\mid z\rangle \nonumber \\
\del_{\bar{z}}f\,&=&\,\frac{1}{\theta}\langle z\mid
\left[\hat{a},\hat{f}\right]\mid z\rangle \label{exactdersym} \eea
This relation is exact in the full algebra $\mathcal{A}_{\theta}$.
Given an operator $\fo$, the derivatives of the symbol
$f\left(\baz,z\right)$ are related to the symbol of the commutator
of $\fo$ with the creation and annihilation operators. A fuzzyfied
version, namely a truncated version, of these operations, can be
introduced defining: \bea
\del_{z}f^{\left(N\right)}_{\theta}&\equiv &\frac{1}{\theta}
\langle z\mid\hat{P}^{\left(N\right)}_{\theta}
\left[\hat{P}^{\left(N\right)}_{\theta}\hat{f}\hat{P}^{\left(N\right)}_{\theta},\hat{a}^{\dagger}\right]
\hat{P}^{\left(N\right)}_{\theta}\mid z\rangle \nonumber \\
\del_{\bar{z}}f^{\left(N\right)}_{\theta}& \equiv
&-\frac{1}{\theta} \langle z\mid\hat{P}^{\left(N\right)}_{\theta}
\left[\hat{P}^{\left(N\right)}_{\theta}\hat{f}\hat{P}^{\left(N\right)}_{\theta},\hat{a}\right]
\hat{P}^{\left(N\right)}_{\theta}\mid z\rangle \label{fuzzder}
\eea It is important to note that this is actually a derivation
(it is a linear operation, satisfying the Leibniz rule) on each
$\mathcal{A}^{\left(N\right)}_{\theta}$. The idea behind this
definition is to consider an element
$\fo^{\left(N\right)}_{\theta}$ of a finite step of the sequence
as a finite rank matrix in a space of infinite dimensional matrices.
The projection is then seen as an embedding of the "truncated"
matrix $\fo^{\left(N\right)}_{\theta}$ as an upper left block
diagonal matrix into an infinite dimensional matrix where the remaining 
infinite number of rows and columns have elements equal to zero. The
way this embedding acts is in the left hand term of the inner
commutator:
$\hat{P}^{\left(N\right)}_{\theta}\hat{f}\hat{P}^{\left(N\right)}_{\theta}$.
Next the "exact" relation (\ref{exactdersym}) for derivatives of
operators in $\mathcal{F}$ - so to say infinite matrices, in this
context
 - is used. This is the meaning of the commutator with operators $\ao$ and $\aod$,
 which are considered in their complete matrix representation, without any "truncation".
Last, since these derivations should map dimensional matrix of
$\mathcal{A}^{\left(N\right)}_{\theta}$ into a finite dimensional
matrix in the same $\mathcal{A}^{\left(N\right)}_{\theta}$, the
image of the commutator should be projected back to the finite
dimensional space generated by the first $N+1$ ket state
$\mid\psi_{n}\ra$. Moreover, this projection is important, because
creation and annihilation operators are ladder operators, and then
they tend to shift $N^{th}$ column's and row's element into the
$\left(N+1\right)^{th}$ column and row of the image matrix.

The first example is the calculation of the derivatives of the
fuzzyfied coordinate functions $z$ and $\bar{z}$.
From the definition, $z$ is the symbol of the annihilation operator, while
$\bar{z}$ is the symbol of the creation operator, that can be written as:
\bea
\hat{a}\,&=&\,\sum_{s=0}^{\infty}\,\sqrt{\left(s+1\right)\theta}\mid\psi_{s}\rangle\langle\psi_{s+1}\mid \nonumber
\\
\hat{a}^{\dagger}\,&=&\,\sum_{k=0}^{\infty}\,\sqrt{\left(k+1\right)\theta}\mid\psi_{k+1}\rangle\langle\psi_{k}\mid
\eea
Projection into $\mathcal{A}^{\left(N\right)}_{\theta}$ gives:
\beq
\hat{a}^{\left(N\right)}_{\theta}=\sum_{s=0}^{N-1}\,
\sqrt{\left(s+1\right)\theta}\mid\psi_{s}\rangle\langle\psi_{s+1}\mid
\eeq
To perform the derivative with respect to $z$ variable, one considers:
\bea
\left[\hat{a}^{\left(N\right)}_{\theta},\hat{a}^{\dagger}\right]&=&
\theta\left[\mid\psi_{0}\rangle\langle\psi_{0}\mid+\sum_{s=1}^{N-1}\mid
\psi_{s}\rangle\langle\psi_{s}\mid-N\mid\psi_{N}\rangle\langle\psi_{N}\mid\right] \nonumber
\\
&=&\theta\left[\sum_{s=0}^{N-1}\mid\psi_{s}\rangle\langle\psi_{s}\mid-
N\mid\psi_{N}\rangle\langle\psi_{N}\mid\right] \nonumber
\\
&=&\theta\left[\sum_{s=0}^{N}\mid\psi_{s}\rangle\langle\psi_{s}\mid-\left(1+N\right)
\mid\psi_{N}\rangle\langle\psi_{N}\mid\right] \eea The first term
of the sum is the projector onto the first $N+1$ basis elements, the
identity on $\mathcal{A}^{\left(N\right)}_{\theta}$, a fuzzy
identity. What is interesting is that this commutator has no terms
'outside' the space we are considering, namely there are no
components on density matrices of order greater than $N$: this
means that in this case there is no need to project it on
$\mathcal{A}^{\left(N\right)}_{\theta}$. The symbol of this
commutator is: \beq
\del_{z}\left(z^{\left(N\right)}_{\theta}\right)=e^{-Nr^{2}}\left[\sum_{s=0}^{N}\,\frac{r^{2s}N^{s}}{s!}\,-\,
\frac{N+1}{N!}r^{2N}N^{N}\right] \eeq In the limit of
$N\rightarrow\infty$ the first term is what has been called
'characteristic function for the disc', while the second converges
to a factor $\pi\delta\left(r-1\right)$. This factor is a radial
$\delta$ selecting the value for $r=1$ with respect to the
Lebesgue measure on the plane: \beq
\lim_{N\rightarrow\infty}\del_{z}\left(z^{\left(N\right)}_{\theta}\right)=Id\left(r\right)-\pi\delta\left(r-1\right)
\eeq To calculate the derivative of $\bar{z}$ with respect to $z$
one needs to consider: $$\hat{a}^{\dagger\left(N\right)}_{\theta}=
\sum_{k=0}^{N-1}\,\sqrt{\left(k+1\right)\theta}\mid\psi_{k+1}\rangle\langle\psi_{k}\mid$$
obtaining: \beq
\left[\hat{a}^{\dagger\left(N\right)}_{\theta},\hat{a}^{\dagger}\right]
=-\theta\sqrt{N\left(N+1\right)}\mid\psi_{N+1}\rangle\langle\psi_{N}
\mid \eeq Since this operator must be projected back to the
algebra $\mathcal{A}^{\left(N\right)}_{\theta}$, one finally has:
\beq \del_{\bar{z}}\left(z^{\left(N\right)}_{\theta}\right)=0 \eeq

\subsection{Fuzzy Laplacian and fuzzy Bessels}

One can consider the problem for the Laplacian in a similar context. From the exact expressions:
\bea
\nabla^2\,f&=&4\del_{\bar{z}}\del_{z}f
\nonumber \\
{\left(\nabla^{2}\,f\right)}\left(\baz,z\right)&=&\frac{4}{\theta^{2}}\la z\mid\left[\hat{a},
\left[\hat{f},\hat{a}^{\dagger}\right]\right]\mid z\ra
\eea
it is possible to define, in each $\mathcal{A}^{\left(N\right)}_{\theta}$:
\beq
\nabla^2\,\fo^{\left(N\right)}_{\theta}\equiv\frac{4}{\theta^{2}}\hat{P}^{\left(N\right)}_{\theta}
\left[\hat{a},\left[\hat{P}^{\left(N\right)}_{\theta}
\hat{f}\hat{P}^{\left(N\right)}_{\theta},\hat{a}^{\dagger}\right]\right]\hat{P}^{\left(N\right)}_{\theta}
\label{lapl}
\eeq

The image of the element of the truncated algebra:
$$\hat{f}^{\left(N\right)}_{\theta}=\sum_{a,b=0}^{N}\,f_{ab}\mid\psi_{a}\rangle\langle\psi_{b}\mid$$
is: \bea
\nabla^{2}\,f^{\left(N\right)}_{\theta}&=&\,4N\,\left[\sum_{s=0}^{N-1}\,\sum_{b=0}^{N-1}\,f_{s+1,b+1}\sqrt{\left(s+1\right)
\left(b+1\right)}\mid\psi_{s}\rangle\langle\psi_{b}\mid+\right.
\nonumber
\\
&-&\sum_{s=0}^{N}\,\sum_{b=0}^{N}\,f_{sb}\left(s+1\right)\mid\psi_{s}\rangle\langle\psi_{b}\mid\,-\,
\sum_{s=0}^{N-1}\,f_{0,s+1}\left(s+1\right)\mid\psi_{0}\rangle\langle\psi_{s+1}\mid+ \nonumber
\\
&+&\sum_{s=0}^{N-1}\,\sum_{b=0}^{N-1}\,f_{sb}\sqrt{\left(s+1\right)\left(b+1\right)}\mid\psi_{s+1}\rangle
\langle\psi_{b+1}\mid+ \nonumber
\\
&-&\left.\sum_{s=0}^{N-1}\,\sum_{b=0}^{N-1}\,f_{s+1,b+1}\,\left(b+1\right)\mid\psi_{s+1}\rangle\langle\psi_{b+1}
\mid\,\right] \label{explapl}\eea

The spectrum of this \emph{fuzzy Laplacian} can be numerically
calculated. Its eigenvalues seem to converge to the spectrum of
the continuum Laplacian for functions on a disc, with boundary
conditions on the edge of the disc of Dirichlet homogeneous kind
\cite{tichonov}. In the continuum case, the eigenvalue problem for
the Laplacian with
 Dirichlet
 homogeneous boundary conditions is solved by the zeroes of the Bessel functions of first kind, namely
 $\lambda$ is an eigenvalue if it solves
the implicit equation \beq J_{n}\left(\sqrt{\lambda}\right)=0 \eeq
where $n$ is the order of the Bessel. In particular, those related
to $J_{0}$ are simply degenerate, the other are doubly degenerate:
so there is a sequence of eigenvalues  labelled by $\lambda_{n,k}$
where $n$ is the order of the Bessel function and $k$ indicates
that it is the $k^{th}$ zero of the function. The spectrum of the
fuzzy Laplacian is in good agreement with the spectrum of the
continuum explained case, even for low values $N$ of the dimension
of truncation, as can be seen in figure~\ref{fuzzydrum}.
\begin{figure}[htbp]
\epsfxsize=2.3 in \centerline{\epsfxsize=2.2
in\epsffile{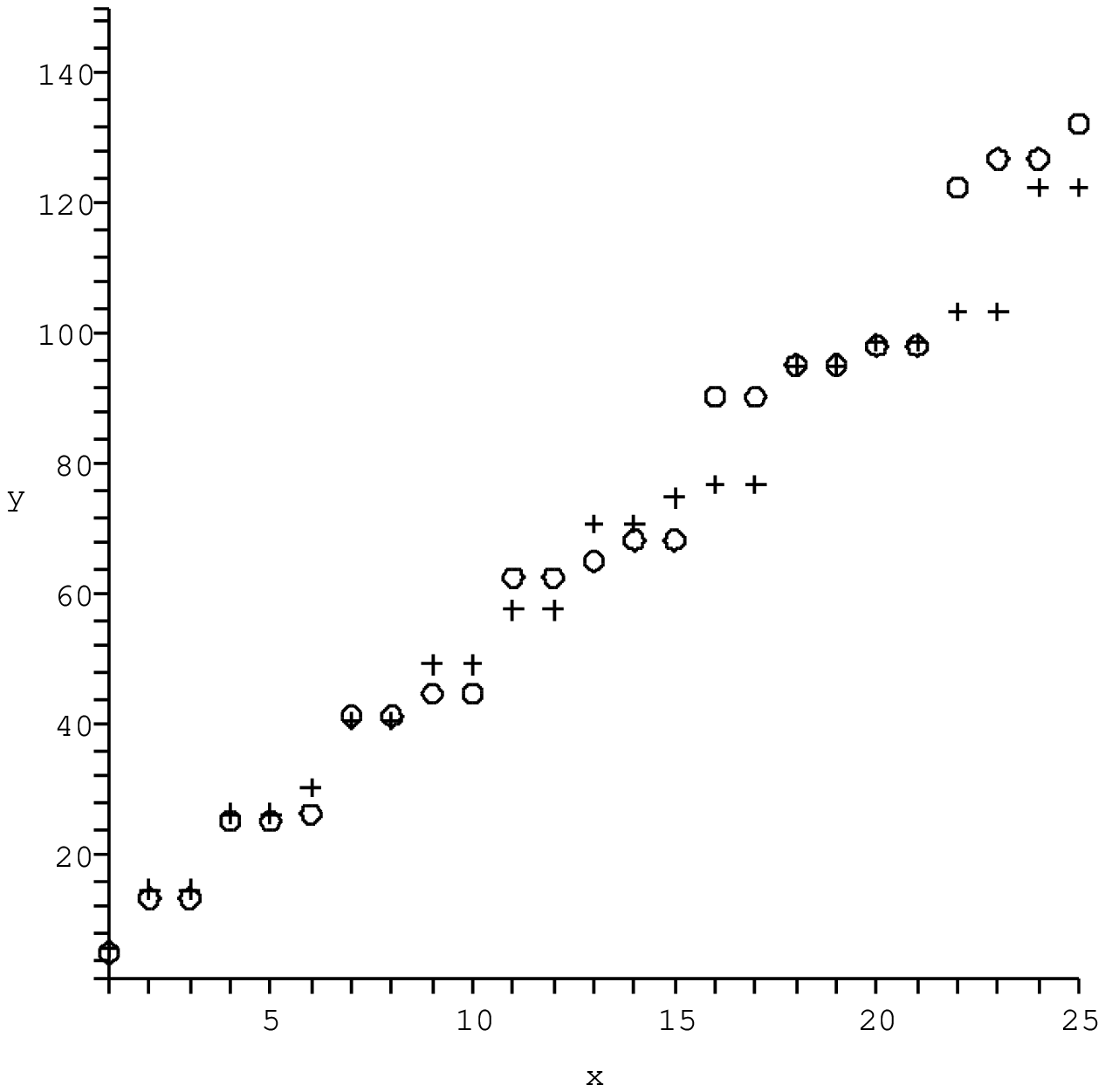}\epsfxsize=2.2
in\epsffile{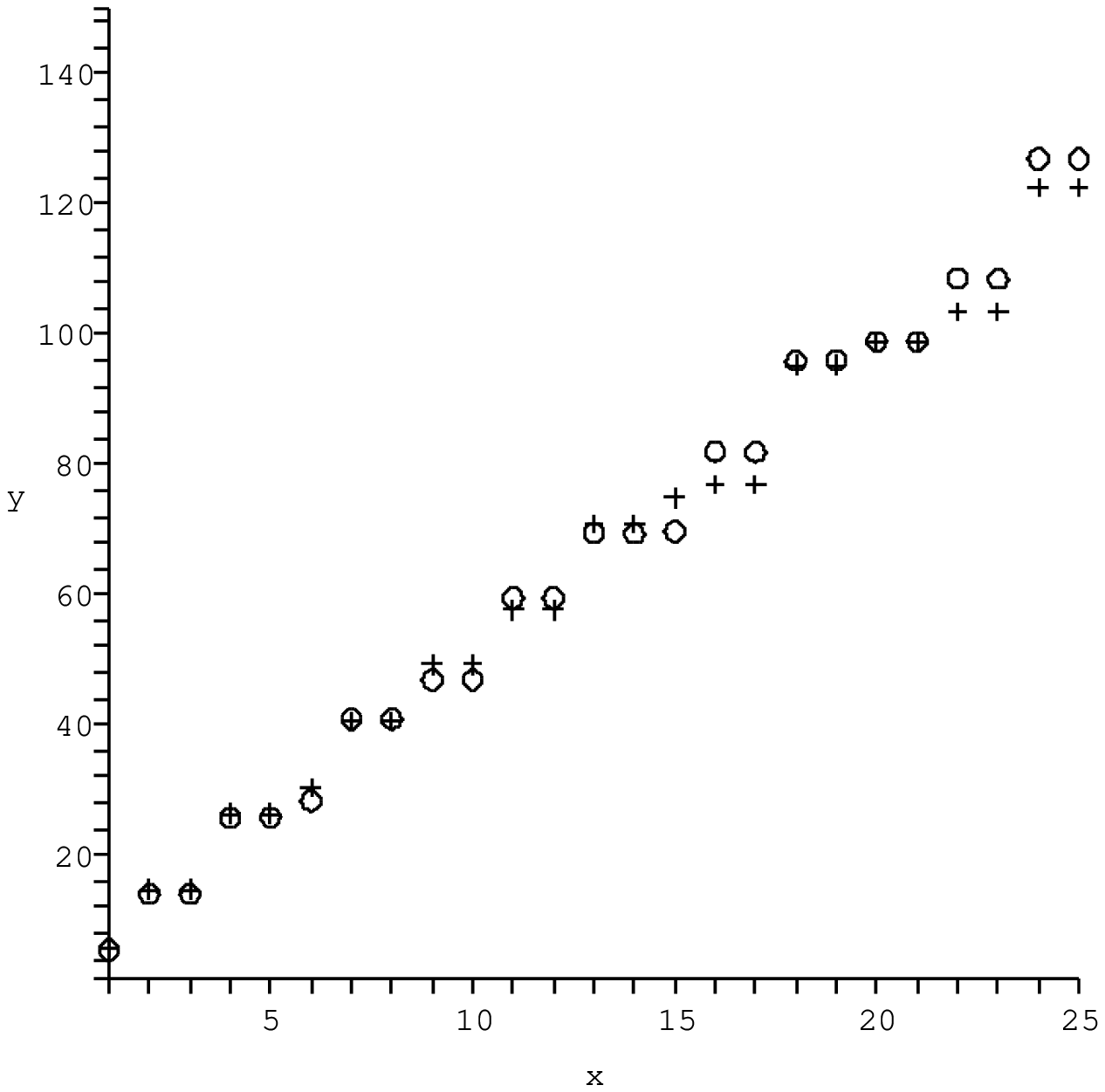}\epsfxsize=2.2
in\epsffile{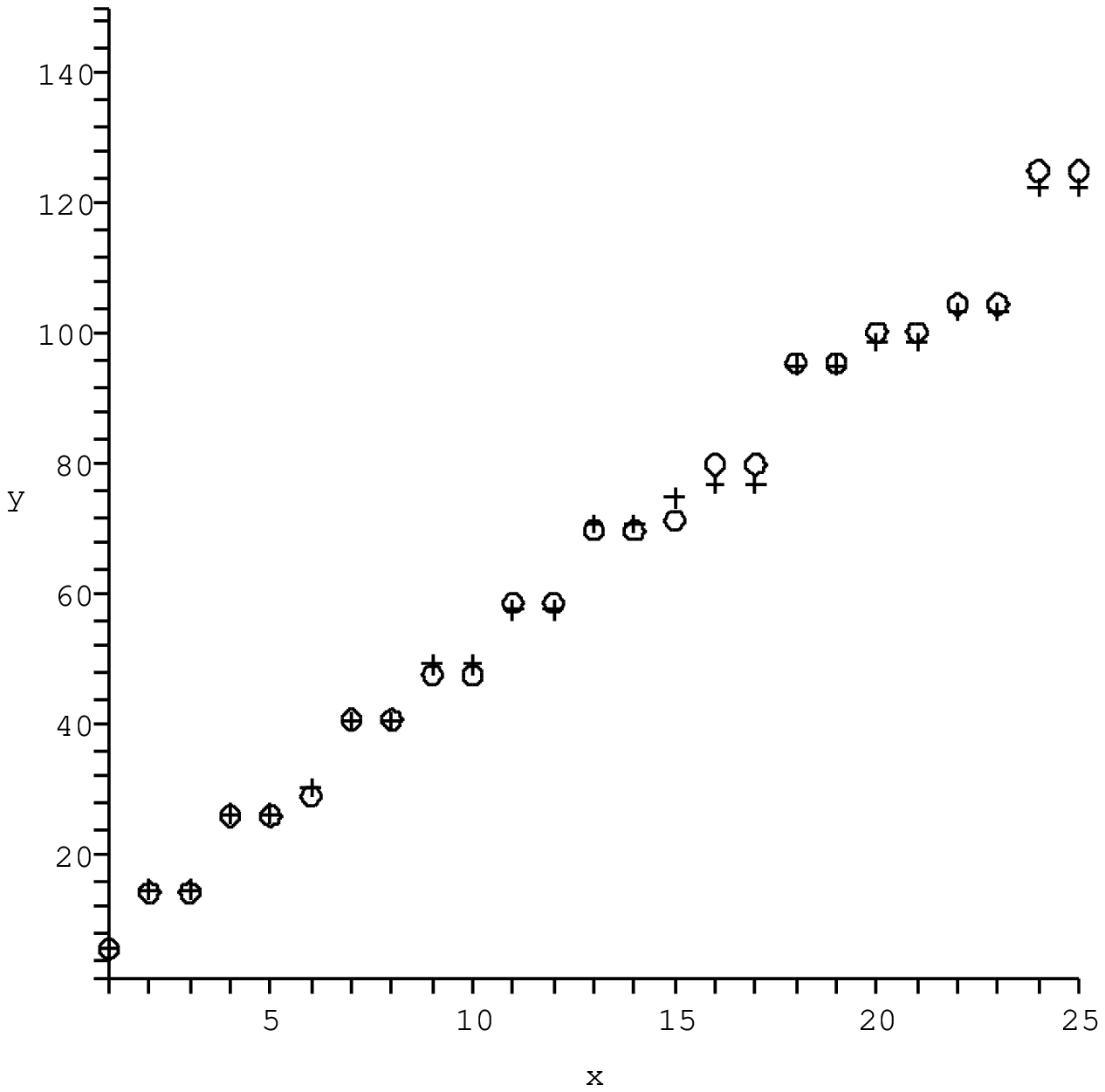}} \caption{\baselineskip=12pt {\it
Comparison of the first eigenvalues of the fuzzy Laplacian
(circles) with those of the continuum Laplacian (crosses) on the
domain of functions with Dirichlet homogeneous boundary
conditions. The orders of truncation are $N=9,19,29$.}}
\bigskip
\label{fuzzydrum}
\end{figure}

What is interesting is that this definition of Laplacian
(\ref{lapl}) gives the expected pattern of non degenerate and
double degenerate eigenvalues. The difference with the case of the
spectrum of the fuzzy laplacian for the fuzzy sphere is that now
the "fuzzy spectrum" is both a cut-off and an approximation of the
continuum spectrum. It is a cut-off because, of course, it is a
finite rank operator. The fact that it is an approximation is
related to the fact that it has been defined using a formalism
whose building blocks are related to a noncompact group, namely
the Heisenberg-Weyl, so that there is no finite dimensional
realization of its generators.

The eigenfunctions for the continuum problem are: \beq
\psi_{n,k}=e^{in\varphi}\left(\frac{\sqrt{\lambda_{\left|n\right|,k}}r}{2}\right)^{|n|}\sum_{k=0}^{\infty}
\frac{\left(-\lambda_{|n|,k}\right)^{k}}{k!\left(|n|+k\right)!}\left(\frac{r}{2}\right)^{2k}=
e^{in\varphi}J_{|n|}\left(\sqrt{\lambda_{\left|n\right|,k}}r\right)
\eeq In this expression $n$ is an integer number, $|n|$ is its
absolute value: this is a way to write eigenfunctions in a compact
form, taking into account the degeneracy of eigenvalues for
$|n|\geq 1$. Now, in the 'fuzzy' approximation, one has a sequence
of eigenvalues $\lambda^{\left(N\right)}_{n,k}$, with $N$
indicating the dimension of the fuzzyfication. The label $n$ runs
from $-N$ to $+N$, while $k$ runs from $1$ to
$N-\left|n\right|+1$. To each eigenvalue one can associate an
eigenmatrix, indicated with
$\hat{\Phi}\left(\lambda^{\left(N\right)}_{n,k}\right)$. Its
symbol is a function of $z,\baz$: \beq
\Phi\left(\lambda^{\left(N\right)}_{n,k}\right)=e^{-N\mid
z\mid^{2}}\sum_{a,b=0}^{N}\Phi_{ab}^{\left(N\right)}\left(\lambda_{n,k}^{\left(N\right)}\right)
\bar{z}^{a}z^{b}\left(\frac{N^{a+b}}{a!b!}\right)^{1/2} \eeq 

Some of them are plotted. In figure~\ref{fuzzygroundstate} it is
plotted the radial shape of the fuzzy Bessel relative to $n=0$ and
$k=1$, so to say the fuzzy ground state of this matrix model. The
comparison shows that this fuzzy ground state converges to the
continuum eigenfunctions $\psi_{0,1}\left(r,\varphi\right)$ for
values of $r$ inside the disc of radius $1$, while it converges to
zero outside the disc.

\begin{figure}[htbp]
\epsfxsize=2.3 in \centerline{\epsfxsize=2.2
in\epsffile{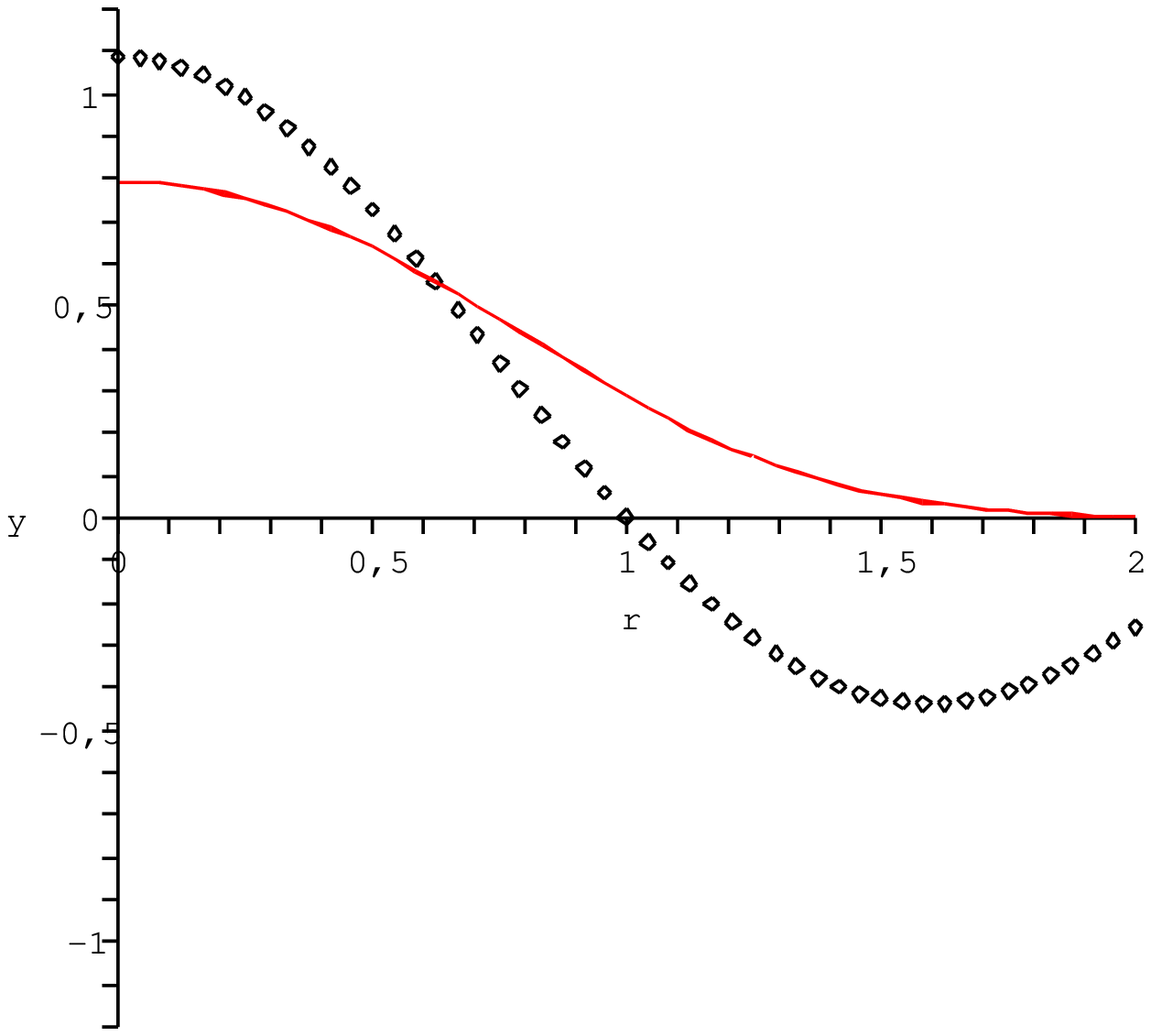}\epsfxsize=2.2
in\epsffile{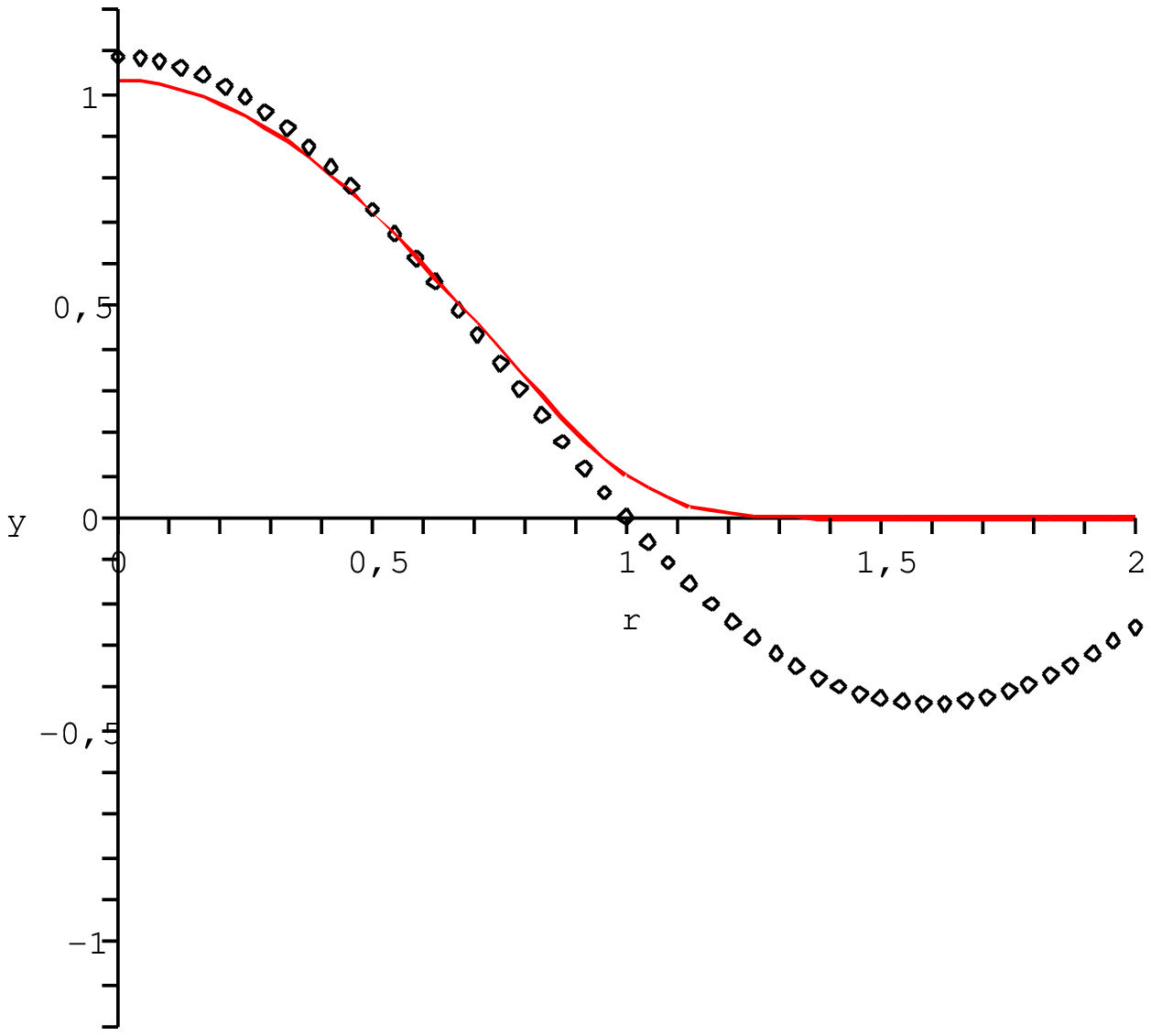}\epsfxsize=2.2
in\epsffile{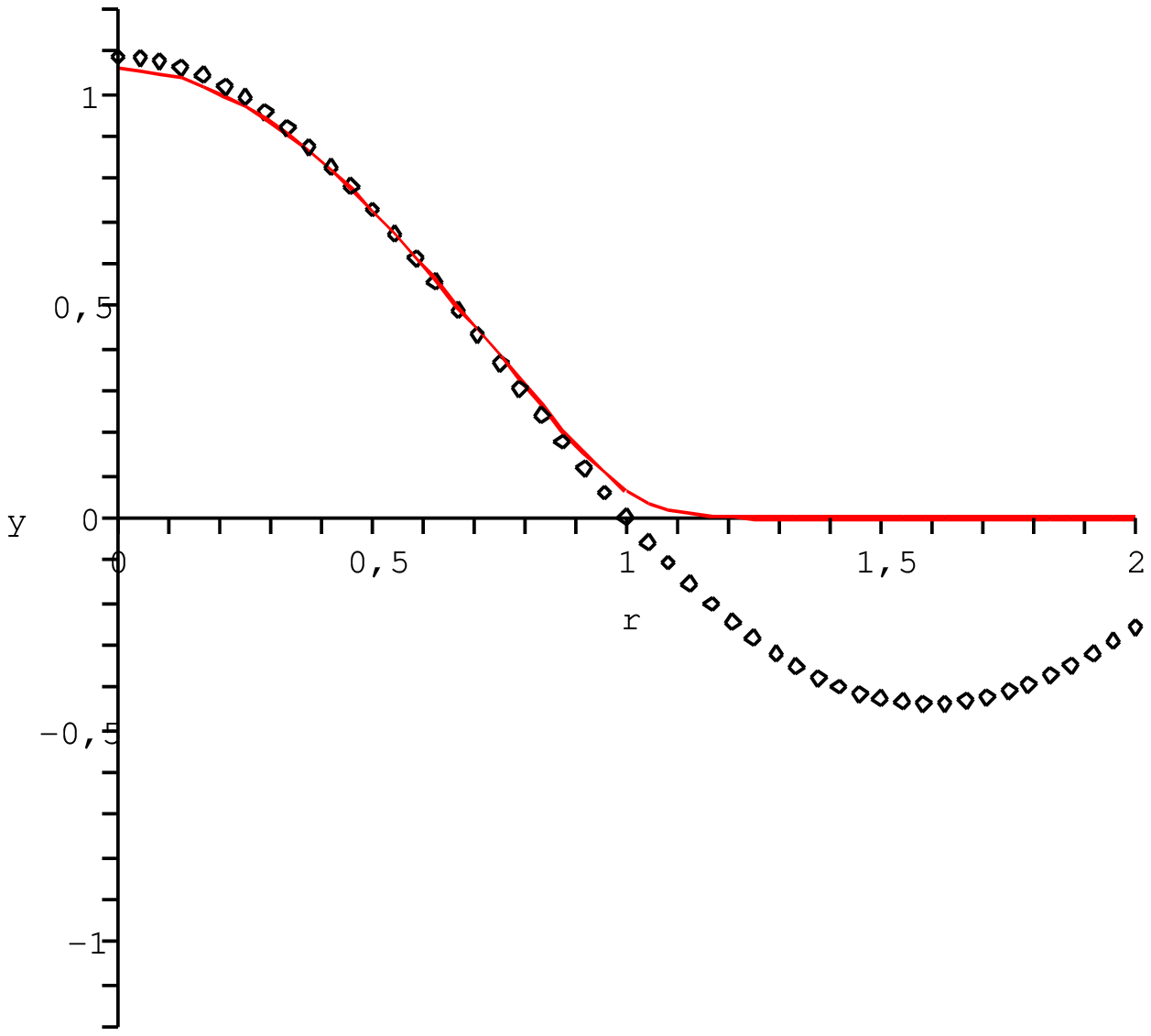}} \caption{\baselineskip=12pt
{\it Comparison of the radial shape for the ground state fuzzy
Bessel $\Phi\left(\lambda_{0,1}^{\left(N\right)}\right)$
(continuum line), the symbol of the eigenmatrix of the fuzzy
Laplacian with respect to the lowest eigenvalue, with
$\psi_{0,1}\left(r,\varphi\right)$, the ground state eigenfunction
of the continuum problem (diamond line), for $N=3,15,30$. The
fuzzy Bessel converges to zero outside the disc of radius $1$.  }}
\bigskip
\label{fuzzygroundstate}
\end{figure}

This behaviour is valid also for eigenstates of
different eigenvalues. The plots are for the symbols
$\Phi\left(\lambda_{0,2}^{\left(N\right)}\right)$ in figure
\ref{FSDS} and $\Phi\left(\lambda_{0,3}^{\left(N\right)}\right)$
in figure \ref{FTDS}. Symbols
$\Phi\left(\lambda_{0,k}^{\left(N\right)}\right)$ and functions
$\psi_{0,k}\left(r,\varphi\right)$ are seen to be radial
functions.

This behaviour induces to define the eigenmatrices of the fuzzy Laplacian as \emph{fuzzy Bessels}.

Since fuzzy Bessels are obtained as eigenstates of an Hermitian
operator, they are defined up to a normalization factor. In view
of their use, the continuum Bessel functions are normalised as:
\beq \int_{0}^{2\pi}\,d\varphi\,\int_{0}^{1}\,r\,dr\,
\left|J_{\left|n\right|}\left(\sqrt{\lambda_{\left|n\right|,k}}r\right)\right|^{2}\,=\,1
\eeq while fuzzy Bessels are normalised by: \beq
\int_{0}^{2\pi}\,d\varphi\,\int_{0}^{\infty}\,r\,dr\,
\left|\Phi\left(\lambda_{\left|n\right|,k}^{\left(N\right)}\right)\right|^{2}\,=\,1
\eeq

\begin{figure}[htbp]
\epsfxsize=2.3 in \centerline{\epsfxsize=2.2
in\epsffile{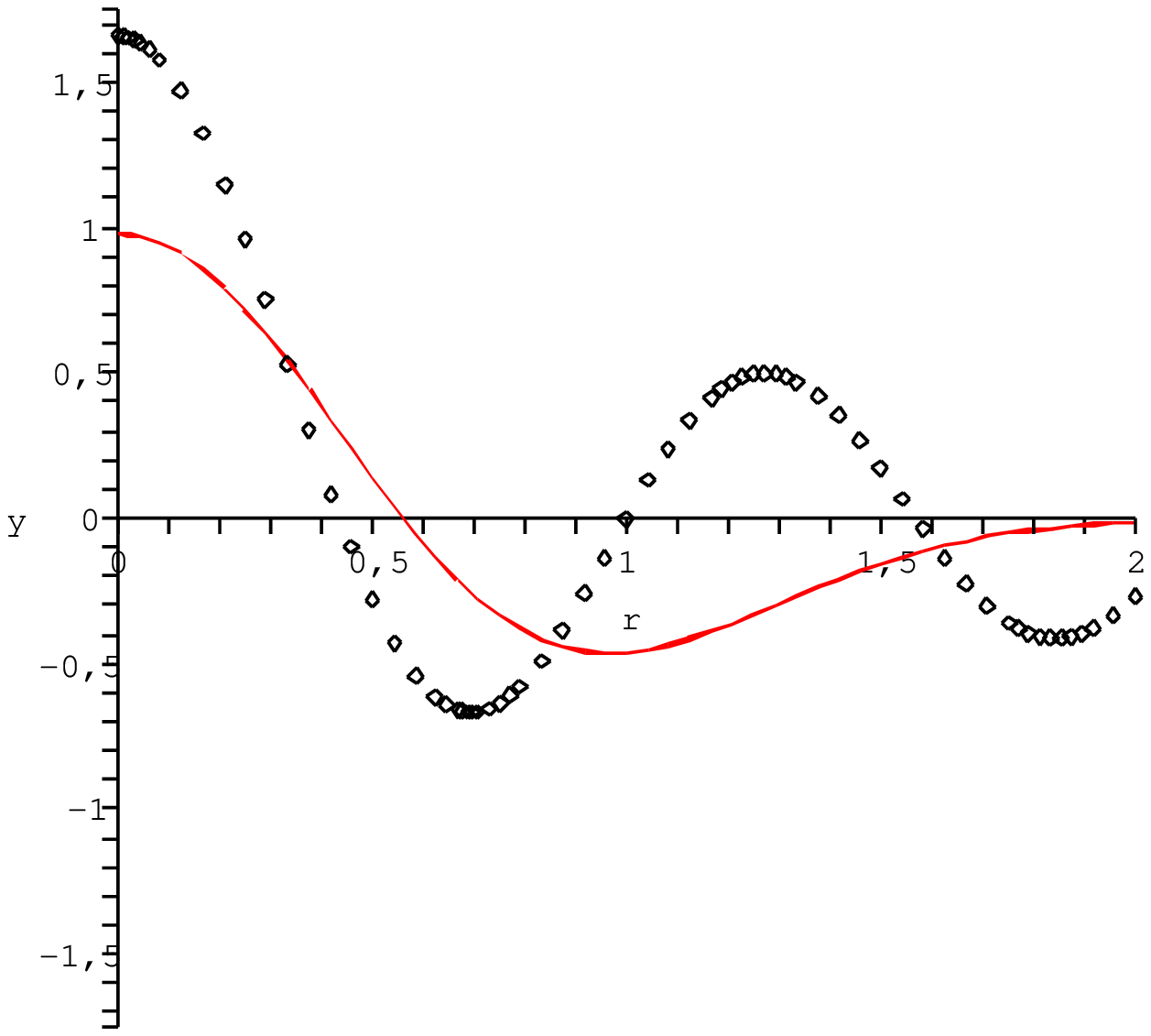}\epsfxsize=2.2
in\epsffile{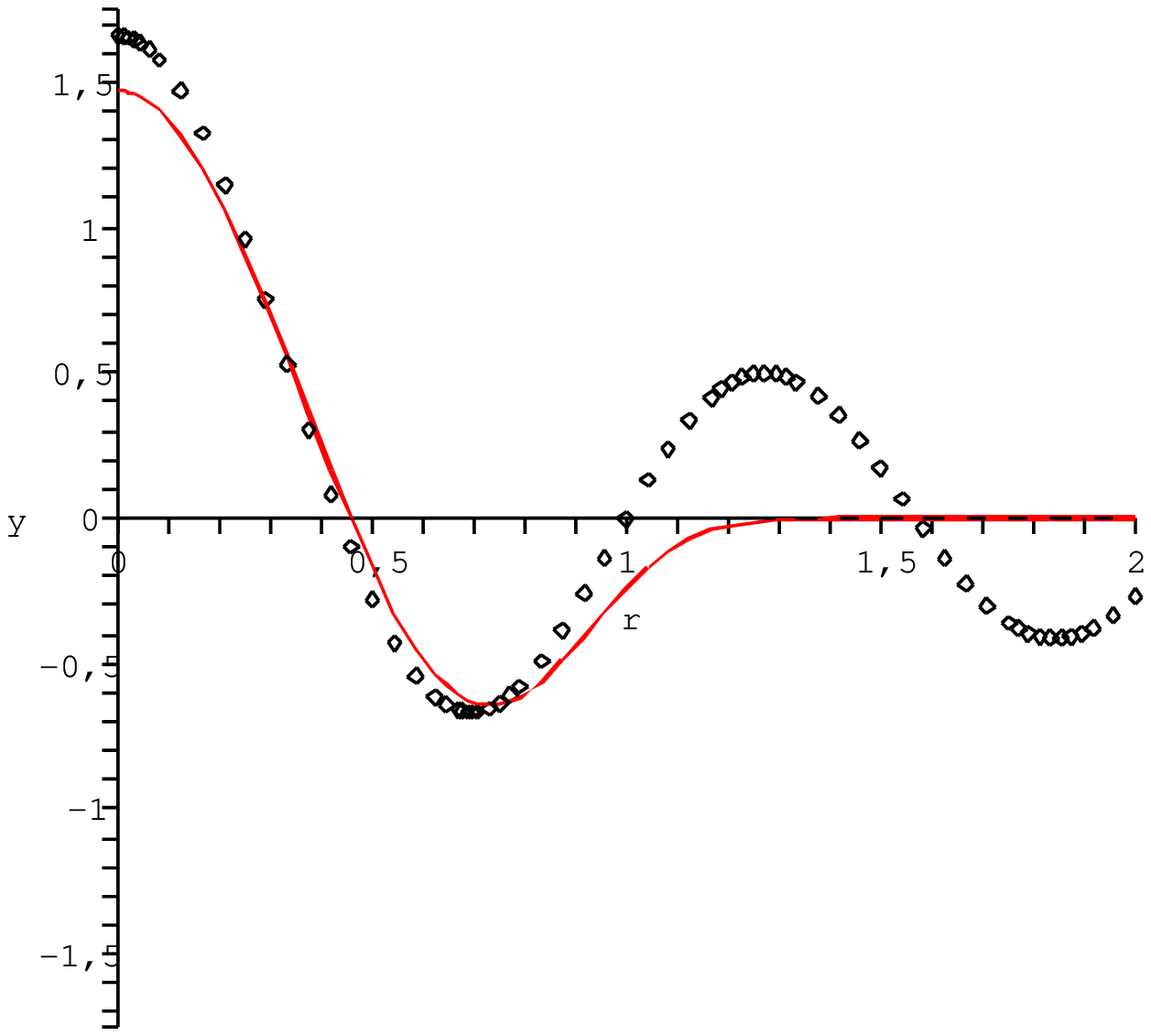}\epsfxsize=2.2
in\epsffile{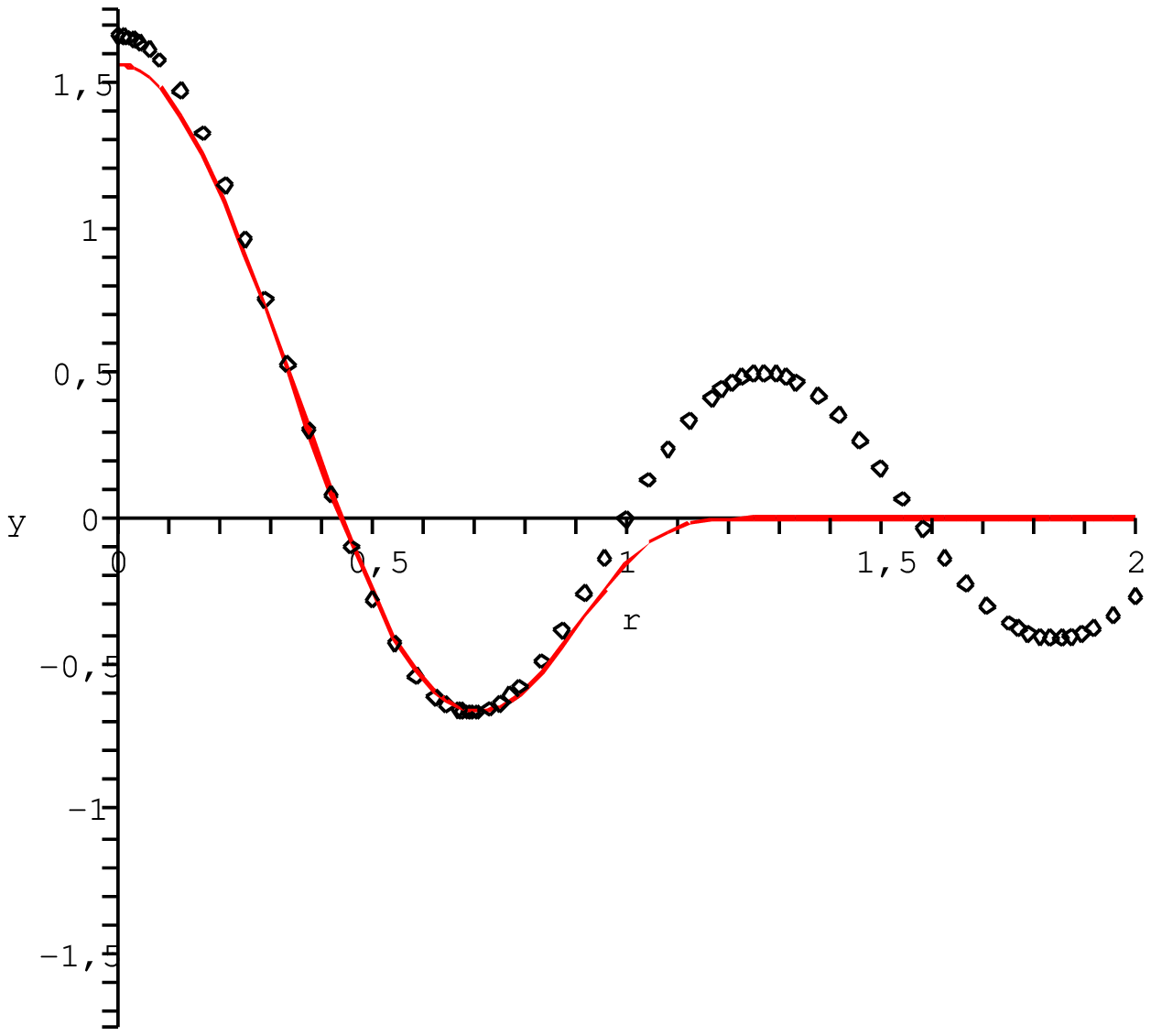}} \caption{\baselineskip=12pt {\it
Comparison of the radial shape for the
$\Phi\left(\lambda_{0,2}^{\left(N\right)}\right)$ fuzzy Bessel
(continuum line), the symbol of the eigenmatrix of the fuzzy
Laplacian with respect to the eigenvalue
$\lambda_{0,2}^{\left(N\right)}$, with the function
$\psi_{0,2}\left(r\right)\left(r,\varphi\right)$ that is the
eigenfunction of the continuum problem with eigenvalue
$\lambda_{0,2}$. Here the orders of truncation are $N=2,15,30$. The
fuzzy Bessel converges to zero outside the disc of radius $1$.}}
\bigskip
\label{FSDS}
\end{figure}

\begin{figure}[htbp]
\epsfxsize=2.3 in \centerline{\epsfxsize=2.2
in\epsffile{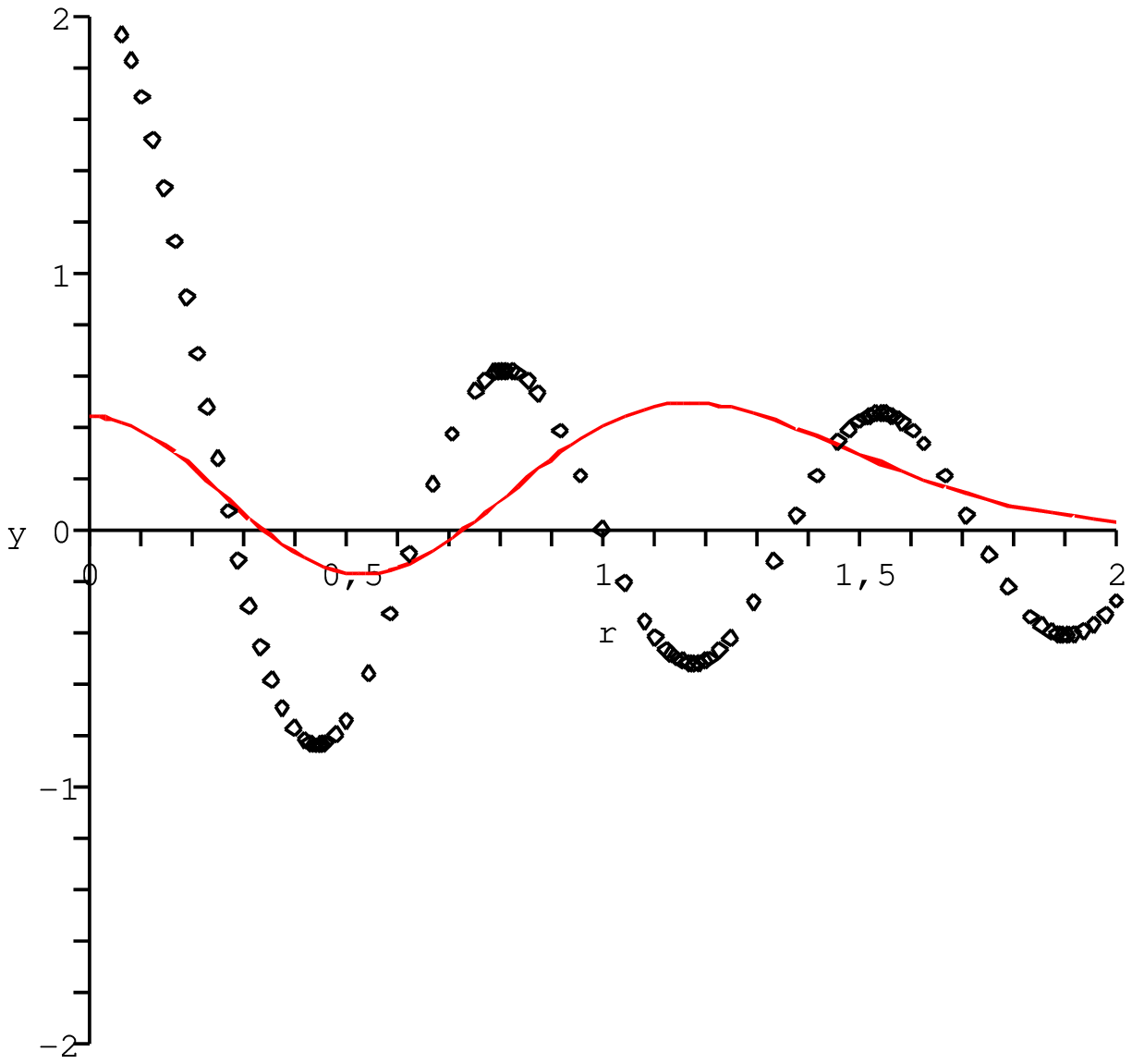}\epsfxsize=2.2
in\epsffile{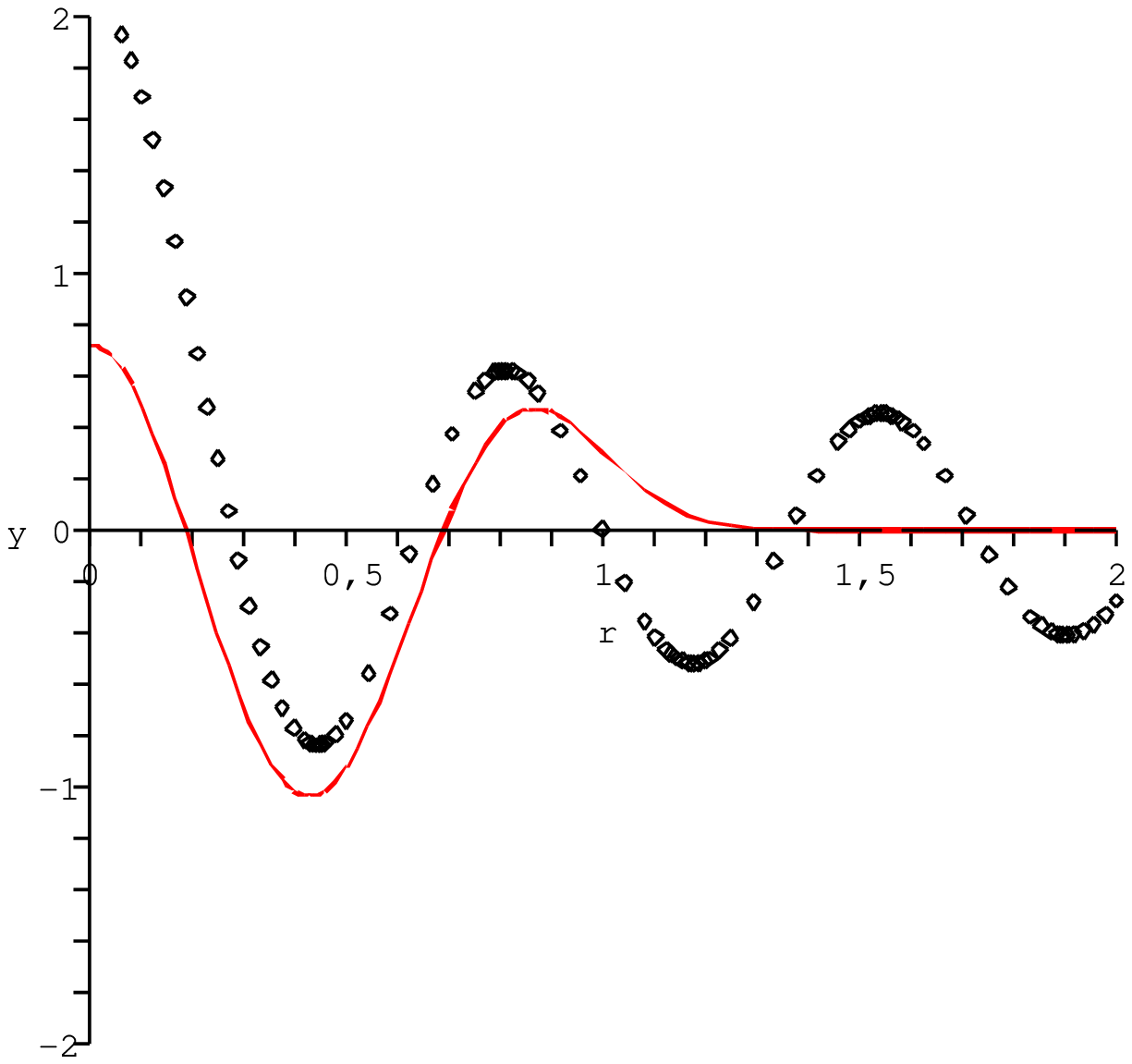}\epsfxsize=2.2
in\epsffile{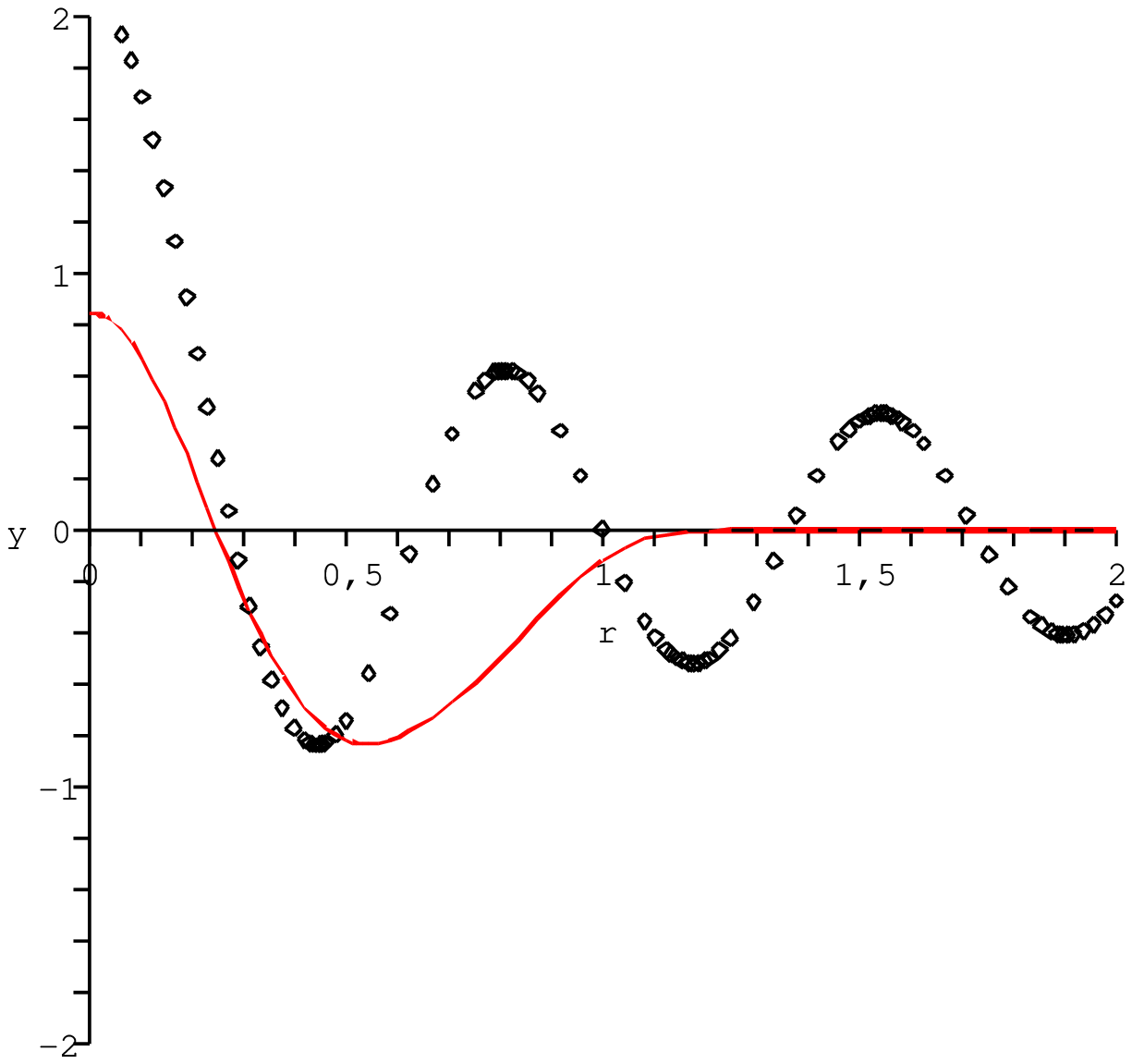}} \caption{\baselineskip=12pt {\it
Comparison of the radial shape for the
$\Phi\left(\lambda_{0,3}^{\left(N\right)}\right)$ fuzzy Bessel
(continuum line), the symbol of the eigenmatrix of the fuzzy
laplacian with respect to the eigenvalue
$\lambda_{0,3}^{\left(N\right)}$, with the function
$\psi_{0,3}\left(r,\varphi\right)$ that is the eigenfunction of
the continuum problem with eigenvalue $\lambda_{0,3}$. Here the
orders of truncation are $N=2,15,30$. The fuzzy Bessel converges to
zero outside the disc of radius $1$.}}
\bigskip
\label{FTDS}
\end{figure}

These functions will play a role similar to the role 'fuzzy
harmonics' play in the fuzzy sphere algebra: they will be seen as
a fuzzy version of the basis of the space of functions on the
disc.

Given a function on the disc, if it is square integrable,
$f\,\in\,\mathcal{L}^{2}\left(D,rdrd\varphi\right)$: \beq
f\left(r,\varphi\right)\,=\,\sum_{n=-\infty}^{+\infty}\,\sum_{k=1}^{\infty}\,
f_{nk}e^{in\varphi}J_{\left|n\right|}
\left(\sqrt{\lambda_{\left|n\right|,k}}r\right) \eeq it is
possible to truncate: \beq
f^{\left(N\right)}\left(r,\varphi\right)=\sum_{n=-N}^{+N}\sum_{k=1}^{N+1-\mid
n\mid}\,f_{nk}e^{in\varphi}J_{\left|n\right|}\left(\sqrt{\lambda_{\left|n\right|,k}}r\right)
\eeq This set of functions is a vector space, but it is no more an
algebra with respect to the pointwise product. The mapping \beq
\hat{f}^{\left(N\right)}_{\theta}\,=\,
\sum_{n=-N}^{+N}\sum_{k=1}^{N+1-\mid n\mid}\,f_{nk}
\hat{\Phi}\left(\lambda_{n,k}^{\left(N\right)}\right) \eeq define
a sequence (indexed by $N$) of finite rank matrix algebras, whose
formal limit is an abelian algebra because of the constraint
$N\theta=1$. This is defined a \emph{fuzzy disc}.

\subsection{Free Field Theory on the Fuzzy Disc: Green's functions}

The formalism developed lends itself readily for matrix
approximations to field theories on a disc \cite{fuzzyaction}. For
the real scalar case described by the action:
\be
S=\int d^2z\; \phi \nabla^2 \phi + \frac{m^2}{2}\phi^2 +V(\phi)  \ee 
a quantum version can be studied using the path integral
formalism. This path integral is ill defined, relying on the
concept of an integral with an infinite dimensional functional
measure. But the path integral can be made rigorous if the space of
admissible configurations for field variables is made finite. Such
is the case with this approximation. The fuzzy version of the
action is:
\be
S_{\theta}^{\left(N\right)}=\frac{1}{\pi} Tr\,\left[ \hat{\Phi}^{\left(N\right)}_{\theta} \nabla^2
\hat{\Phi}^{\left(N\right)}_{\theta} + \frac{m^2}{2}
\hat{\Phi}^{\left(N\right)}_{\theta}\cdot\hat{\Phi}^{\left(N\right)}_{\theta} +
V(\hat{\Phi}^{\left(N\right)}_{\theta})  \right] \ee 
A first analysis is the calculation of
the two points Green function for the free massless scalar theory.
In this case the path integral may be explicitly performed, yielding just the
inverse of the Laplacian with Dirichlet boundary conditions which
has been already computed. The fuzzy Laplacian is a map from
$\mathcal{A}^{\left(N\right)}_{\theta}$ to itself, so it can be
seen as a linear operator acting on the space
$\complex^{\left(N+1\right)^{2}}$, so to say a matrix belonging to
$\mathbb{M}_{\left(N+1\right)^{2}}$. Its inverse will be a matrix
belonging to the same space, that can be mapped into a two points
symbol via:
\be
G^{(N)}_\theta (z,z')= 4 \sum^N_{m,n,p,q =0}
\frac{e^{-{|z|^2+|z'|^2\over \theta}} (\nabla^{2})^{-1}_{mnpq}\bar z^p
z^q z'^m \bar z'^n }{\sqrt{p!q!m!n!\theta^{m+n+p+q}}}
\label{fuzzygre}
\ee
This expression can be evaluated numerically (figure~\ref{greenplot})
\begin{figure}[htbp]
\epsfxsize=2.5 in \centerline{\epsfxsize=2.5
in\epsffile{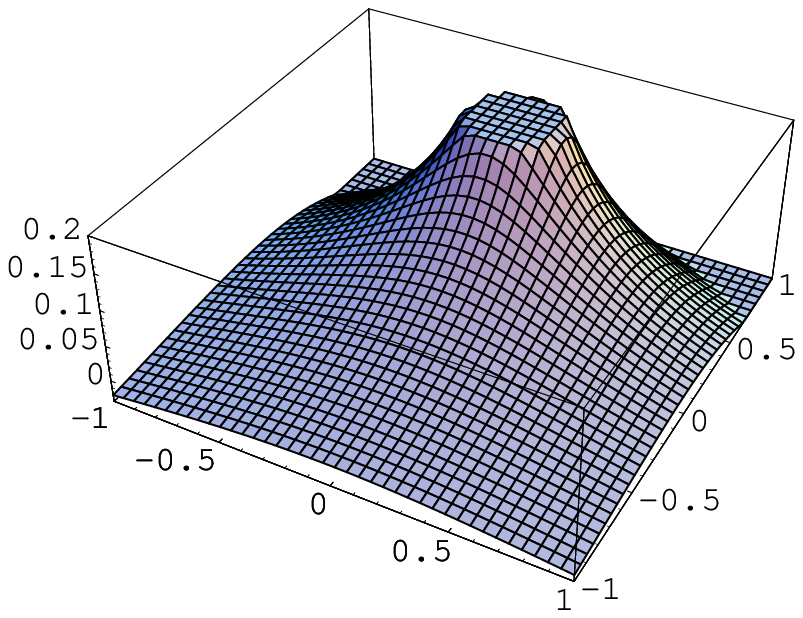} \epsfxsize=2.5
in\epsffile{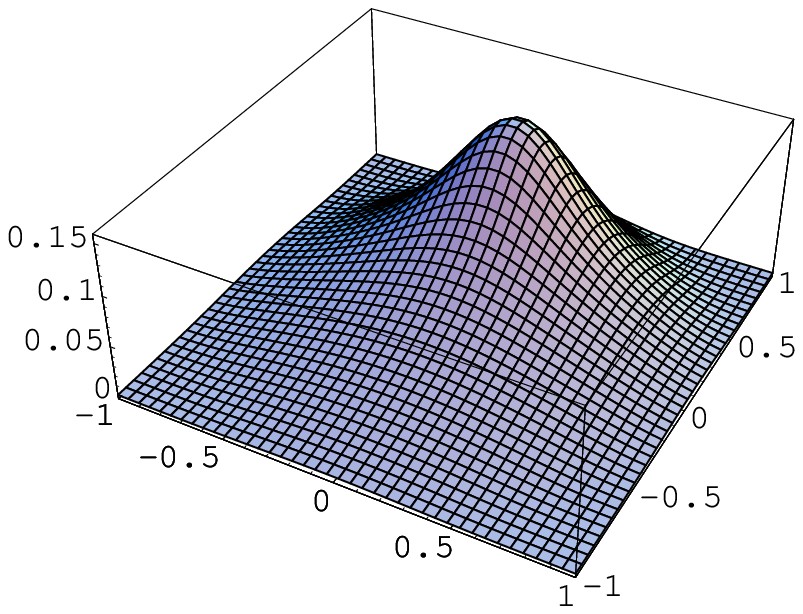}} \caption{\baselineskip=12pt {\it Comparison between the Green's
function $G(z,z')$ as function of $z$ for $z'=i/2$ On the left the
exact function (with the singularity at $z=z'$ truncated), on the
right the approximated function for $N=20$.
}}
\bigskip
\label{greenplot}
\end{figure}
and compared with the exact case, known from the classic theory of electrodynamics:
\be
G(z,z')= {1 \over 2 \pi} \ln \frac{|z-z'|}{|z'|z|-z|z|^{-1}|}
\label{exactgre}
\ee

The agreement is quite remarkable already for a
small value of $N$. The logarithmic divergence has been
smoothened by the ultraviolet cutoff, but apart from that the two
functions are quite similar. The choice of different values of
$z'$ gives similar pictures.

It is very interesting to note that, since fuzzy Laplacian is
represented by an hermitian finite dimensional matrix, its inverse
can be written in terms of a spectral decomposition. If a matrix,
say $M$, is hermitian, then its components satisfy the condition
$M_{ab}=M_{ba}^{*}$ in terms of transposition and complex
conjugation. Eigenvalue problem gives a number of real eigenvalues
equal to the dimension of the space on which the matrix acts: \beq
\sum_{b}M_{ab}v_{b}^{\left(k\right)}\,=\,\lambda^{\left(k\right)}v_{a}^{\left(k\right)}
\eeq here $v_{a}^{\left(k\right)}$ indicates the $a^{th}$
components of the eigenvector relative to the eigenvalue
$\lambda^{\left(k\right)}$. The inverse, if it exists, of the
matrix $M$ is a matrix $G$ whose components can be written as:
\beq
G_{sq}\,=\,\sum_{k}\,v_{s}^{\left(k\right)}v_{q}^{\left(k\right)}/\lambda^{\left(k\right)}
\eeq This line can be translated in the specific problem under
analysis. The notion of eigenvector of components
$v_{a}^{\left(N\right)}$ with eigenvalue
$\lambda^{\left(k\right)}$ goes into that of fuzzy Bessel
$\widehat{\Phi}\left(\lambda^{\left(N\right)}_{n,k}\right)$
matrix, from which it is immediate to obtain the symbols. The
fuzzy Green function becomes: \beq
G^{\left(N\right)}_{\theta}\left(z,z^{\prime}\right)\,=\,
\sum_{n=-N}^{+N}\sum_{k=1}^{N+1-\left|n\right|}
\frac{\psi_{n,k}^{\left(N\right)}\left(z^{\prime}\right)^{*}
\psi_{n,k}^{\left(N\right)}\left(z\right)}
{\lambda^{\left(N\right)}_{|n|,k}} \eeq

This expression is exactly the same obtained by Madore in the case of a field theory on a fuzzy sphere: the role of
fuzzy harmonics is now played by fuzzy Bessels.

\appendix

\addcontentsline{toc}{chapter}{{Appendices}}

\chapter{}

\section{An elementary introduction to the theory of
$C^{*}$-algebras}\label{appendixcalg}

This appendix is meant to be an introduction to some
algebraic notions mentioned in the text. All the topics here
recollected are covered, in more complete form, in
\cite{landsmann} and \cite{NCG}.

Let $\mathcal{V}$ a vector space defined over $\complex$, the
field of complex numbers. A norm on the vector space $\mathcal{V}$
is a map:
$$\parallel\,\cdot\,\parallel\,:\,\mathcal{V}\,\mapsto\,\real$$
such that:
\begin{itemize}
\item
$\parallel\,v\,\parallel\,\geq\,0\,\,\,\,\forall\,v\,\in\,\mathcal{V}\,;\,
\parallel\,v\,\parallel\,=\,0\,\Leftrightarrow\,v\,=\,0$
\item
$\parallel\,\lambda
v\,\parallel\,=\,\left|\,\lambda\,\right|\,\parallel\,v\,\parallel\,;\,\forall\,\lambda\,\in\,\complex$
and $v\,\in\,\complex$
\item
$\parallel\,v+u\,\parallel\,\leq\,\parallel\,v\,\parallel\,+\,\parallel\,u\,\parallel\,$
(triangle inequality)
\end{itemize}
A norm on $\mathcal{V}$ defines a metric $d$ on $\mathcal{V}$ by
$d\left(v,u\right)\,\equiv\,\parallel\,v-u\,\parallel.$ A vector
space with a norm which is complete in the associated metric (in
the sense that every Cauchy sequence converges) is called a Banach
space. On a Banach space $\mathrm{B}$ a functional is a linear
map: $$\rho\,:\,\mathrm{B}\,\mapsto\,\complex$$ The norm of such a
functional is defined by: \beq
\parallel\rho\parallel\,\equiv\,\sup\,\{\left|\rho\left(v\right)\right|/\parallel
v\parallel\,;v\,\in\,\mathrm{B}\}\eeq A linear functional is
continuous if and only if it is bounded. The dual $\mathrm{B}^{*}$
space of a Banach space is the space of all functionals on
$\mathrm{B}$: it is an example of a Banach space.

On a vector space it is possible to define an inner product, a
map:
$$\la\cdot\mid\cdot\ra\,:\,\mathcal{V}\times\mathcal{V}\,\mapsto\,\complex$$
such that:
\begin{itemize}
\item it is bilinear in both entries
\item
$\la\,v\,\mid\,v\,\ra\,\geq\,0\,\,\,\forall\,v\,\in\,\mathcal{V}$.
Moreover $\la\,v\,\mid\,v\,\ra\,=\,0\,\,\Leftrightarrow\,v\,=\,0$
\item
it is skew-hermitian\footnote{The overline on the complex numbers
indicates the complex conjugation.} :
$\overline{\la\,v\,\mid\,u\,\ra}\,=\,\la\,u\,\mid\,v\,\ra\,\,\,\forall\,u\,\,\,v\,\in\,\mathcal{V}$
\end{itemize}
From these axioms, it is possible to derive the Cauchy-Schwarz
inequality:
$$\left|\la\,v\,\mid\,u\,\ra\right|^{2}\,\leq\,\la\,v\,\mid\,v\,\ra\,\la\,u\,\mid\,u\,\ra$$
An Hilbert space $\hil$ is a vector space with an inner product
which is complete in the associated norm: it is an example of a
Banach space. It can be proved to be completely characterised by
its dimension, i.e. the cardinality of an arbitrary orthogonal
basis.

On a Banach space $\mathrm{B}$ a bounded operator is a linear map:
$$\Ao\,:\,\mathrm{B}\,\mapsto\,\mathrm{B}$$ whose norm is defined
by: \beq
\parallel\Ao\parallel\,\equiv\,\sup\,\{\parallel\Ao\,v\parallel/\parallel
v\parallel\,;\,v\,\in\,\mathrm{B}\} \label{opnorm}\eeq The space
$\mathcal{B}\left(\mathrm{B}\right)$ of bounded operators on a
Banach space, with this definition of norm, is itself a Banach
space.

An algebra is a vector space with an associative bilinear
operation, the multiplication, which is compatible with the linear
structure. If $AB$ denotes such a multiplication: \beq
\left(A+B\right)C\,=\,AC\,+\,BC\,\,\,\,\,\,\,A\left(B+C\right)\,=\,AB\,+\,AC
\eeq A Banach space is defined to be a Banach algebra if it is an
algebra such that the multiplication is separately continuous in
each variable: \beq
\parallel\,AB\,\parallel\,\leq\,\parallel\,A\,\parallel\,\parallel\,B\,\parallel
\eeq The Banach space of bounded operators on a Banach space
$\mathcal{B}\left(\mathrm{B}\right)$ is equipped with such a
multiplication considering the product as the standard composition
of operators: so it becomes a Banach algebra.

An involution on an algebra is a real-linear map
$A\,\mapsto\,A^{*}$, satisfying, for every pair of algebra
elements and every complex scalar:
\begin{itemize}
\item
$A^{**}\,=\,A$
\item
$\left(AB\right)^{*}\,=\,B^{*}A^{*}$
\item
$\left(\lambda A\right)^{*}\,=\,\overline{\lambda}A^{*}$
\end{itemize}
A $*$-algebra is an algebra with an involution. The Banach algebra
$\mathcal{B}\left(\hil\right)$ of bounded operators on a Hilbert
space is a $*$-algebra if the involution is the operation
$\Ao\,\mapsto\,\Ao^{\dagger}$ that maps the operator $\Ao$ into
its adjoint $\Ao^{\dagger}$. Moreover, for this $*$-algebra there
is a relation between the involution and the notion of norm, that
is the operator norm. It can be proved that: \beq
\parallel\,\Ao\,\parallel\,=\,\parallel\,\Ao^{\dagger}\,\parallel
\eeq The importance of a relation between the notion of involution
and that of norm in a Banach algebra is such to motivate the
introduction of the notion of $C^{*}$-algebra.

A $C^{*}$-algebra is a Banach, complex, $*$-algebra, whose norm
satisfies, for every pair of elements: \beqa
\parallel\,AB\,\parallel&\leq&\parallel\,A\,\parallel\,\parallel\,B\,\parallel
\nonumber \\
\parallel\,AA^{*}\,\parallel&=&\parallel\,A\,\parallel^{2}
\eeqa The algebra of bounded operators on a Hilbert space is a
$C^{*}$-algebra.

A morphism between two $C^{*}$-algebras $\mathcal{A}$ and
$\mathcal{A}^{\prime}$ is a complex-linear map
$\phi\,:\,\mathcal{A}\,\mapsto\,\mathcal{A}^{\prime}$ such that
$\forall\,A\,,\,B\,\in\,\mathcal{A}$: \beqa
\phi\left(AB\right)&=&\phi\left(A\right)\,\phi\left(B\right)
\nonumber \\
\phi\left(A^{*}\right)&=&\left(\phi\left(A\right)\right)^{*} \eeqa
An isomorphism is a bijective morphism. Two $C^{*}$-algebras are
isomorphic if there is an isomorphism between them.

A unit in a $C^{*}$-algebra (or in a Banach algebra) is an element
$\idop$ such that, for every element $A$ of this algebra one has:
$$A\,=\,A\idop\,=\idop A$$ For a $C^{*}$-algebra\footnote{This
result is not valid for a Banach algebra, where the condition that
the norm of the identity element is $1$ must be put as a definition.}
this also implies that the norm of the identity is $1$. With a
unit, an algebra is called unital: an important result in this
context is that a nonunital Banach algebra can always be extended
to a unital Banach algebra.

A state on a $C^{*}$-algebra $\mathcal{A}$ is a linear map
$\sigma\,:\,\mathcal{A}\,\mapsto\,\complex$ which is positive (it
means that $\forall\,A\,\in\,\mathcal{A}$, one has
$\sigma\left(A^{*}A\right)\geq 0$) and of norm $1$. The set of
states $\mathcal{S}\left(\mathcal{A}\right)$ is given the weak
$*$-topology if the convergence of a sequence
$\{\sigma_{n}\}\,\in\,\mathcal{S}\left(\mathcal{A}\right)$ is
defined by $\lim_{n}\,\sigma_{n}\,\rightarrow\,\sigma$ if and only
if $\lim_{n}\,\sigma_{n}\left(A\right)\,=\,\sigma\left(A\right)$
for every $A\,\in\,\mathcal{A}$. Equipped with this topology, the
set of states $\mathcal{S}\left(\mathcal{A}\right)$ is a convex
set. Pure states are those which cannot be expressed as a convex
combination of states. If the $C^{*}$-algebra is unital, the set
of states can be proved to be compact.

If $X$ is a compact Hausdorff space, let $C\left(X\right)$ be the
space of all the complex continuous functions on $X$. If the
involution is given by the usual complex conjugation for elements
on $\complex$, and the norm is the so called $\sup$-norm: \beq
\parallel f\parallel_{\infty}\,\equiv\,\sup_{x\in
X}\,\left|f\left(x\right)\right| \label{supnorm}\eeq then
$C\left(X\right)$ is a unital $C^{*}$-algebra. If the space $X$ is
locally compact (each point has a compact neighborhood), then
$C_{0}\left(X\right)$ is the space of complex continuous functions
vanishing at infinity (this means that for each $\epsilon > 0$ there is a compact
subset $K\subset X$ such that
$\left|f\left(x\right)\right|<\epsilon$ for all $x$ outside $K$).
Equipped with the usual involution, and the $\sup$-norm, the space
$C_{0}\left(X\right)$ is a nonunital $C^{*}$-algebra.

The commutative GNS theorem proves that these two examples are
paradigmatic. For every commutative $C^{*}$-algebra $\mathcal{A}$,
there exists a locally compact space $X$ such that $\mathcal{A}$
is isometrically isomorphic to the $C^{*}$-algebra
$C_{0}\left(X\right)$. Here isometrically means that this
isomorphism preserves the norm of the elements of the algebras. If
the $\mathcal{A}$ is unital, then $X$ can be proved to be compact.
Moreover, $X$ is proved to be homeomorphic to the set of pure
states of the $C^{*}$-algebra $\mathcal{A}$. One could say that
any commutative $C^{*}$-algebra can be realised as the
$C^{*}$-algebra of complex valued functions over a locally compact
Hausdorff space.

The space $\mathcal{B}\left(\hil\right)$ of bounded operators on a
complex separable Hilbert space, with the involution given by the
adjoint operation, and the norm given by \ref{opnorm}, is a unital
$C^{*}$-algebra. The GNS theorem proves that, for every
noncommutative $C^{*}$-algebra $\mathcal{A}$, there exists an
Hilbert space $\hil$ such that $\mathcal{A}$ is isomorphic to a
norm closed $*$-subalgebra of $\mathcal{B}\left(\hil\right)$. The
constructive GNS proof of this theorem proceeds defining, starting
from a state $\sigma\,\in\,\mathcal{A}$, a separable Hilbert space
$\hil_{\sigma}$, and a representation (a $*$-morphism):
$$\pi_{\sigma}\,:\,\mathcal{A}\,\mapsto\,\mathcal{B}\left(\hil_{\sigma}\right)$$
This representation is cyclic: this means that in $\hil_{\sigma}$
there is a cyclic vector $\psi_{\sigma}$, that is a vector such
that the span of $\pi_{\sigma}\left(A\right)\cdot\psi_{\sigma}$
for all $A\,\in\,\mathcal{A}$, coincides with the whole
$\hil_{\sigma}$. With GNS representations
$\left(\pi_{\sigma},\hil_{\sigma}\right)$ it is possible to
introduce a universal representation $\pi_{\mathcal{U}}$, given by
their direct sum, on the space: \beq
\hil_{\mathcal{U}}\,\equiv\,\oplus_{\sigma\,\in\,\mathcal{S}\left(\mathcal{A}\right)}\,\hil_{\sigma}
\eeq It can be proved that the universal representation
$\left(\pi_{U},\hil_{U}\right)$ defines a $C^{*}$-algebra
isomorphism between $\mathcal{A}$ and
$\mathcal{B}\left(\hil_{\mathcal{U}}\right)$. In this approach, a
state $\sigma\,\in\,\mathcal{S}\left(\mathcal{A}\right)$ is a pure
state if and only if its associated GNS representation
$\left(\pi_{\sigma},\hil_{\sigma}\right)$ is irreducible. It is
then represented as a ray of the vector space $\hil_{\sigma}$.

\section{Fourier symplectic transform}\label{digrSFT} In the
standard approach, Fourier analysis starts considering an
integrable function on a vector space, equipped with the
translationally invariant Lebesgue measure,
$f\,\in\,\mathcal{L}^{1}\left(\real^{m},dz\right)$, so that
Fourier transform of $f$ is defined by: \beq
\check{f}\left(w\right)\,=\,\int\frac{dz}{\left(2\pi\right)^{m/2}}\,f\left(z\right)e^{-iw\cdot
z} \eeq This map can be inverted, modulo a measurability condition
for the the local variation of the function $f$: the inverse, the
Fourier antitransform, is given by: \beq
f\left(z\right)\,=\,\int\frac{dw}{\left(2\pi\right)^{m/2}}\,\check{f}\left(w\right)e^{iz\cdot
w} \eeq Plancherel theorem shows the way this map can be extended
to the space of square integrable functions, and that, on this
Hilbert space, it defines a unitary operator. It is usually said
that the two variables, in this example $z$ and $w$, are Fourier
conjugate. From a more geometrical point of view, this map is
defined via a unitary representation of the translation group
$\left(\real^{m},+\right)$. The integral kernel in the expressions
above is a phase factor, that defines a representation of the
unitary action of the linear functional $w$ on the vector $z$.
This kind of harmonic analysis perfectly fits with the general
theory of noncompact abelian groups: the dual space of such groups
coincides with the group itself. So,
$\left(\real^{m}\right)^{*}\,=\,\real^{m}$. The action of the $w$
coordinate on the $z$ coordinate is written in terms of a scalar
product, and the expression $z\cdot w$ can be seen as the image of
a symmetric 2-form on the pair of vectors $z$ and $w$, following
the natural identification.

In the setting described in the text, there is actually a vector space,
equipped with a symplectic, translationally invariant, 2-form,
$\left(L\simeq\real^{2n},\omega\right)$. It seems natural to
perform an harmonic analysis for this homogeneous space using
this very structure. Such a symplectic form can always be brought in
the canonical, Darboux, form, represented by a matrix
$\tilde{\omega}$\footnote{In the usual identification of
$\real^{2n}$ with the symplectic phase space of certain classical
dynamics $T^{*}\real^{n}$, with coordinates $\left(q^{a},p_{b}\right)$,
the matrix representing the canonical symplectic form is
$\tilde{\omega}=dq^{a}\,\wedge\,dp_{a}$.}, via an invertible transformation.
Moreover, such a $T$ is not uniquely determined by $\omega$. The composition of
$T$ with an arbitrary symplectic tranformation gives another transformation,
still reducing $\omega$ in the canonical form:
\beq
T^{t}\tilde{\omega} T\,=\,\omega
\eeq
The symplectic Fourier transform \cite{folland} is ($\left|T\right|$ is the determinant of $T$):
\beq
\tilde{f}\left(w\right)\,=\,\int\frac{dz}{\left(2\pi\right)^{n}}\left|T\right|
\,f\left(z\right)e^{-i\omega\left(z,w\right)}
\eeq This can be inverted:
\beq
f\left(z\right)\,=\,\int\frac{dw}{\left(2\pi\right)^{n}}\left|T\right|
\,\tilde{f}\left(w\right)e^{i\omega\left(z,w\right)}
\eeq
In this definition it is encoded the ambiguity in the realization of $T$.
This definition can be seen to be covariant for symplectic transformation. If $\Phi$ is a symplectic transformation in $\real^{2n}$, then it induces a transformation (the push-forward) in the set of functions on that space:
\beq
\Phi_{*}f\,=f\circ\Phi^{-1}
\eeq
or equivalently:
\beq
\widetilde{\left(\Phi_{*}f\right)}\,=\,\Phi_{*}\left(\tilde{f}\right)
\eeq

\section{Generalised coherent states}\label{CSappendix}

In the main text the concept of generalised coherent states has been extensively used. It has been used to describe the introduction of
a set of quantizer operators in chapter one, and to define, following the work of Berezin on quantization, maps from operators to
functions (symbols) on the sphere and on the plane in chapter three. The aim of this appendix is to briefly recollect
the main definitions and results, to make easier the reading of the main text. The main reference is \cite{perelomov}.

The analysis starts considering a Lie group $G$, and $\Uo\left(g\right)$ a unitary irreducible representation of this group on a Hilbert
space $\hil$. Chosen a \emph{fiducial} vector $\mid\psi_{0}\ra$ in $\hil$, one obtains a set of vectors for each element of the group,
acting on it with $\Uo\left(g\right)$:
\beq
\mid\psi_{g}\ra\,\equiv\,\Uo\left(g\right)\mid\psi_{0}\ra
\eeq
Two such vectors are considered \emph{equivalent} if they correspond, quantum-mechanically, to the same state, i.e. if they differ
by a phase. So $\mid\psi_{g}\ra\simeq\mid\psi_{g^{\prime}}\ra$ if $\mid\psi_{g}\ra\,=e^{i\phi\left(g,g^{\prime}\right)}
\mid\psi_{g^{\prime}}\ra$. This condition is equivalent to $\Uo\left(g^{\prime -1}g\right)\mid\psi_{0}\ra\,=
e^{i\phi\left(g,g^{\prime}\right)}\,\mid\psi_{0}\ra$. If $H$ is the subgroup of $G$ whose elements are represented, by $\Uo$, as
operators whose action on the fiducial vector is just a multiplication by a phase, then the equivalence relation is among points of $G$,
and the quotient is the space $G/H$. If $H$ is maximal, then it is called isotropy subgroup for the state $\mid\psi_{0}\ra$. Choosing a
representative $g\left(x\right)$ in each equivalence class $x\,\in\,X=G/H$ (which is a cross section of the fiber bundle $G$ with base
$X$) defines a set of vectors on $\hil$, depending, clearly, on $G$ and $\mid\psi_{0}\ra$. This set of states is called a
\emph{system of coherent states} for $G$. As it has been presented in section \ref{rieffelsection}, the state corresponding to the vector
$\mid x\ra$ may be considered as the range of a rank one projector in $\hil$. Thus, the system of generalised coherent states determines
a set of one dimensional subspaces in $\hil$, parametrised by points of the homogeneous space $X=G/H$. An evolution of this analysis
drives naturally to the issue of overcompleteness for the system of coherent states, mentioned in (\ref{HWWcomplete}),
(\ref{HWcomplete}), (\ref{SU2complete}).

\chapter{}

\section{Product among symbols in the Weyl-Wigner isomorphism}

In this appendix there are recollected the calculations performed
following the definition of the Weyl-Wigner isomorphism between a
subset of functions in $\mathcal{F}\left(G\times\tilde{G}\right)$
($G$ is a compact simple Lie group, and $\tilde{G}$ is a discrete
set, whose values label the UIR's of the group, and the elements
of the basis chosen in each space for each \emph{inequivalent}
representation) and the set of Hilbert-Schmidt operators on the
space $\mathcal{H}=\mathcal{L}^{2}\left(G,d\mu\right)$.

In this Hilbert space, one can consider a set of generalized
states that constitute a basis, such that, if
$\psi\in\mathcal{H}$, then: $$\langle
g\mid\psi\rangle=\psi\left(g\right)$$ and satisfy an uncountable
form of completeness relation: $$\int\,d\mu\,\mid g\rangle\langle
g\mid\,=1 \,\,\,\,\,\,\langle g\mid
g^{\prime}\rangle\,=\,\delta\left(g^{\prime}g^{-1}\right)$$

Two sets of operators are defined in such a way that: \beq
\left(\hat{V}\left(g^{\prime}\right)\psi\right)\left(g\right)=\psi\left(g^{\prime-1}g\right)
\eeq \beq \left(\hat{U}\left(jmn\right)\psi\right)\left(g\right)=
D^{j}_{mn}\left(g\right)\psi\left(g\right) \eeq where
$D^{j}_{mn}\left(g\right)$ are the matrix elements of
the representative of group element $g$ in the representation
labelled by $j$.

Quantizer operators are (\ref{quantizers}):
\beq
\hat{W}\left(g,jmn\right)=\sum_{j^{\prime},m^{\prime},n^{\prime}}N_{j^{\prime}}\int\,d\mu^{\prime}
\hat{U}\left(j^{\prime}n^{\prime}m^{\prime}\right)\hat{V}\left(g^{\prime}\right)
D^{j}_{mn}\left(g^{\prime}\right)D^{j^{\prime}}_{m^{\prime}n^{\prime}}\left(g^{-1}s_{0}
\left(g^{\prime-1}\right)\right) \eeq then the isomorphism is
realized by:\beq
\hat{A}=\sum_{j,m,n}N_{j}\int_{G}d\mu\,A\left(g,jmn\right)\hat{W}\left(g,jnm\right)
\eeq \beq A\left(g,jmn\right)=Tr\,\hat{A}\hat{W}\left(g,jmn\right)
\eeq 
The product among symbols is (eq.\ref{Gstarprod}): \beq
\left(A*B\right)\left(g,\gamma\right)=Tr\hat{A}\hat{B}\hat{W}\left(g,\gamma\right)
\eeq 
\beq
\left(A*B\right)\left(g,\gamma\right)=\sum_{\tilde{\gamma}\,\check{\gamma}}
N_{\tilde{j}}N_{\check{j}}\int_{G}d\tilde{\mu}\int_{G}d\check{\mu}\,
A\left(\tilde{g},\tilde{\gamma}\right)B\left(\check{g},\check{\gamma}\right)\left[Tr\,
\hat{W}\left(g,\gamma\right)\hat{W}\left(\tilde{g},\tilde{\underline{\gamma}}\right)
\hat{W}\left(\check{g},\check{\underline{\gamma}}\right)\right]
\eeq As it has been stressed in the main text, this product is non
local, and the integral kernel is given by the trace term between
square brackets. To analyse this term, the first step is to study
the possibility of a kind of inversion of (\ref{quantizers}). From
the fact that $D$ functions are matrix elements of a UIR, one has:
\beqa
D^{j^{\prime}}_{m^{\prime}n^{\prime}}\left(g^{-1}s_{0}\left(g^{\prime-1}\right)\right)&=&
\sum_{s=1}^{N_{j^{\prime}}}
D^{j^{\prime}}_{m^{\prime}s}\left(g^{-1}\right)D^{j^{\prime}}
_{sn^{\prime}}\left(s_{0}\left(g^{\prime-1}\right)\right)
\nonumber \\ &=&
\sum_{s=1}^{N_{j^{\prime}}}\left(D^{j^{\prime}}_{sm^{\prime}}\left(g\right)\right)^{*}
D^{j^{\prime}}
_{sn^{\prime}}\left(s_{0}\left(g^{\prime-1}\right)\right)\eeqa and
then \beqa
\hat{W}\left(g,\gamma\right)=\sum_{\gamma^{\prime}}N_{j^{\prime}}\int_{G}\,dg^{\prime}&{}&
\hat{U}\left(j^{\prime},n^{\prime},m^{\prime}\right)\hat{V}\left(g^{\prime}\right)\cdot
\nonumber
\\ &{}&
\cdot D^{j}_{mn}\left(g^{\prime}\right)
\sum_{s=1}^{N_{j^{\prime}}}\left(D^{j^{\prime}}_{sm^{\prime}}\left(g\right)\right)^{*}
D^{j^{\prime}}
_{sn^{\prime}}\left(s_{0}\left(g^{\prime-1}\right)\right)
\nonumber\eeqa Since $D$ functions are an orthonormal basis for
$\mathcal{H}$, one can put: \beqa
\int_{G}dg\,\hat{W}\left(g,\gamma\right)D^{j^{\prime\prime}}
_{n^{\prime\prime}m^{\prime\prime}}\left(g\right)&=&\sum_{\gamma^{\prime}}\int_{G}dg^{\prime}\int_{G}dg\,
\hat{U}\left(j^{\prime},n^{\prime},m^{\prime}\right)\hat{V}\left(g^{\prime}\right)
D^{j}_{mn}\left(g^{\prime}\right)\cdot\nonumber \\
&\cdot&
\sum_{s=1}^{N_{j^{\prime}}}\left(D^{j^{\prime}}_{sm^{\prime}}\left(g\right)\right)^{*}
N_{j^{\prime}} D^{j^{\prime\prime}}
_{n^{\prime\prime}m^{\prime\prime}}\left(g\right)D^{j^{\prime}}
_{sn^{\prime}}\left(s_{0}\left(g^{\prime-1}\right)\right)\nonumber\eeqa
Integration on $dg$ in the RHS gives: \beq
\int_{G}dg\,\hat{W}\left(g,\gamma\right)D^{j^{\prime\prime}}
_{n^{\prime\prime}m^{\prime\prime}}\left(g\right)=\sum_{n^{\prime}}\int_{G}dg^{\prime}
\hat{U}\left(j^{\prime\prime},n^{\prime},m^{\prime\prime}\right)\hat{V}\left(g^{\prime}\right)
D^{j}_{mn}\left(g^{\prime}\right)
D^{j^{\prime\prime}}
_{n^{\prime\prime}n^{\prime}}\left(s_{0}\left(g^{\prime-1}\right)\right)
\eeq Again using orthonormality of $D$ functions: \beqa
\sum_{\gamma}\int_{G}dg\,\hat{W}\left(g,\gamma\right)&\cdot&D^{j^{\prime\prime}}
_{n^{\prime\prime}m^{\prime\prime}}\left(g\right)\left(D^{j}_{mn}\left(\tilde{g}\right)\right)^{*}N_{j}=
\nonumber
\\&\cdot&\sum_{n^{\prime}}\sum_{\gamma}\int_{G}dg^{\prime}
\hat{U}\left(j^{\prime\prime},n^{\prime},m^{\prime\prime}\right)\hat{V}\left(g^{\prime}\right)\cdot\nonumber
\\
&\cdot& D^{j}_{mn}\left(g^{\prime}\right)
\left(D^{j}_{mn}\left(\tilde{g}\right)\right)^{*}N_{j}
D^{j^{\prime\prime}}
_{n^{\prime\prime}n^{\prime}}\left(s_{0}\left(g^{\prime-1}\right)\right)
\eeqa Summation over discrete indices $\gamma$ gives a
$\delta\left(\tilde{g}g^{\prime-1}\right)$ factor in RHS: \beqa
\sum_{\gamma}\int_{G}dg\,\hat{W}\left(g,\gamma\right)&\cdot&D^{j^{\prime\prime}}
_{n^{\prime\prime}m^{\prime\prime}}\left(g\right)\left(D^{j}_{mn}\left(\tilde{g}\right)\right)^{*}N_{j}=
\nonumber \\ &=&\sum_{n^{\prime}}
\hat{U}\left(j^{\prime\prime},n^{\prime},m^{\prime\prime}\right)\hat{V}\left(\tilde{g}\right)
D^{j^{\prime\prime}}
_{n^{\prime\prime}n^{\prime}}\left(s_{0}\left(\tilde{g}^{-1}\right)\right)
\eeqa Now one can even saturate the index $n^{\prime\prime}$: \beqa
\sum_{n^{\prime\prime}}
\sum_{\gamma}\int_{G}dg\,\hat{W}\left(g,\gamma\right)&\cdot&
D^{j^{\prime\prime}}
_{n^{\prime\prime}m^{\prime\prime}}\left(g\right)\left(D^{j}_{mn}\left(\tilde{g}\right)\right)^{*}N_{j}
D^{j^{\prime\prime}}_{kn^{\prime\prime}}\left(s_{0}\left(\tilde{g}\right)\right)=\nonumber
\\ &=& \sum_{n^{\prime}}
\hat{U}\left(j^{\prime\prime},n^{\prime},m^{\prime\prime}\right)\hat{V}\left(\tilde{g}\right)
D^{j^{\prime\prime}}
_{kn^{\prime}}\left(e\right)=\nonumber \\ &=&
\hat{U}\left(j^{\prime\prime},k,m^{\prime\prime}\right)\hat{V}\left(\tilde{g}\right)
\eeqa the last equality comes from
$D^{j^{\prime\prime}}
_{kn^{\prime}}\left(e\right)=\delta_{kn^{\prime}}$ since $e$ is
the identity of the group $G$. Finally one has: \beq
\hat{U}\left(\tilde{j},\tilde{n},\tilde{m}\right)\hat{V}\left(\tilde{g}\right)=\sum_{\gamma}\int_{G}dg
\,\hat{W}\left(g,\gamma\right)D^{\tilde{j}}_{\tilde{n}\tilde{m}}\left(s_{0}\left(\tilde{g}\right)g\right)
\left(D^{j}_{mn}\left(\tilde{g}\right)\right)^{*}N_{j}\label{UVopinverse}\eeq
This can be seen as a sort of antitransform of (\ref{quantizers}).

The second step of the analysis just gives the composition
properties of $\hat{U}$and $\hat{V}$ operators: \beq
\hat{U}\left(j^{\prime},n^{\prime},m^{\prime}\right)
\hat{U}\left(j^{\prime\prime},n^{\prime\prime},m^{\prime\prime}\right)=\sum_{JNM,\lambda}
C^{j^{\prime},j^{\prime\prime},J\,\lambda}
_{n^{\prime}m^{\prime},n^{\prime\prime}m^{\prime\prime},NM}\hat{U}\left(J,N,M\right)\eeq
\beq
\hat{U}\left(j^{\prime},n^{\prime},m^{\prime}\right)\hat{V}\left(g^{\prime}\right)
\hat{U}\left(j^{\prime\prime},n^{\prime\prime},m^{\prime\prime}\right)\hat{V}\left(g^{\prime\prime}\right)
=\sum_{k=1}^{N_{j^{\prime\prime}}}\sum_{JNM,\lambda}
D^{j^{\prime\prime}}_{n^{\prime\prime}k}\left(g^{\prime-1}\right)
C^{j^{\prime},j^{\prime\prime},J\,\lambda}
_{n^{\prime}m^{\prime},km^{\prime\prime},NM}\hat{U}
\left(J,N,M\right)\hat{V}\left(g^{\prime}g^{\prime\prime}\right)
\eeq
where the meaning of the coefficient has been explained in the main text.

The third step is to study the composition properties in the set of
$\hat{W}$. From the last relations, it is easy to see that:
\beqa
\hat{W}\left(g,\gamma\right)\hat{W}\left(\tilde{g},\tilde{\gamma}\right)&=&
\sum_{\gamma^{\prime}\,\gamma^{\prime\prime}}N_{j^{\prime}}N_{j^{\prime\prime}}\int_{G}dg^{\prime}
\int_{G}dg^{\prime\prime}\sum_{k=1}^{N_{j^{\prime\prime}}}\sum_{JNM,\lambda}\cdot\nonumber
\\
&\cdot& D^{j^{\prime\prime}}_{n^{\prime\prime}k}\left(g^{\prime-1}\right)
C^{j^{\prime},j^{\prime\prime},J\,\lambda}
_{n^{\prime}m^{\prime},km^{\prime\prime},NM}
\hat{U}\left(J,N,M\right)\hat{V}\left(g^{\prime}g^{\prime\prime}\right)\cdot\nonumber
\\
&\cdot& D^{j}_{mn}\left(g^{\prime}\right)
D^{j^{\prime}}_{m^{\prime}n^{\prime}}\left(g^{-1}s_{o}\left(g^{\prime-1}\right)\right)
D^{\tilde{j}}_{\tilde{m}\tilde{n
}}\left(g^{\prime\prime}\right)
D^{j^{\prime\prime}}_{m^{\prime\prime}n^{\prime\prime}}\left(\tilde{g}^{-1}s_{o}
\left(g^{\prime\prime-1}\right)
\right)\nonumber
\eeqa
Substitution of (\ref{UVopinverse}) into the RHS of the previous
relation gives:
\beqa
\hat{W}\left(g,\gamma\right)\hat{W}\left(\tilde{g},\tilde{\gamma}\right)&=&
\sum_{\gamma^{\prime}\,\gamma^{\prime\prime}\,\gamma^{\prime\prime\prime}}
N_{j^{\prime}}N_{j^{\prime\prime}}N_{j^{\prime\prime\prime}}
\int_{G}dg^{\prime}
\int_{G}dg^{\prime\prime}\int_{G}dg^{\prime\prime\prime}
\sum_{k=1}^{N_{j^{\prime\prime}}}\sum_{JNM,\lambda}\cdot\nonumber
 D^{j^{\prime\prime}}_{n^{\prime\prime}k}\left(g^{\prime-1}\right)
\\
&\cdot&
C^{j^{\prime},j^{\prime\prime},J\,\lambda}
_{n^{\prime}m^{\prime},km^{\prime\prime},NM}
\hat{W}\left(g^{\prime\prime\prime},\gamma^{\prime\prime\prime}\right)
D^{J}_{NM}\left(s_{o}\left(g^{\prime}g^{\prime\prime}\right)g^{\prime\prime\prime}\right)
\left(D^{j^{\prime\prime\prime}}_{m^{\prime\prime\prime}n^{\prime\prime\prime}}
\left(g^{\prime}g^{\prime\prime}\right)\right)^{*}\cdot\nonumber
\\
&\cdot& D^{j}_{mn}\left(g^{\prime}\right)
D^{j^{\prime}}_{m^{\prime}n^{\prime}}\left(g^{-1}s_{o}\left(g^{\prime-1}\right)\right)
\cdot\nonumber
\\
&\cdot&D^{\tilde{j}}_{\tilde{m}\tilde{n
}}\left(g^{\prime\prime}\right)
D^{j^{\prime\prime}}_{m^{\prime\prime}n^{\prime\prime}}\left(\tilde{g}^{-1}
s_{o}\left(g^{\prime\prime-1}\right) \right)\label{mulWW}
\eeqa
The definition (\ref{Gstarprod}) indicates that the problem is evaluating the trace of the product of three
$\hat{W}$ operators. One has:
\beqa
Tr\left[\hat{W}\left(g,\gamma\right)\hat{W}\left(\tilde{g},\tilde{\gamma}\right)
\hat{W}\left(\check{g},\check{\gamma}\right)\right]&=&
\sum_{\gamma^{\prime}\,\gamma^{\prime\prime}\,\gamma^{\prime\prime\prime}}
N_{j^{\prime}}N_{j^{\prime\prime}}N_{j^{\prime\prime\prime}}
\int_{G}dg^{\prime}
\int_{G}dg^{\prime\prime}\int_{G}dg^{\prime\prime\prime}
\sum_{k=1}^{N_{j^{\prime\prime}}}\sum_{JNM,\lambda}\cdot\nonumber
\\
&\cdot& D^{j^{\prime\prime}}_{n^{\prime\prime}k}\left(g^{\prime-1}\right)
C^{j^{\prime},j^{\prime\prime},J\,\lambda}
_{n^{\prime}m^{\prime},km^{\prime\prime},NM}
D^{J}_{NM}\left(s_{o}\left(g^{\prime}g^{\prime\prime}\right)g^{\prime\prime\prime}\right)
\cdot\nonumber
\\
&\cdot&
\left(D^{j^{\prime\prime\prime}}_{m^{\prime\prime\prime}n^{\prime\prime\prime}}
\left(g^{\prime}g^{\prime\prime}\right)\right)^{*}
D^{(j}_{mn}\left(g^{\prime}\right)
D^{j^{\prime}}_{m^{\prime}n^{\prime}}\left(g^{-1}s_{o}\left(g^{\prime-1}\right)\right)
\cdot\nonumber
\\
&\cdot& D^{\tilde{j}}_{\tilde{m}\tilde{n
}}\left(g^{\prime\prime}\right)
D^{j^{\prime\prime}}_{m^{\prime\prime}n^{\prime\prime}}\left(\tilde{g}^{-1}
s_{o}\left(g^{\prime\prime-1}\right)
\right)\cdot\nonumber
\\
&\cdot&
Tr\left[\hat{W}\left(g^{\prime\prime\prime},\gamma^{\prime\prime\prime}\right)
\hat{W}\left(\check{g},\check{\gamma}\right)\right]
\eeqa
Since:
\beq
Tr\left[\hat{W}\left(g^{\prime\prime\prime},\gamma^{\prime\prime\prime}\right)
\hat{W}\left(\check{g},\check{\gamma}\right)\right]=\frac{1}{N_{j^{\prime\prime\prime}}}
\delta_{j^{\prime\prime\prime}\check{j}}\delta_{m^{\prime\prime\prime}\check{n}}
\delta_{n^{\prime\prime\prime}\check{m}}\delta\left(\check{g}^{-1}g^{\prime\prime\prime}\right)
\eeq
So one obtains the final expression for the integral kernel of the star product (\ref{Gstarprod}):
\beqa
Tr\left[\hat{W}\left(g,\gamma\right)\hat{W}\left(\tilde{g},\tilde{\underline{\gamma}}\right)
\hat{W}\left(\check{g},\check{\underline{\gamma}}\right)\right]&=&
\sum_{\gamma^{\prime}\,\gamma^{\prime\prime}}\sum_{\Gamma,\lambda}
\int_{G}dg^{\prime}\int_{G}dg^{\prime\prime}\sum_{k=1}^{N_{j^{\prime\prime}}}
N_{j^{\prime}}N_{j^{\prime\prime}}D^{j^{\prime\prime}}
_{n^{\prime\prime}k}\left(g^{\prime-1}\right)
\cdot\nonumber
\\
&\cdot&
C^{j^{\prime},j^{\prime\prime},J\,\lambda}
_{n^{\prime}m^{\prime},km^{\prime\prime},NM}
D^{J}_{NM}\left(s_{o}\left(g^{\prime}g^{\prime\prime}\right)\check{g}\right)
\left(D^{\check{j}}_{\check{m}\check{n}}\left(g^{\prime}g^{\prime\prime}\right)\right)^{*}\cdot\nonumber
\\
&\cdot& D^{j}_{mn}\left(g^{\prime}\right)
D^{j^{\prime}}_{m^{\prime}n^{\prime}}\left(g^{-1}s_{o}\left(g^{\prime-1}\right)\right)
D^{\tilde{j}}_{\tilde{n}\tilde{m
}}\left(g^{\prime\prime}\right)\cdot\nonumber
\\
&\cdot&
D^{j^{\prime\prime}}_{m^{\prime\prime}n^{\prime\prime}}\left(\tilde{g}^{-1}
s_{o}\left(g^{\prime\prime-1}\right)
\right)
\eeqa
(here $\Gamma$ is a short for $\left(J,N,M\right)$)

%
%

\begin{flushright}
{{\Large\bf acknowledgements}}
\end{flushright}

{\bf to my travelmates...}

Vorrei ringraziare Beppe e Fedele - il prof.Marmo e il prof.Lizzi - perch\`e nella mia rivoluzione culturale sono stati un grande timoniere e un piccolo timoniere: un grazie enorme per tutte le volte in cui mi sono rivolto a loro cercando una figura materna, e loro si sono assunti la responsabilit\`a di rivolgersi a me assumendo un ruolo paterno. 

Vorrei ringraziare Bala - il prof.Balachandran -  che mi ha accolto a Syracuse, invitandomi a discutere di fisica con lui nella stanza 
316, e raccontandomi meravigliose storie della sua terra. 

Vorrei ringraziare Patrizia - la Dr.Vitale -  per la gentilezza che ha intessuto nei nostri dialoghi, e poi Zac, Alberto, Franco - il prof.Zaccaria, il prof.Simoni, il Dr.Ventriglia - per il sorriso e la semplicit\`a con la quale mi hanno ascoltato. 

Vorrei ringraziare Rodolfo e GianFausto - il prof.Figari e il prof.Dell'Antonio - per la naturalezza con la quale mi hanno incoraggiato. 

Vorrei ringraziare Gianni - il prof.Landi - per avermi dato la libert\`a di raccontargli le mie storie tre anni fa, all'inizio di una avventura, e poi poche settimane fa, negli stessi luoghi, al compimento di questa stessa avventura. 

Vorrei ringraziare Al, Pepe, Joe, Mukunda - il prof.Stern, il prof.Gracia-Bondia, il prof.Varilly, il prof.Mukunda - per la sincerit\`a con la quale mi hanno narrato le loro storie, durante i loro soggiorni a Napoli.

Vorrei ringraziare Giovanna, Rebecca, Alessandro e Antonella. Loro nei loro nomi, nei loro sorrisi. 

Vorrei ringraziare tutti i miei compagni di viaggio dottorandi del Dipartimento, perch\`e insieme il nostro grande spazio \`e divenuto una agor\`a, e Guido, per l'umilt\`a con la quale ha semplificato la mia interazione con l'amministrazione~della burocrazia universitaria.

To all of you...many many thanks!

\end{document}